\documentclass[11pt,letterpaper,twoside]{article}

\usepackage[T1]{fontenc}
\usepackage[utf8]{inputenc}
\usepackage{lmodern}
\usepackage[margin=1in,headheight=14pt,headsep=14pt]{geometry}
\usepackage{bbm}
\usepackage{amssymb,mathtools,bm,amsthm}
\usepackage{natbib}
\bibpunct[, ]{(}{)}{,}{a}{}{,}

\usepackage{booktabs,array,enumitem,chngcntr}
\usepackage{algorithm,algpseudocode}
\usepackage{graphicx,subcaption,float}
\usepackage{microtype,url,xcolor,extarrows}
\usepackage{multirow,makecell,colortbl,pifont,xspace}
\usepackage{titlesec,fancyhdr}
\usepackage[hidelinks,hypertexnames=false]{hyperref}
\usepackage{cleveref}

\newtheoremstyle{preprintresult}{7pt}{7pt}{\itshape}{\parindent}{\scshape}{.}{0.5em}{}
\theoremstyle{preprintresult}
\newtheorem{theorem}{Theorem}
\newtheorem{proposition}{Proposition}
\newtheorem{corollary}{Corollary}
\newtheorem{lemma}{Lemma}
\newtheorem{assumption}{Assumption}
\newtheoremstyle{preprintmodel}{7pt}{7pt}{\normalfont}{\parindent}{\scshape}{.}{0.5em}{}
\theoremstyle{preprintmodel}
\newtheorem{model}{Model}

\newcommand{\Halmos}{\unskip\nobreak\hfill\ensuremath{\square}\par}

\makeatletter
\renewcommand{\normalsize}{%
  \@setfontsize\normalsize{11pt}{17pt}%
  \abovedisplayskip 7pt plus 2pt minus 2pt
  \belowdisplayskip 7pt plus 2pt minus 2pt
  \abovedisplayshortskip 0pt plus 1pt
  \belowdisplayshortskip 5pt plus 1pt minus 1pt
  \let\@listi\@listI
}
\normalsize
\providecommand*{\theHALG@line}{}
\renewcommand*{\theHALG@line}{\thealgorithm.\arabic{ALG@line}}
\renewcommand{\@maketitle}{%
  \newpage\null\vskip8pt
  \begin{center}
  {\fontsize{20}{23}\selectfont\@title\par}
  \vskip12pt
  {\normalsize\@author\par}
  \end{center}\vskip12pt
}
\makeatother
\renewenvironment{abstract}{\par\begingroup\fontsize{10}{14}\selectfont\noindent\textbf{Abstract.}\ }{\par\endgroup\medskip}
\titleformat{\section}{\large\bfseries}{\thesection.}{0.6em}{}
\titleformat{\subsection}{\normalsize\bfseries}{\thesubsection.}{0.6em}{}
\titleformat{\subsubsection}[runin]{\normalsize\bfseries}{\thesubsubsection.}{0.6em}{}
\titlespacing*{\section}{0pt}{16pt plus 3pt minus 2pt}{7pt}
\titlespacing*{\subsection}{0pt}{12pt plus 2pt minus 2pt}{5pt}
\titlespacing*{\subsubsection}{\parindent}{9pt plus 2pt minus 1pt}{0.5em}
\fancypagestyle{plain}{\fancyhf{}\fancyfoot[C]{\thepage}}
\hypersetup{
  pdftitle={Distributions or Quantiles? A Convex Variational Theory of Worst-Case Risk},
  pdfauthor={Jianjun Gao, Jonathan Li, Zhen Wang, Yang Zhang},
  pdfsubject={Worst-case spectral risk under distributional ambiguity}
}

\title{Distributions or Quantiles? A Convex Variational Theory of Worst-Case Risk}
\author{Jianjun Gao\textsuperscript{a}, Jonathan Yu-Meng Li\textsuperscript{b}, Zhen Wang\textsuperscript{a}, and Yang Zhang\textsuperscript{a}\\[6pt]
{\fontsize{9}{12}\selectfont
\textsuperscript{a}School of Information Management and Engineering,\\
Shanghai University of Finance and Economics, China\\[3pt]
\textsuperscript{b}Telfer School of Management, University of Ottawa, Canada\\[5pt]
Jianjun Gao: \href{mailto:gao.jianjun@shufe.edu.cn}{gao.jianjun@shufe.edu.cn};
Jonathan Yu-Meng Li: \href{mailto:Jonathan.Li@telfer.uottawa.ca}{Jonathan.Li@telfer.uottawa.ca}\\
Zhen Wang: \href{mailto:wangzhen@sufe.edu.cn}{wangzhen@sufe.edu.cn};
Yang Zhang: \href{mailto:2024214707@stu.sufe.edu.cn}{2024214707@stu.sufe.edu.cn}\endgraf}}
\date{}

\begin{document}
\maketitle
\begin{abstract}
Worst-case risk evaluation is central to finance and operations research, yet existing analytical results remain fragmented and largely confined to ambiguity sets with specific structures. Using spectral risk measures as the underlying class, we develop a convex variational theory that unifies these analyses and extends them to richer ambiguity models. The theory rests on a representation principle: use the quantile or distribution formulation that preserves convexity of the full ambiguity set. In both representations, the resulting optimality conditions provide a systematic route to identify the worst-case law and determine its risk value. This approach recovers classical central-moment bounds and yields closed-form or semi-closed-form solutions for practically motivated models involving Wasserstein ambiguity with tail stress, two-sided expected shortfall (ES) bands, asymmetric partial moments, geometric-moment bounds, and exponential- or Gaussian-reference Kullback--Leibler (KL) neighborhoods. Beyond these analytical results, the same theory yields projection-based algorithms with fixed-grid convergence under the stated assumptions. In a high-dimensional portfolio application, we establish an exact projection identity for mean-anchored joint KL ambiguity under a Gaussian reference. The resulting standardized worst-case ES coefficient applies to every portfolio, reducing robust portfolio selection to an exact second-order cone program after a single calibration.
\end{abstract}
\noindent{\small\textit{Keywords:} Worst-case risk; spectral risk measures; convex variational analysis; distributional ambiguity; distributionally robust optimization.}
\par\medskip\noindent\rule{\textwidth}{0.4pt}\par

\section{Introduction}
Central to finance and operations research is the problem of quantifying risk, a task greatly advanced by the modern theory of coherent and law-invariant risk measures \citep{artzner1999coherent,kusuoka2001law}. The key challenge, however, is that these measures must often be evaluated when the underlying distribution is only partially known, giving rise to worst-case risk: the highest risk level over all distributions consistent with the available information. This challenge has led to an extensive analytical literature on worst-case risk across ambiguity sets defined by moment and shape information \citep{ghaoui2003worst,popescu2005semidefinite,zhu2009worst,natarajan2010tractable,delage2010distributionally,li2018closed,shao2023extreme}, Wasserstein neighborhoods \citep{pichler2013evaluations,gao2017wasserstein,bernard2024robust}, and divergence constraints \citep{ben2013robust,hu2012kullback,glasserman2014robust,lam2016robust}.

Despite these advances, explicit characterizations of worst-case laws often rely on bounds or constructions tailored to particular ambiguity structures. General optimization methods can accommodate richer combinations of distributional information, but deriving explicit extremal laws for these combinations remains challenging. A Wasserstein neighborhood, for example, can represent uncertainty around a data-based reference distribution, while an expected shortfall (ES) floor imposes a prescribed tail stress. Moment information can likewise be supplemented by a two-sided ES band, encoding both a minimum stress severity and a maximum plausible tail risk \citep{bcbs2017stress,breuer2009plausible}. Exact partial moments provide another example: their equality constraints are affine in a density but generally nonconvex in a quantile. Such ambiguity sets combine statistical evidence with stress scenarios and expert or regulatory judgments and motivate the paired variational formulations developed here.

We address this challenge by developing a unified convex variational theory of worst-case risk, using spectral risk measures (SRMs) as the underlying class. This class includes expected loss, ES, mixtures of ES, and measures generated by smooth spectra, all of which admit quantile and distribution representations \citep{acerbi2002spectral,artzner1999coherent,kusuoka2001law}. The theory rests on a representation principle: use the quantile or distribution formulation that preserves convexity of the full ambiguity set. Quantile space preserves convexity for rank-based and tail constraints and one-dimensional Wasserstein neighborhoods, whereas distribution space does so for exact generalized-moment constraints, relative-entropy budgets, and density-divergence neighborhoods. Some restrictions, however, including means and upper bounds on convex moment functionals, are convex in both representations. Here the appropriate choice depends on the accompanying information. For example, the same mean constraint is naturally handled in quantile space when paired with ES or Wasserstein restrictions, and in distribution space when paired with exact partial moments or density divergences. The representation is therefore chosen for the ambiguity set as a whole, providing a coherent rule for combining heterogeneous information.

With convexity secured by this choice, we make convex variational analysis operational for worst-case risk in both representations. Under the stated regularity and constraint qualifications, we derive necessary and sufficient function-space optimality systems that characterize the extremal quantile or distribution and hence the worst-case value. Crucially, these systems do more than verify a conjectured extremal law: stationarity identifies its functional form, the ambiguity conditions calibrate its parameters, and normal-cone conditions enforce the defining shape of a quantile or cumulative distribution function (CDF). In quantile space, pointwise stationarity either yields a valid quantile directly or the normal cone determines where monotonicity binds. In distribution space, stationarity determines the CDF on its increasing region, with boundary signs certifying its support, or yields a global density-ratio equation for positive-density candidates under Kullback--Leibler (KL) ambiguity with a strictly positive entropy multiplier.

These systems recover existing results and extend the analysis to richer ambiguity models within the same variational framework. In quantile space, they recover classical central-moment bounds and yield closed-form or semi-closed-form solutions for 2-Wasserstein ambiguity with ES stress and mean--variance (MV) ambiguity with a two-sided ES band. The ES-band model is especially revealing: a floor that excludes the MV optimizer preserves monotonicity of the stationary candidate, whereas a cap that excludes it activates the normal cone, forces a flat segment in the worst-case quantile, and therefore creates an atom in the extremal law. In distribution space, the systems yield explicit CDF constructions for asymmetric partial moments and geometric moments (Section~\ref{subsec:geometric_moment}), as well as two-piece density tilts for ES under mean-anchored KL ambiguity with exponential or Gaussian references. Across these cases, the infinite-dimensional search reduces to identifying the extremal shape and solving a finite-dimensional calibration problem, rather than constructing a new bound for each ambiguity set.

When no explicit analytical solution is available, the same systems remain constructive: their normal-cone inclusions provide the basis for projected primal--dual procedures. Isotonic regression enforces quantile monotonicity, while density-simplex projection enforces distributional admissibility. We give proximal methods of multipliers for smooth convex discretizations and prove fixed-grid convergence when a Karush--Kuhn--Tucker (KKT) point exists and the inner optimality errors are summable.

Beyond risk evaluation, the variational theory extends to high-dimensional distributionally robust optimization. We establish that a mean-anchored joint KL ball around a multivariate Gaussian reference projects exactly, with unchanged radius, onto every portfolio-loss direction. This identity reduces the distributional inner problem for every portfolio to the same standardized univariate worst-case ES problem. Its ambiguity coefficient is therefore calibrated only once, after which robust ES portfolio selection reduces to a second-order cone program (SOCP). The result shows how a structural characterization of worst-case laws can convert a high-dimensional distributionally robust decision problem into a tractable deterministic optimization model.

\begin{table}[htp!]
\centering
\footnotesize
\setlength{\tabcolsep}{3pt}
\renewcommand{\arraystretch}{0.9}
\caption{Univariate ambiguity models, support restrictions, and analytical solution mechanisms.}
\begin{tabular}{@{} l c c l l @{}}
\toprule
\textbf{Ambiguity set} & \textbf{Space} & \textbf{Support} & \textbf{Mechanism} & \textbf{Result} \\
\midrule
Central moment (benchmark) & Quantile & Not fixed & Pointwise stationarity & Theorem~\ref{theo-1,k} \\
Wasserstein and ES stress & Quantile & Not fixed & Pointwise stationarity & Theorem~\ref{theo-cw} \\
\shortstack[l]{Mean, variance budget,\\two-sided ES band} & Quantile & Not fixed & Stationarity/normal cone & Theorem~\ref{thm:band-general} \\
Partial moments & Distribution & Bounded $[a,b]$ & Stationarity/boundary signs & Theorem~\ref{theo-LU} \\
Geometric bound & Distribution & Bounded $[a,b]$ & Stationarity/boundary signs & Theorem~\ref{theo-log} \\
\shortstack[l]{Mean-anchored KL:\\exponential reference} & Distribution & Bounded $[0,b]$ & Global stationarity & Theorem~\ref{theo-KL} \\
\shortstack[l]{Mean-anchored KL:\\Gaussian reference} & Distribution & Unbounded $\mathbb{R}$ & Global stationarity & Theorem~\ref{theo-KL-normal} \\
\bottomrule
\end{tabular}
\par\smallskip
\parbox{\textwidth}{\footnotesize Support refers to the admissible laws; ``not fixed'' means no finite support bounds are imposed. The exponential KL model also has a conditional extension to $[0,\infty)$ in Section~\ref{PRRM-case-KL}.}
\end{table}

Complete proofs and algorithmic details are collected in Appendices~\ref{appendix:preliminaries}--\ref{appendix:analytical_examples} at the end of the paper.

\subsection{Related Literature}

One related stream studies worst-case distortion risk measures and other law-invariant risk functionals. Alongside sharp moment bounds for coherent and concave distortion risks \citep{li2018closed,cornilly2018upper}, \citet{li2018range} combine convex order and joint mixability to obtain sharp range value-at-risk bounds for individual and aggregate risks under moment and shape information. \citet{shao2023distortion} unify extreme-case distortion risk under MV and symmetry information and apply their bounds to optimization with uncertain moments. \citet{shao2023extreme} extend the analysis to higher-order absolute central moments. Their distortion-envelope constructions provide an important analytical foundation: they identify extremal laws for a broad class of risk criteria and explain when different criteria share the same sharp bounds. These results supply the central-moment benchmark for our variational analysis.

Related work develops tractable reformulations for distributionally robust distorted expectations and nonconvex distortion functionals \citep{cai2025distributionally,liu2022inf,mao2025model}. Isotonic projection provides a powerful constructive approach: \citet{bernard2024robust} derive sharp distortion-risk bounds under combined MV--Wasserstein constraints, and \citet{pesenti2022reverse} construct nearest distributions satisfying specified risk stresses. These results make monotonicity an explicit part of extremal-law construction. Beyond coherent risk, \citet{pesenti2025optimizing} establish equivalence with concave-envelope optimization under suitable concentration-closure conditions, substantially broadening the class of distortion riskmetrics amenable to worst-case analysis.

Related structural results characterize extremal quantiles under MV--Wasserstein constraints \citep{liu2025robust} and Wasserstein penalties with moment constraints \citep{du2025robust}. \citet{li2026extreme} extend quantile-based variational analysis to distorted utility under moment equalities, combining nonsmooth first-order conditions with normal-cone geometry and isotonic projection. This broadens the constructive treatment of extremal laws to nonsmooth utility functions. Our emphasis is complementary: by retaining coherent spectral risk, we preserve convexity and develop paired quantile- and distribution-space optimality systems across heterogeneous ambiguity sets. For the two-sided ES band, the normal-cone correction has an isotonic-projection interpretation; our analysis gives the full classification of active constraints, plateau boundaries, and parameter equations, as discussed in Section~\ref{sub-sec-quantile-normal-cone}.

A second stream is the classical generalized moment problem, which optimizes the expectation of a fixed payoff over distributions satisfying moment restrictions, sometimes augmented by shape information. In its standard form, the objective and moment constraints are linear in the probability law, so duality yields semidefinite, conic, or semi-infinite reformulations and often finitely supported extremal laws \citep{popescu2005semidefinite,bertsimas2005optimal,natarajan2010tractable}. This linear structure also supports analytical characterizations of extremal laws through primal--dual optimality conditions \citep{guo2022unified}. The CDF representation of a general SRM, however, is nonlinear in the distribution. Our variational theory works directly with quantiles or CDFs in the representation where the full problem remains convex and derives optimality systems for the extremal quantile, CDF, or density.

A third stream is distributionally robust optimization (DRO), which embeds an inner worst-case distribution in a decision problem. Scarf's distribution-free newsvendor is the canonical moment-based example, with subsequent extensions incorporating richer moment and asymmetry information \citep{scarf1958min,natarajan2018asymmetry,das2021heavy}. Modern high-dimensional DRO develops tractable reformulations for moment, Wasserstein, and divergence ambiguity through duality over probability measures \citep{delage2010distributionally,mohajerin2018data,gao2017wasserstein,ben2013robust,kuhn2025distributionally}. For divergence ambiguity, \citet{glasserman2014robust} develop robust risk measurement through relative entropy and exponential changes of measure, including additional moment restrictions. Our paper is complementary: the variational systems target the worst-case law directly, and the portfolio application shows how its structure can in turn eliminate a high-dimensional inner problem before the outer decision model is solved. For the Gaussian model, the exact projection retains the joint mean constraint and KL radius.

A complementary literature addresses uncertainty in risk preferences. \citet{wang2020robust} optimize over ambiguous risk spectra and develop tractable approximations with error bounds; \citet{wang2023preference} extend preference-robust analysis to distortion functions, including inverse-S-shaped preferences. Together, these studies provide a systematic way to incorporate incomplete risk preferences into optimization. From a statistical perspective, \citet{wang2021quantitative} derive quantitative stability bounds for risk estimators under perturbations of the data-generating law. Our problem instead fixes the spectrum and characterizes the worst-case loss distribution subject to heterogeneous ambiguity information.

\textbf{Notation.}
For $x\in\mathbb{R}$, write $(x)_+\triangleq\max\{x,0\}$, and let $\mathbf{1}_A$ denote the indicator of an event or condition $A$. We write $\mathbb{E}_P[\cdot]$ for expectation under $P$, omitting the subscript when the law is clear. The abbreviation a.e. means almost everywhere with respect to Lebesgue measure unless another measure is specified. For an interval $I$ and $1\leq p<\infty$, $L^p(I)$ is the space of measurable functions $h$ with finite norm $\|h\|_p\triangleq(\int_I|h(x)|^p\,\mathrm{d}x)^{1/p}$; functions equal a.e. are identified. For \(p\in(1,\infty)\), let \(q=p/(p-1)\) denote its conjugate exponent. Further functional-analytic definitions used in the proofs are collected in Appendix~\ref{appendix:functional_notation}.

\section{A Representation Principle for Worst-Case Risk}\label{sec:formulation}

We ground our analysis of worst-case risk in SRMs \citep{acerbi2002spectral}, a flexible class of law-invariant coherent risk measures \citep{kusuoka2001law}. Their risk spectrum specifies how weights vary across loss quantiles, allowing different degrees and patterns of aversion to large losses to be represented within the same class. These measures represent risk-averse preferences in Yaari's dual theory of choice \citep{yaari1987dual}. They also admit natural quantile and distribution representations, allowing worst-case risk to be formulated and analyzed in either space.

We begin with the quantile representation. Let $\big(\Omega,\mathcal{A},\mathbb{P}\big)$ be an atomless probability space, and let a loss $X$ have CDF $F(x)\triangleq\mathbb{P}(X\leq x)$ and left-continuous quantile $Q(u)\triangleq\inf\{x:F(x)\geq u\}$, $u\in(0,1)$. Given a nonnegative, nondecreasing spectrum $\phi$ satisfying $\int_0^1\phi(u)\,\mathrm{d}u=1$, the corresponding SRM is defined in quantile form by
\begin{align}
R_\phi(X)\triangleq \int_0^1 \phi(u)Q(u) \,\mathrm{d}u, \label{eq:srm_def}
\end{align}
whenever $\int_0^1\phi(u)|Q(u)|\,\mathrm{d}u<\infty$.\footnote{We use $R_\phi(X)$, $R_\phi(F)$, and $R_\phi(Q)$ interchangeably for the same loss law, with the same convention for $\mathrm{ES}_\alpha$. We write $\mathrm{ES}_\alpha(X;P)$ when the underlying probability measure $P$ needs to be specified.} The constant spectrum gives expected loss, $\phi(u)=(1-\alpha)^{-1}\mathbf{1}_{\{u\geq\alpha\}}$ gives ES at level $\alpha\in(0,1)$, and nondecreasing step spectra give finite mixtures of ES \citep{acerbi2002spectral}.

The same risk measure admits a CDF representation on the whole real line. Define $\Phi(t)\triangleq\int_0^t\phi(u)\,\mathrm{d}u$. Under the integrability condition above, the two-sided Choquet representation is
\begin{align}
R_\phi(F)=\int_0^\infty\big(1-\Phi(F(x))\big)\,\mathrm{d}x-\int_{-\infty}^0\Phi(F(x))\,\mathrm{d}x,
\label{eq:F_objective_unbounded}
\end{align}
with both integrals finite \citep{wang1995choquet,wang2000choquet}. This identity applies to unbounded losses and to distributions with atoms. For a law supported on a finite interval $[a,b]$, it simplifies to
\begin{align}
R_\phi(F)=b-\int_a^b\Phi(F(x))\,\mathrm{d}x.
\label{eq:srm_def_distribution}
\end{align}
If $F$ is also absolutely continuous with density $f=F'$, then $R_\phi(F)=\int_a^b x\phi(F(x))f(x)\,\mathrm{d}x$. Appendix~\ref{app:min-rep} explains these identities and their connection to positive-part minimization representations.

Let $\mathcal{X}$ be an ambiguity class of losses satisfying the integrability condition above and consistent with the available distributional information. For a fixed spectrum $\phi$, the worst-case spectral risk measure (WCSRM) is
\begin{align}
\mathcal{R}_{\phi}(\mathcal{X})\triangleq \sup_{X\in\mathcal{X}} R_\phi(X). \tag{WCSRM}
\end{align}
The class need not have a common bounded support; finiteness of the worst-case value depends on the restrictions defining $\mathcal{X}$. An optimizer, when it exists, identifies an adverse law attaining the tightest compatible bound.

The two representations give the same SRM for each law and the same worst-case value over the same family of laws. They need not, however, lead to the same convex structure. For example, with the mean fixed at $\mu$, an exact variance constraint $\int_0^1(Q(u)-\mu)^2\,\mathrm{d}u=\sigma^2$ is generally nonconvex in $Q$, whereas $\int_{\mathbb{R}}(x-\mu)^2\,\mathrm{d}F(x)=\sigma^2$ is affine in $F$. Working directly with CDFs thus provides a route to global optimality conditions and explicit worst-case laws for constraints that do not fit the convex quantile formulation. Equation~\eqref{eq:srm_def} is linear in $Q$, whereas convexity of $\Phi$ makes \eqref{eq:F_objective_unbounded} concave in $F$. We therefore adopt the following representation principle: formulate the full ambiguity set in whichever representation preserves its convexity. The resulting quantile- and distribution-space formulations follow.

\subsection{Convex Formulation in Quantile Space}\label{subsec:quantile_formulation}

Quantile space preserves convexity for information expressed through probability ranks, tail averages, and one-dimensional transport. Fix $p\in(1,\infty)$ and its conjugate exponent $q$, and suppose $Q\in L^p[0,1]$ and $\phi\in L^q[0,1]$. The valid quantile functions form the cone $\mathcal{Q}_0\triangleq\{Q\in L^p[0,1]:Q\text{ is a.e. nondecreasing on }[0,1]\}$.

We keep this monotonicity restriction separate and encode the remaining ambiguity information through affine equalities and convex integral inequalities. Let $\mathcal{I}_1$ and $\mathcal{I}_2$ be finite index sets, let $h_i\in L^q[0,1]$ for $i\in\mathcal{I}_1$, and let $H_i:[0,1]\times\mathbb{R}\to\mathbb{R}$ be measurable in $u$ and convex in its second argument for $i\in\mathcal{I}_2$. Define
\begin{align}
\mathcal{Q} \triangleq \left\{Q\in\mathcal{Q}_0~\middle|~
\begin{aligned}
&\int_0^1 h_i(u)Q(u)\,\mathrm{d}u=c_i, &&i\in\mathcal{I}_1,\\
&\int_0^1 H_i(u,Q(u))\,\mathrm{d}u\leq d_i, &&i\in\mathcal{I}_2
\end{aligned}\right\}, \label{eq:Q_ambiguity}
\end{align}
where $c_i,d_i\in\mathbb{R}$. The equality integrals are finite by H\"older's inequality, and the inequality integrals are assumed to be well defined in $(-\infty,+\infty]$ for $Q\in\mathcal{Q}_0$.

The functions $h_i$ and $H_i$ accommodate a broad range of distributional information. For example, $h_i(u)\equiv1$ with $c_i=\mu$ specifies the mean, while $H_i(u,y)=|y-\mu|^\kappa$, with $\kappa>1$, imposes a central-moment budget when the mean is fixed at $\mu$. Taking $H_i(u,y)=(1-\alpha)^{-1}\mathbf{1}_{\{u\geq\alpha\}}y$ gives an ES upper bound (and, with both the integrand and the threshold negated, a lower bound). For one-dimensional distributions with CDFs $F,G$ and quantiles $Q,Q_G\in\mathcal{Q}_0$, the $p$-Wasserstein distance is
\begin{align*}
W_p(F,G)=\left(\int_0^1|Q(u)-Q_G(u)|^p\,\mathrm{d}u\right)^{1/p}=\|Q-Q_G\|_p.
\end{align*}
Thus, a $p$-Wasserstein neighborhood of radius $\epsilon$ around $G$ is obtained by setting $H_i(u,y)=|y-Q_G(u)|^p$ and $d_i=\epsilon^p$.

The set $\mathcal{Q}$ is convex, and the corresponding worst-case problem is
\begin{align}
\sup_{Q\in\mathcal{Q}} \quad & \int_0^1 \phi(u)Q(u) \,\mathrm{d}u, \tag{Q-WCSRM} \label{prob:Q_WCSRM}
\end{align}
where the objective is linear in $Q$. We study three ambiguity models motivated by risk assessment and stress testing.

\begin{model}[Central-moment ambiguity]\label{ex:ambiguity-moments}
Given a mean $\mu$ and a central-moment budget $m_\kappa>0$ for some $\kappa>1$, the restrictions are $\int_0^1 Q(u)\,\mathrm{d}u=\mu$ and $\int_0^1 |Q(u)-\mu|^\kappa\,\mathrm{d}u\leq m_\kappa$. The first constraint is affine in $Q$ and the second is convex. This is the classical benchmark for distribution-free bounds on coherent and distortion risk measures \citep{li2018closed,shao2023extreme}.
\end{model}

\begin{model}[Wasserstein ambiguity with ES stress]\label{ex:ambiguity-wasserstein}
A $p$-Wasserstein neighborhood limits deviations from a reference distribution, while an ES floor imposes a minimum level of tail stress. In quantile space, the transport restriction is a convex $L^p$ distance constraint and the ES floor is affine in $Q$. This is useful when a loss model fitted to historical data must be assessed under a more severe tail scenario. Related ES stresses based on Wasserstein distance are studied by \citet{pesenti2022reverse}. Our analytical solution below specializes to $p=2$.
\end{model}

\begin{model}[Two-sided ES bands]\label{ex:ambiguity-es-band}
A two-sided ES band supplements a specified mean and a variance budget with lower and upper bounds on ES. The floor rules out tails that are too mild for the prescribed stress, whereas the cap imposes a credibility limit on tail severity \citep{bcbs2017stress,breuer2009plausible}. The mean and ES restrictions are affine in $Q$, and the variance budget is convex.
\end{model}

Section~\ref{sec:cases-Q} develops these three models, starting with the central-moment benchmark and then treating Wasserstein ambiguity with ES stress and the two-sided ES band. An upper bound on a convex moment functional fits \eqref{eq:Q_ambiguity}, whereas an exact nonlinear-moment equality generally does not.

\subsection{Convex Formulation in Distribution Space}\label{subsec:distribution_formulation}

We then turn to distribution space, which preserves convexity for exact generalized moments and density divergences. To introduce the main ideas of the CDF-based formulation, we first consider distributions supported on a bounded interval $[a,b]$, where $a,b\in\mathbb{R}$ and $a<b$. Section~\ref{subsec:F_optimality} develops the optimality conditions and extends the sufficient conditions to unbounded supports. For $r\in(1,\infty)$, we work in the Sobolev space $W^{1,r}[a,b] \triangleq \{F\in L^r[a,b] \mid F^{\prime} \in L^r[a,b]\text{ in the weak sense}\}$. Each $F\in W^{1,r}[a,b]$ has an absolutely continuous representative. Writing $f \triangleq F'$, the valid CDFs form the convex set
\begin{align*}
\mathcal{F}_0\triangleq \big\{F\in W^{1,r}[a,b]~\bigm|~F(a)=0,\ F(b)=1,\ F'(x)\geq 0\text{ a.e. on }[a,b] \big\}.
\end{align*}

Let $\mathcal{I}_1$ and $\mathcal{I}_2$ be finite index sets. For $j\in\mathcal{I}_1$, let $k_j\in W^{1,r'}[a,b]$, where $r'$ is conjugate to $r$. For $j\in\mathcal{I}_2$, let $K_j:[a,b]\times[0,\infty)\to\mathbb{R}$ be measurable in its first argument and convex in its second; differentiability in the second argument on $(0,\infty)$ is imposed only when the variational system is derived. Define
\begin{align}
\mathcal{F}\triangleq\left\{F\in\mathcal{F}_0~\middle|~
\begin{aligned}
&\int_a^b k_j(x)f(x)\,\mathrm{d}x=c_j, &&j\in\mathcal{I}_1,\\
&\int_a^b K_j(x,f(x))\,\mathrm{d}x\leq d_j, &&j\in\mathcal{I}_2
\end{aligned}\right\}, \label{eq:F_ambiguity}
\end{align}
where $c_j,d_j\in\mathbb{R}$. The equality integrals are finite under the stated Sobolev conditions; the inequality integrals are assumed to be well defined in $(-\infty,+\infty]$ for $F\in\mathcal{F}_0$. The set $\mathcal{F}$ is convex because the constraints are affine or convex in $f=F'$, which depends linearly on $F$.

The functions $k_j$ and $K_j$ describe moment information and restrictions on the density itself. For example, $k_j(x)=x$ with $c_j=\mu$ fixes the mean, while $k_j(x)=(\mu-x)_+^2$ or $k_j(x)=(x-\mu)_+^2$ specifies a lower or upper second-order partial moment. These equalities are affine in $f$, even when $k_j$ is nonlinear in $x$. On $[a,b]\subset(0,\infty)$, $K_j(x,y)=-y\ln(w_0x)$, with $w_0>0$, encodes a lower bound on $\mathbb{E}[\ln(w_0X)]$. Density divergence constraints also fit: $K_j(x,y)=y\ln(y/g(x))$, with $K_j(x,0)=0$, gives a KL budget relative to a reference density $g>0$ on $[a,b]$.

Using the bounded-support specialization \eqref{eq:srm_def_distribution}, the corresponding worst-case problem is
\begin{align}
\sup_{F\in\mathcal{F}} \quad & b-\int_a^b\Phi(F(x))\,\mathrm{d}x, \tag{F-WCSRM} \label{prob:F_WCSRM}
\end{align}
where the objective is concave in $F$. We study the following three ambiguity models, with the KL model also considered on unbounded support.

\begin{model}[Exact partial moments]\label{ex:ambiguity-partial-moments}
Exact lower and upper partial moments of order at least one supplement a prescribed mean with separate information on deviations below and above it. These restrictions are affine in $f$, whereas the corresponding exact nonlinear-moment equalities are generally nonconvex in $Q$ \citep{fishburn1977mean,bawa1975optimal,gao2014dynamic,natarajan2018asymmetry,cesarone2025semivariance}. Section~\ref{subsec:F_endogenous} develops the second-order case.
\end{model}

\begin{model}[Mean and geometric-moment bound]\label{ex:ambiguity-geometric}
For a loss $X$ supported on $[a,b]\subset(0,\infty)$, a lower bound on $\mathbb{E}[\ln(w_0X)]$, with $w_0>0$, supplements a prescribed mean with information on a multiplicative scale. This restriction is affine in $f$ and also defines a convex constraint in $Q$. Section~\ref{subsec:geometric_moment} gives the worst-case CDF and its scalar calibration for an exponential spectrum; Appendix~\ref{appendix:geometric_moment} provides the proof.
\end{model}

\begin{model}[Mean-anchored Kullback--Leibler ambiguity]\label{ex:ambiguity-kl}
A mean-anchored KL neighborhood limits divergence from a reference law while fixing the mean at a prescribed value. The mean restriction is affine and the KL budget is convex in $f$, so the ambiguity set is convex in distribution space \citep{ben2013robust,hu2012kullback,glasserman2014robust}.
We develop two instances: an exponential reference with candidate distributions restricted to a bounded interval in Section~\ref{PRRM-case-KL}, and a Gaussian reference with candidate distributions on $\mathbb{R}$ in Section~\ref{PRRM-case-KL-Gaussian}. The latter uses the unbounded-support extension in Section~\ref{subsec:F_optimality}.
\end{model}

When comparing Problems~\eqref{prob:Q_WCSRM} and~\eqref{prob:F_WCSRM}, the admissible families of laws must be matched. In particular, the Sobolev restriction excludes atomic laws, which may attain the extrema of classical moment problems \citep{bertsimas2005optimal,guo2022unified}. Some constraints, including prescribed means and upper bounds on convex moment functionals, are convex in both representations, so the accompanying restrictions determine the choice. Sections~\ref{sec:cases-Q} and~\ref{sec:cases-F} develop the corresponding optimality systems.

\section{Convex Variational Analysis in Quantile Space}\label{sec:cases-Q}
The quantile formulation in Section~\ref{subsec:quantile_formulation} separates the integral ambiguity constraints from quantile monotonicity. Using the functional KKT conditions in Appendix~\ref{appendix-functionals}, we derive an optimality system that handles both. In the models below, pointwise stationarity either gives a nondecreasing candidate directly or requires a correction determined by the normal cone. We express that correction through cumulative conditions, which identify where the worst-case quantile can increase and where it must be flat. A projection algorithm treats smooth instances for which no explicit solution is available.

\subsection{General Variational Optimality System}

Before stating the optimality system, we introduce two assumptions. The first ensures that a worst-case quantile exists.

\begin{assumption}\label{assump:Q_regular}
The ambiguity set $\mathcal{Q}$ defined in \eqref{eq:Q_ambiguity} is nonempty and weakly compact in $L^p[0,1]$.
\end{assumption}
Since $\phi\in L^q[0,1]$, the objective is weakly continuous, so this assumption ensures attainment. A coercive constraint together with weak closedness can ensure weak compactness; Appendix~\ref{proof_prop_Q_WCSRM_sol} gives sufficient conditions. To characterize an optimizer, we next impose conditions that ensure the existence of Lagrange multipliers for the ambiguity constraints and justify the first variations.

\begin{assumption}\label{assump:Q_dual}
The following conditions hold.
\begin{enumerate}[label=(\roman*)]
\item The generalized Slater--Robinson constraint qualification relative to $\mathcal{Q}_0$, stated in Appendix~\ref{appendix-functionals}, holds for the affine equalities and convex integral inequalities defining $\mathcal{Q}$.
\item For each $i\in\mathcal{I}_2$, $H_i(u,\cdot)$ is differentiable and convex for a.e. $u$. There are nonnegative $a_i\in L^1[0,1]$, $b_i\in L^q[0,1]$ and constants $A_i,B_i\geq0$ such that, for a.e. $u\in[0,1]$ and every $y\in\mathbb{R}$,
\begin{align*}
|H_i(u,y)|\leq a_i(u)+A_i|y|^p,\qquad
\left|\frac{\partial H_i(u,y)}{\partial y}\right|\leq b_i(u)+B_i|y|^{p-1}.
\end{align*}
\end{enumerate}
\end{assumption}
Condition (i) ensures that optimality can be expressed with ordinary Lagrange multipliers, while condition (ii) makes the integral constraints finite and continuous on $L^p$ and justifies differentiation under the integral, with derivatives in $L^q$. Appendix~\ref{appendix-functionals} gives the full functional-analytic formulation.

It remains to represent the restriction that $Q$ be nondecreasing. Unlike the integral constraints defining the ambiguity set, monotonicity is not captured by a scalar multiplier. Recall that $\mathcal{Q}_0$ is the cone of admissible quantile functions. For $Q\in\mathcal{Q}_0$, its normal cone is
\begin{align*}
N_{\mathcal{Q}_0}(Q)
\triangleq
\left\{\pi\in L^q[0,1]~\middle|~
\langle\pi,V-Q\rangle\leq0,\ \forall V\in\mathcal{Q}_0\right\},
\end{align*}
where $\langle\pi,V\rangle\triangleq\int_0^1\pi(u)V(u)\,\mathrm{d}u$ is the dual pairing between $L^q[0,1]$ and $L^p[0,1]$. Geometrically, $\pi\in N_{\mathcal{Q}_0}(Q)$ means that its pairing with every admissible move $V-Q$ is nonpositive; it records the first-order correction created by monotonicity. In particular, $0\in N_{\mathcal{Q}_0}(Q)$ for every admissible $Q$. For a strictly increasing $Q$, the normal cone contains only zero; a nonzero normal element can occur only if $Q$ has a flat segment (Proposition~\ref{prop:ncone}).

\begin{proposition}\label{prop-Q-WCSRM-sol}
Suppose Assumptions~\ref{assump:Q_regular} and~\ref{assump:Q_dual} hold. Then \eqref{prob:Q_WCSRM} attains its optimum. A function $Q^*\in\mathcal{Q}$ is optimal if and only if there exist multipliers $\gamma_i^*\in\mathbb{R}$, $i\in\mathcal{I}_1$, and $\lambda_i^*\geq0$, $i\in\mathcal{I}_2$, satisfying complementary slackness, $\lambda_i^* \cdot \left(\int_0^1 H_i(u,Q^*(u))\,\mathrm{d}u-d_i\right)=0$ for $i\in\mathcal{I}_2$,
and the normal-cone stationarity condition
\begin{align}
g^*\in N_{\mathcal{Q}_0}(Q^*),\label{optimal-q-norm}
\end{align}
where the residual $g^*(\cdot)\in L^q[0,1]$ is defined by
\begin{align*}
g^*(u)=\phi(u)-\sum_{i\in\mathcal{I}_1}\gamma_i^* h_i(u)-\sum_{i\in\mathcal{I}_2}\lambda_i^*\left.\frac{\partial H_i(u,y)}{\partial y}\right|_{y=Q^*(u)}\qquad \text{for a.e. }u\in[0,1].
\end{align*}
\end{proposition}

The stationarity residual $g^*$ has a direct interpretation. The first term, $\phi$, is the marginal weight placed on each quantile level by the risk measure; the remaining terms are the marginal shadow costs of the ambiguity constraints. Condition~\eqref{optimal-q-norm} says that any imbalance between them must be supported by quantile monotonicity. Section~\ref{sub-sec-quantile-direct-solution} studies models in which pointwise stationarity, $g^*=0$, yields a nondecreasing candidate. Section~\ref{sub-sec-quantile-normal-cone} treats cases in which a nonzero normal-cone term is needed to enforce monotonicity.

\subsection{Direct Solutions from Pointwise Stationarity} \label{sub-sec-quantile-direct-solution}

Because $0$ belongs to every normal cone, the stationarity condition holds whenever its residual $g^*(\cdot)$ vanishes a.e. Together with feasibility and complementary slackness, this gives the following sufficient condition.

\begin{corollary}\label{corollary-normal-QRRM}
Suppose Assumption~\ref{assump:Q_dual}(ii) holds. Let $\widehat{Q}\in\mathcal{Q}$ be feasible for \eqref{prob:Q_WCSRM}. If multipliers $\hat{\gamma}_i\in\mathbb{R}$, $i\in\mathcal{I}_1$, and $\hat{\lambda}_i\geq0$, $i\in\mathcal{I}_2$, satisfy complementary slackness and
\begin{align}
\phi(u)-\sum_{i\in\mathcal{I}_1}\hat{\gamma}_i h_i(u)-\sum_{i\in\mathcal{I}_2}\hat{\lambda}_i\left.\frac{\partial H_i(u,y)}{\partial y}\right|_{y=\widehat{Q}(u)}=0,\qquad \text{for a.e. }u\in[0,1],
\end{align}
then $\widehat{Q}$ is optimal for \eqref{prob:Q_WCSRM}. Neither Assumption~\ref{assump:Q_regular} nor the constraint qualification in Assumption~\ref{assump:Q_dual}(i) is required.
\end{corollary}

When pointwise stationarity yields a quantile parameterized by the multipliers, Corollary~\ref{corollary-normal-QRRM} reduces verification of optimality to feasibility and complementary slackness.

\subsubsection{Central-Moment Ambiguity.}\label{subsec:central_moment}
We first apply Corollary~\ref{corollary-normal-QRRM} to the central-moment model (Model~\ref{ex:ambiguity-moments}), recovering the sharp bounds of \citet{shao2023extreme}. Pointwise stationarity identifies the extremal quantile as an increasing transformation of the risk spectrum, giving a variational derivation of this established benchmark. The ambiguity set is
\begin{align}
\mathcal{Q}_{1\kappa} \triangleq \left\{Q\in\mathcal{Q}_0~\middle|~\int_0^1 Q(u)\,\mathrm{d}u=\mu,\quad \int_0^1 |Q(u)-\mu|^\kappa\,\mathrm{d}u\leq m_\kappa\right\}. \label{eq:Q_1kappa}
\end{align}
The subscript $1\kappa$ records the mean and central-moment order. Here $\mathcal{Q}_0$ is the nondecreasing cone in $L^\kappa[0,1]$, and $\mathrm{sgn}(t)=\mathbf{1}_{\{t>0\}}-\mathbf{1}_{\{t<0\}}$.

\begin{theorem}\label{theo-1,k}
Let $\mu\in\mathbb{R}$, $\kappa>1$, and $m_\kappa>0$. Suppose $\phi\in L^{\kappa/(\kappa-1)}[0,1]$ is nonnegative, nondecreasing, normalized by $\int_0^1\phi(u)\,\mathrm{d}u=1$, and not a.e.\ constant. The equation
\begin{align*}
\int_0^1 |\phi(u)-\gamma_1|^{1/(\kappa-1)}\mathrm{sgn}\big(\phi(u)-\gamma_1\big)\,\mathrm{d}u=0
\end{align*}
has a unique real solution $\gamma_1^*$. Set
\begin{align*}
\lambda_1^*=\frac{1}{\kappa}
\left(\frac{\int_0^1|\phi(u)-\gamma_1^*|^{\kappa/(\kappa-1)}\,\mathrm{d}u}{m_\kappa}\right)^{(\kappa-1)/\kappa}>0.
\end{align*} Then the unique worst-case quantile over $\mathcal{Q}_{1\kappa}$ is
\begin{align*}
Q^*(u)=\mu+\left(\frac{1}{\kappa\lambda_1^*}\right)^{\frac{1}{\kappa-1}}|\phi(u)-\gamma_1^*|^{\frac{1}{\kappa-1}}\mathrm{sgn}\big(\phi(u)-\gamma_1^*\big),
\end{align*}
and the worst-case value is
\begin{align*}
\mathcal{R}_{\phi}(\mathcal{Q}_{1\kappa})
=\mu+m_\kappa^{1/\kappa}\mathbin{\cdot}
\left(\int_0^1|\phi(u)-\gamma_1^*|^{\kappa/(\kappa-1)}\,\mathrm{d}u\right)^{(\kappa-1)/\kappa}.
\end{align*}
\end{theorem}

The binding moment constraint explains why this moment-budget formulation gives the same quantile and value as the exact-moment formulation in \citet{shao2023extreme}. For $\kappa=2$ and $m_2=\sigma^2>0$, write $\mathcal{Q}_{12}$ for this ambiguity set. The centering equation gives $\gamma_1^*=1$. Writing $\sigma_\phi=(\int_0^1\phi(u)^2\,\mathrm{d}u-1)^{1/2}>0$, we obtain the MV benchmark of \citet{li2018closed}:
\begin{align}
Q^*(u)=\mu+\frac{\sigma}{\sigma_\phi}\big(\phi(u)-1\big),
\qquad
\mathcal{R}_\phi(\mathcal{Q}_{12})=\mu+\sigma\sigma_\phi.
\label{eq:preprint_MV_benchmark}
\end{align}
For the ES spectrum $\phi(u)=\mathbf{1}_{\{u\geq\alpha\}}/(1-\alpha)$, with $\alpha\in(0,1)$, this reduces to
\begin{align}
\sup_{Q\in\mathcal{Q}_{12}}\mathrm{ES}_\alpha(Q)
=\mu+\sigma\sqrt{\frac{\alpha}{1-\alpha}}.
\label{eq:preprint_MV_ES_bound}
\end{align}
The extremal quantile takes the two values $\mu-\sigma\sqrt{(1-\alpha)/\alpha}$ below $\alpha$ and $\mu+\sigma\sqrt{\alpha/(1-\alpha)}$ above it, up to its immaterial value at the threshold. Thus the stationarity calculation recovers the general MV bound and its familiar ES (also called CVaR) specialization. These benchmarks will also identify when the ES constraints in Section~\ref{sub-sec-quantile-normal-cone} change the extremal law. Appendix~\ref{proof_theo_1_k} gives the complete proof of Theorem~\ref{theo-1,k}.

Beyond this benchmark, we derive new analytical characterizations of the worst-case quantiles for the Wasserstein--ES and two-sided ES-band models. The former follows from pointwise stationarity, whereas the latter requires a normal-cone correction in the cap-active regime.

\subsubsection{Wasserstein Ambiguity with ES Stress.} For Model~\ref{ex:ambiguity-wasserstein}, take $p=q=2$, so $\mathcal{Q}_0$ is the nondecreasing cone in $L^2[0,1]$. Since $W_2(F,G)=\|Q-Q_G\|_2$ for the corresponding CDFs $F$ and $G$, the ambiguity set is
\begin{align}
\mathcal{Q}_{\mathrm{W}\text{-}\mathrm{ES}}\triangleq \left\{Q\in\mathcal{Q}_0~\middle|~\frac{1}{1-\alpha}\int_\alpha^1 Q(u)\,\mathrm{d}u\geq c,\quad \|Q-Q_G\|_2\leq\epsilon\right\},
\label{eq:Q_cw}
\end{align}
where $\alpha\in(1/2,1)$, $\epsilon>0$ is the Wasserstein radius, $c$ is the ES stress floor, and $Q_G$ is the reference quantile. Write $c_G\triangleq(1-\alpha)^{-1}\int_\alpha^1Q_G(u)\,\mathrm{d}u$ for the reference ES. For $Q_G\in\mathcal{Q}_0$, the set is nonempty exactly when $c\leq c_G+\epsilon/\sqrt{1-\alpha}$; strict inequality gives the required Slater condition (Lemma~\ref{lem:nonempty_Qcw}). Accordingly, we study $c_G\leq c<c_G+\epsilon/\sqrt{1-\alpha}$, where the ES floor is at least the reference ES and the ambiguity set remains strictly feasible. Applying Corollary~\ref{corollary-normal-QRRM} gives the following worst-case quantile and risk value.

\begin{theorem}\label{theo-cw}
Consider $\mathcal{Q}_{\mathrm{W}\text{-}\mathrm{ES}}$ in \eqref{eq:Q_cw}, with $Q_G\in\mathcal{Q}_0\subset L^2[0,1]$, $\epsilon>0$, and $\alpha\in(1/2,1)$. Let $\phi\in L^2[0,1]$ be a nonnegative, nondecreasing spectrum with $\int_0^1\phi(u)\,\mathrm{d}u=1$, and suppose $c_G\leq c<c_G+\epsilon/\sqrt{1-\alpha}$. Define
\begin{align}
A&\triangleq\int_\alpha^1\phi(u)\,\mathrm{d}u,
&B&\triangleq\int_0^1\phi(u)^2\,\mathrm{d}u,
&c^\dagger&\triangleq c_G+\frac{\epsilon A}{(1-\alpha)\sqrt{B}}.
\end{align}
The threshold $c^\dagger$ is the ES attained by the worst-case quantile without the stress constraint. The unique optimal quantile over $\mathcal{Q}_{\mathrm{W}\text{-}\mathrm{ES}}$ is
\begin{align}
Q^*(u)=Q_G(u)+\frac{\phi(u)+\frac{\lambda_c^*}{1-\alpha}\mathbf{1}_{\{u\geq\alpha\}}}{2\lambda_w^*},
\label{eq:Q-cw-candidate}
\end{align}
where $\lambda_c^*\geq0$ and $\lambda_w^*>0$ are the multipliers for the ES floor and the squared Wasserstein constraint $\|Q-Q_G\|_2^2\leq\epsilon^2$, respectively. They are determined as follows:
\begin{enumerate}[label=(\roman*)]
\item If $c_G\leq c\leq c^\dagger$, then $\lambda_c^*=0$ and $\lambda_w^*=\sqrt{B}/(2\epsilon)$.
\item If $c^\dagger<c<c_G+\epsilon/\sqrt{1-\alpha}$, set $D\triangleq(1-\alpha)(c-c_G)/\epsilon$. Then
\begin{align}
\lambda_c^*=-A+\sqrt{A^2-(1-\alpha)\frac{A^2-D^2B}{1-\alpha-D^2}},\qquad
\lambda_w^*=\frac{A+\lambda_c^*}{2D\epsilon}.
\label{eq:lambda-cw}
\end{align}
\end{enumerate}
The corresponding worst-case value is
\begin{align*}
\mathcal{R}_{\phi}(\mathcal{Q}_{\mathrm{W}\text{-}\mathrm{ES}})
=\int_0^1 \phi(u)Q_G(u)\,\mathrm{d}u+\frac{1}{2\lambda_w^*}\left(B+\frac{\lambda_c^*}{1-\alpha}A\right).
\end{align*}
\end{theorem}

The displacement from $Q_G$ in \eqref{eq:Q-cw-candidate} is a positive multiple of the spectrum plus a nonnegative multiple of the upper-tail indicator. Both terms are nondecreasing, so adding this displacement to $Q_G$ gives a valid quantile. If $c\leq c^\dagger$, the Wasserstein worst case already meets the stress requirement and $\lambda_c^*=0$. If $c>c^\dagger$, the stress and Wasserstein constraints both bind, and their two scalar equations yield \eqref{eq:lambda-cw}.

\begin{figure}[H]
\centering
\captionsetup[subfigure]{font=small,skip=2pt,justification=centering}
\begin{subfigure}[t]{0.49\textwidth}
\centering
\includegraphics[width=\linewidth]{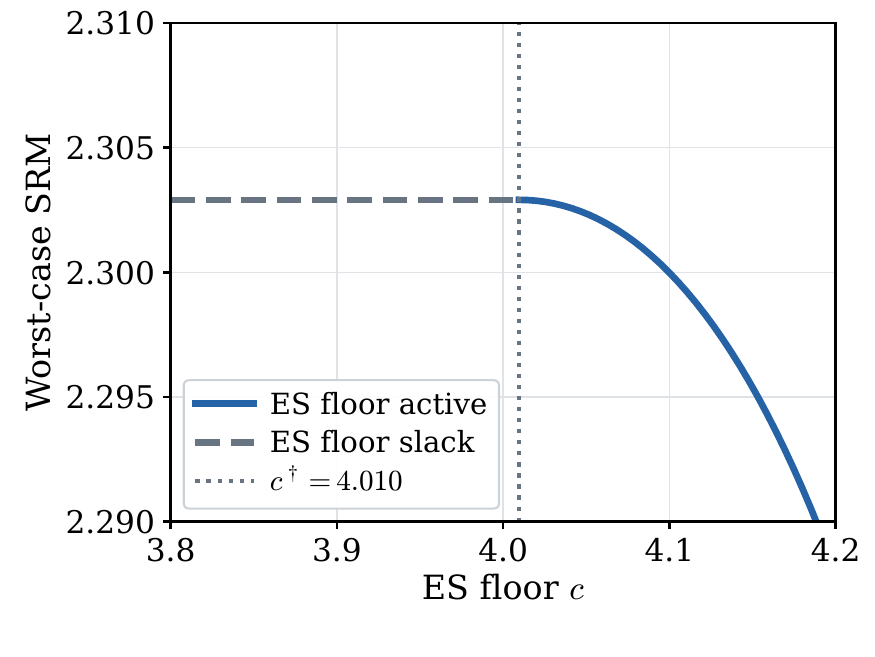}
\caption{Varying the ES floor.}
\end{subfigure}\hfill
\begin{subfigure}[t]{0.49\textwidth}
\centering
\includegraphics[width=\linewidth]{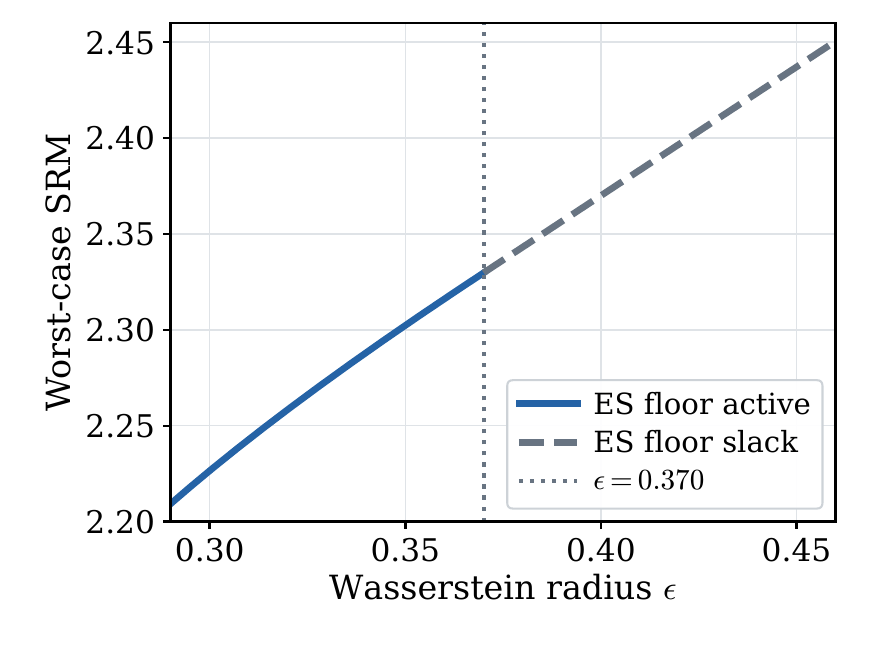}
\caption{Varying the Wasserstein radius.}
\end{subfigure}
\caption{Wasserstein ambiguity with ES stress. \textbf{(a)} For $\epsilon=0.35$, the ES constraint activates when $c$ exceeds $c^\dagger$. \textbf{(b)} For $c=4.05$, increasing $\epsilon$ eventually makes the ES constraint slack.}
\label{fig:W_CVaR_Case}
\end{figure}

To illustrate how the ES floor and Wasserstein radius affect the worst-case SRM, we consider the exponential reference $Q_G(u)=-\ln(1-u)$ and the power spectrum $\phi(u)=\beta u^{\beta-1}$, with $\beta=3$ and $\alpha=0.9$. The results are shown in Figure~\ref{fig:W_CVaR_Case}. In panel (a), with $\epsilon=0.35$, the worst-case SRM decreases once the ES floor exceeds $c^\dagger$. In panel (b), with $c=4.05$, enlarging the Wasserstein radius increases the worst-case SRM and eventually makes the ES floor slack.
Without the ES floor, the problem reduces to the Wasserstein-ball model studied by \citet[Section~4.2]{bernard2024robust}. The threshold $c^\dagger$ identifies when the added stress requirement changes the extremal law: beyond it, the law has higher ES but lower risk under the prescribed spectrum than the benchmark at the same radius. Thus, greater tail-stress severity and greater distributional uncertainty have different effects on worst-case risk.

\subsection{Monotonicity Corrections from the Normal Cone}\label{sub-sec-quantile-normal-cone}

For the central-moment and Wasserstein--ES models, pointwise stationarity produces a nondecreasing candidate. This need not hold in general. A constraint may push a higher quantile below a lower one, in which case the stationary candidate lies outside $\mathcal{Q}_0$ even though the worst-case problem remains well posed. The missing ingredient is precisely the nonzero normal-cone term in Proposition~\ref{prop-Q-WCSRM-sol}. We now show how this term determines the required monotonicity correction. We begin with its characterization in the general $L^p$ setting of Section~\ref{subsec:quantile_formulation}.

\begin{proposition}\label{prop:ncone}
Let $1<p<\infty$, let $q$ be its conjugate exponent, and let $\mathcal{Q}_0$ be the nondecreasing cone in $L^p[0,1]$. For $\pi\in L^q[0,1]$, define $\Pi(u)\triangleq\int_0^u\pi(v)\,\mathrm{d}v$ for $u\in[0,1]$. For $Q\in\mathcal{Q}_0$, let $\mathrm{d}Q$ be its locally finite nonnegative Lebesgue--Stieltjes measure on $(0,1)$. Then $\pi\in N_{\mathcal{Q}_0}(Q)$ if and only if
\begin{align}
\Pi(u)\geq0\quad\forall u\in[0,1],
\qquad \Pi(1)=0,
\qquad \int_{(0,1)}\Pi(u)\,\mathrm{d}Q(u)=0.
\end{align}
The last integral is over $(0,1)$, with no endpoint masses, and is understood as a nonnegative integral in $[0,\infty]$.
\end{proposition}

Proposition~\ref{prop:ncone} turns the abstract normal-cone inclusion into three checkable statements about the cumulative residual $\Pi$. It must never be negative, it must return to zero at $1$, and it must satisfy a complementarity relation with the increments of $Q$. Because both $\Pi$ and $\mathrm{d}Q$ are nonnegative, the last condition implies that $Q$ can increase only where $\Pi=0$; wherever $\Pi>0$, the quantile must be constant. The cumulative residual therefore identifies where monotonicity binds and determines any flat region rather than assuming one in advance. A related characterization appears independently in \citet{li2026extreme} for extreme-case distorted utilities under moment equalities. We next use this characterization to analyze the two-sided ES-band model.

\textbf{Two-sided ES bands (Model~\ref{ex:ambiguity-es-band}):} For this model, again take $p=q=2$ and let $\mathcal{Q}_0$ denote the nondecreasing cone in $L^2[0,1]$. Supplementing the mean and variance restrictions with an ES floor and cap gives
\begin{align}
\mathcal{Q}_{12\text{-}\mathrm{ES}}^{\pm}\triangleq\left\{Q\in\mathcal{Q}_0~\middle|~
\int_0^1 Q(u)\,\mathrm{d}u=\mu,
\int_0^1 (Q(u)-\mu)^2\,\mathrm{d}u\leq\sigma^2,
c_\ell\leq\frac{1}{1-\alpha}\int_\alpha^1 Q(u)\,\mathrm{d}u\leq c_u
\right\},\label{eq:Q_12c_band}
\end{align}
where $\mu\in\mathbb{R}$ is the specified mean, $\sigma\geq0$ determines the variance budget $\sigma^2$, $\alpha\in(1/2,1)$, and $c_\ell<c_u$. The subscript $12$ records the mean and variance constraints. In stress testing, this band specifies an admissible range for tail risk: the floor sets a minimum stress level, while the cap limits the severity of the scenarios considered. Taking $c_u=+\infty$ recovers a one-sided ES stress test.

Write $M_\alpha\triangleq\mu+\sigma\sqrt{\alpha/(1-\alpha)}$ for the sharp mean--variance upper bound on $\mathrm{ES}_\alpha$ in \eqref{eq:preprint_MV_ES_bound} \citep{li2018closed,shao2023extreme}. The ambiguity set is nonempty if and only if $\mu\leq c_u$, $c_\ell\leq M_\alpha$, and $c_\ell<c_u$; strict inequalities give the Slater condition when $\sigma>0$ (Lemma~\ref{lem:nonempty_Q12c_band}). Since $c_\ell<c_u$, the floor and cap cannot bind simultaneously.

Assume $\sigma>0$ and let $\phi$ be continuous and strictly increasing on $[0,1]$, with $\int_0^1\phi(u)\,\mathrm{d}u=1$. Strict increase is used in the cap-active regime to prove that the variance multiplier is positive and that the plateau classification is exhaustive. The uniqueness conclusion below is restricted to this class of spectra; the general optimality system in Proposition~\ref{prop-Q-WCSRM-sol} still permits nondecreasing spectra with flat segments.

To compare the ES band with the MV benchmark, let $\sigma_\phi\triangleq\left(\int_0^1\phi(u)^2\,\mathrm{d}u-1\right)^{1/2}$ and $A_0\triangleq\left(1/(1-\alpha)\right)\int_\alpha^1\big(\phi(u)-1\big)\,\mathrm{d}u>0$. The MV worst-case quantile is $Q^\circ(u)\triangleq\mu+(\sigma/\sigma_\phi)\big(\phi(u)-1\big)$, with ES level
\begin{align}
\bar{c} \triangleq\mathrm{ES}_\alpha(Q^\circ)=\mu+\frac{\sigma A_0}{\sigma_\phi}.\label{eq:cbar-gen}
\end{align}
Thus $\bar{c}$ determines whether the ES band changes the MV solution. Let $\gamma_1\in\mathbb{R}$ and $\lambda_1,\lambda_\ell,\lambda_u\geq0$ denote the multipliers for the mean, variance, floor, and cap, respectively. To calibrate a binding floor, define
\begin{align*}
\psi_{\lambda_\ell}(u)\triangleq\phi(u)-1+\lambda_\ell\mathbin{\cdot}\left(\frac{\mathbf{1}_{\{u\geq\alpha\}}}{1-\alpha}-1\right)
\end{align*}
and
\begin{align}
T(\lambda_\ell) \triangleq\mu+\sigma\, \frac{A_0+\lambda_\ell\frac{\alpha}{1-\alpha}}{\sqrt{\sigma_\phi^2+2\lambda_\ell A_0+\lambda_\ell^2\frac{\alpha}{1-\alpha}}}.
\end{align}
The function $T(\lambda_\ell)$ gives the ES of the candidate obtained by centering and scaling $\psi_{\lambda_\ell}$ to have mean $\mu$ and variance $\sigma^2$.

For $\bar{c}<c_\ell<M_\alpha$, a binding ES floor adds an upward jump at $\alpha$ to the stationary candidate and preserves monotonicity. For $\mu<c_u<\bar{c}$, a binding cap instead creates a downward jump; the normal-cone condition in Proposition~\ref{prop:ncone} replaces it with a plateau containing $\alpha$.

\begin{theorem}\label{thm:band-general}
Consider \eqref{prob:Q_WCSRM} over $\mathcal{Q}_{12\text{-}\mathrm{ES}}^{\pm}$ in \eqref{eq:Q_12c_band}, with $\mu\in\mathbb{R}$, $\sigma>0$, $\alpha\in(1/2,1)$, and $c_\ell<c_u$. Suppose this set is nonempty and $\phi:[0,1]\to[0,\infty)$ is continuous and strictly increasing, with $\int_0^1\phi(u)\,\mathrm{d}u=1$. With $M_\alpha$, $Q^\circ$, $\bar{c}$, $\psi_{\lambda_\ell}$, $T$, and the multipliers as defined above, the unique optimal quantile $Q^*$ is characterized as follows.
\begin{enumerate}[label=(\roman*)]
\item \emph{Benchmark feasible} $(c_\ell\leq\bar{c}\leq c_u)$: $Q^*=Q^\circ$; the ES multipliers can be chosen as zero.
\item \emph{Floor active} $(\bar{c}<c_\ell<M_\alpha)$: the unique $\lambda_\ell^*>0$ satisfying $T(\lambda_\ell^*)=c_\ell$ gives
\begin{equation}
\label{eq:Q-floor}
Q^*(u)=\mu+\sigma\frac{\psi_{\lambda_\ell^*}(u)}{\|\psi_{\lambda_\ell^*}\|_2}.
\end{equation}

\item \emph{Cap active} \label{cases:Q-cap} $(\mu<c_u<\bar{c})$:
the floor is slack, $\lambda_u^*>0$, and $\lambda_1^*>0$, so the variance binds. Exactly one of the following three cases applies.

\begin{enumerate}[label=\rm(\alph*),leftmargin=0pt]

\item \emph{Interior plateau} \label{cases:Q-cap-a}($0<u_1<\alpha<u_2<1$):
there exist $u_1,u_2$ such that
\begin{align}
Q^*(u)=\begin{cases}
\mu+\dfrac{\phi(u)-\gamma_1^*}{2\lambda_1^*},
  & u\in[0,u_1),\\[4pt]
\bar{q}, & u\in[u_1,u_2],\\[4pt]
\mu+\dfrac{\phi(u)-\gamma_1^*
  -\lambda_u^*/(1-\alpha)}{2\lambda_1^*},
  & u\in(u_2,1],
\end{cases}
\label{eq:Q-cap-general}
\end{align}
where $\bar{q}=\mu+(\phi(u_1)-\gamma_1^*)/(2\lambda_1^*)$,
and the plateau endpoints satisfy
\begin{align}
\phi(u_2)-\phi(u_1)
  &=\frac{\lambda_u^*}{1-\alpha},
    \label{eq:W-gen}\\
\int_{u_1}^{u_2}\!\big(\phi(u)-\phi(u_1)\big)\,
  \mathrm{d}u
  &=\big(\phi(u_2)-\phi(u_1)\big)(u_2-\alpha).
    \label{eq:EA-gen}
\end{align}
Together with the mean, binding variance, and binding cap,
these equations form a finite system for the parameters.
The worst-case distribution has an atom of mass $u_2-u_1$
at $\bar{q}$.

\item \emph{Right-boundary plateau} \label{cases:Q-cap-b}($0<u_1<\alpha<u_2=1$):
for a given $c_u$, the third branch in \eqref{eq:Q-cap-general} vanishes and
\begin{align}
Q^*(u)=\begin{cases}
\mu+\dfrac{\phi(u)-\gamma_1^*}{2\lambda_1^*},
  & u\in[0,u_1),\\[4pt]
\bar{q}, & u\in[u_1,1].
\end{cases}
\label{eq:Q-cap-boundary}
\end{align}
The cap condition gives $\bar{q}=c_u$, the width condition
\eqref{eq:W-gen} is replaced by the sign condition
$\phi(1)-\phi(u_1)\le\lambda_u^*/(1-\alpha)$, and
$\lambda_u^*$ is determined by
\begin{align}\label{eq:lamu-boundary}
\lambda_u^*
  =\int_{u_1}^{1}\!\big(\phi(u)-\phi(u_1)\big)\,
    \mathrm{d}u.
\end{align}
The remaining parameters $(u_1,\gamma_1^*,\lambda_1^*)$ satisfy the mean and binding variance together with $\gamma_1^*=\phi(u_1)-2\lambda_1^*\mathbin{\cdot}(c_u-\mu)$, which matches the left branch to the cap level.
The atom mass is $1-u_1$. Whether sub-case \ref{cases:Q-cap-b} arises depends on the spectrum.

\item \emph{Left-boundary plateau} ($u_1=0<\alpha<u_2<1$):
\begin{align}
Q^*(u)=\begin{cases}
\bar{q},&u\in[0,u_2],\\[4pt]
\mu+\dfrac{\phi(u)-\gamma_1^*-\lambda_u^*/(1-\alpha)}{2\lambda_1^*},&u\in(u_2,1],
\end{cases}
\end{align}
where $\bar{q}=\mu+(\phi(u_2)-\gamma_1^*-\lambda_u^*/(1-\alpha))/(2\lambda_1^*)$. The endpoint conditions are $\phi(u_2)-\phi(0)\leq\lambda_u^*/(1-\alpha)$ and
\begin{align}
\lambda_u^*&=\frac{1-\alpha}{\alpha}\int_0^{u_2}\big(\phi(u_2)-\phi(u)\big)\,\mathrm{d}u.
\label{eq:lamu-left}
\end{align}
The mean, binding variance, and binding cap determine the remaining parameters $(u_2,\gamma_1^*,\lambda_1^*)$. The atom at $\bar{q}$ has mass $u_2$.
\end{enumerate}
\end{enumerate}
If $c_\ell=M_\alpha$, the unique feasible quantile $Q^*$ equals $\mu-\sigma\sqrt{(1-\alpha)/\alpha}$ on $[0,\alpha)$ and $M_\alpha$ on $[\alpha,1]$.
If $c_u=\mu$, the unique feasible quantile is $Q^*(u)=\mu$ a.e.
\end{theorem}

To derive the parameter equations, define the net tail multiplier $\Lambda\triangleq\lambda_\ell-\lambda_u$. The stationarity residual and, for $\lambda_1>0$, the candidate obtained by setting it to zero are
\begin{align}
g^*(u)&=\phi(u)-\gamma_1-2\lambda_1\mathbin{\cdot}\big(Q^*(u)-\mu\big)
+\frac{\Lambda}{1-\alpha}\mathbf{1}_{\{u\geq\alpha\}}, \\
\widehat{Q}(u)&=\mu+\frac{1}{2\lambda_1}\left(\phi(u)-\gamma_1
+\frac{\Lambda}{1-\alpha}\mathbf{1}_{\{u\geq\alpha\}}\right). \label{eq:cand-gen}
\end{align}
With both ES multipliers zero, \eqref{eq:cand-gen} gives $Q^\circ$. For a binding floor in $\bar{c}<c_\ell<M_\alpha$, $\Lambda=\lambda_\ell>0$; enforcing the mean and binding variance gives \eqref{eq:Q-floor}, so the zero-residual certificate applies.

For fixed multipliers with $\lambda_1>0$, completing the square in the Lagrangian identifies its minimizer over $\mathcal{Q}_0$ as the $L^2$ isotonic projection of $\widehat{Q}$. This connects the normal-cone correction to the constructive projection methods of \citet{bernard2024robust,pesenti2022reverse}: the ES multipliers modify the function being projected, while monotonicity determines its flat regions. Theorem~\ref{thm:band-general} resolves the resulting two-sided ES-band model by classifying all active regimes and calibrating both interior and boundary plateaus.

In the cap-active regime, $\Lambda=-\lambda_u<0$. For an interior plateau, branch matching and the projection's equal-area condition give \eqref{eq:W-gen} and \eqref{eq:EA-gen}. Solve this system on $0<u_1<\alpha<u_2<1$; if it has no admissible solution, check both boundary systems and their sign conditions in Theorem~\ref{thm:band-general}. Which boundary occurs depends on the spectrum.

For the power spectrum $\phi(u)=\beta u^{\beta-1}$, the endpoint equations simplify further. The following corollary identifies which boundary can occur and gives an explicit transition threshold when $\beta=2$.

\begin{corollary}\label{cor:band-power}
Under the assumptions of Theorem~\ref{thm:band-general}, let $\phi(u)=\beta u^{\beta-1}$ with $\beta>1$. Then
\begin{align*}
\sigma_\phi=\frac{\beta-1}{\sqrt{2\beta-1}},
\qquad
\bar{c}=\mu+\sigma\,\frac{\alpha\mathbin{\cdot}(1-\alpha^{\beta-1})\sqrt{2\beta-1}}{(1-\alpha)(\beta-1)}.
\end{align*}
In the interior cap-active case, \eqref{eq:W-gen}--\eqref{eq:EA-gen} reduce to
\begin{align}
\beta\mathbin{\cdot}\big(u_2^{\beta-1}-u_1^{\beta-1}\big)
  &=\frac{\lambda_u^*}{1-\alpha},
    \label{eq:W-power}\\
\big(u_2^\beta-u_1^\beta\big)
  -\beta u_1^{\beta-1}\mathbin{\cdot}(u_2-u_1)
  &=\beta\mathbin{\cdot}\big(u_2^{\beta-1}-u_1^{\beta-1}\big)
    (u_2-\alpha). \label{eq:EA-power}
\end{align}
For general $\beta>1$, the interior equal-area equation can be written as
\begin{align}
\alpha=\frac{\beta-1}{\beta}\,
\frac{u_2^\beta-u_1^\beta}{u_2^{\beta-1}-u_1^{\beta-1}}.
\label{eq:power-endpoint-mean}
\end{align}
As the interior plateau widens, it reaches the right boundary first if $\beta\mathbin{\cdot}(1-\alpha)<1$, and the left boundary first if $\beta\mathbin{\cdot}(1-\alpha)>1$. In the latter case, the left-boundary equations specialize to
\begin{align}
\lambda_u^*=\frac{(1-\alpha)(\beta-1)}{\alpha}u_2^\beta,
\qquad u_2\geq\frac{\beta\alpha}{\beta-1}.
\label{eq:power-left}
\end{align}
If $\beta\mathbin{\cdot}(1-\alpha)=1$, neither one-sided boundary case occurs for $\mu<c_u<\bar{c}$; the interior endpoints approach $0$ and $1$ as $c_u\downarrow\mu$.

For $\beta=2$, restrict attention to the cap-active regime $\mu<c_u<\bar{c}$. The interior plateau is symmetric about $\alpha$, with $u_{1,2}=\alpha\mp\lambda_u^*/(4(1-\alpha))$.
Sub-case~\ref{cases:Q-cap}\ref{cases:Q-cap-a} holds
if and only if $\lambda_u^*<4(1-\alpha)^2$,
equivalently $c_u>c_u^\ddagger$, where $c_u^\ddagger=\mu+\sigma\sqrt{3(2\alpha-1)/(7-6\alpha)}$ and the corresponding $u_1^\ddagger=2\alpha-1$. For $\mu<c_u\le c_u^\ddagger$,
sub-case~\ref{cases:Q-cap}\ref{cases:Q-cap-b} applies.
\end{corollary}

\begin{figure}[H]
\centering
\captionsetup[subfigure]{font=small,skip=2pt,justification=centering}
\begin{subfigure}[t]{0.49\textwidth}
\centering
\includegraphics[width=\linewidth]{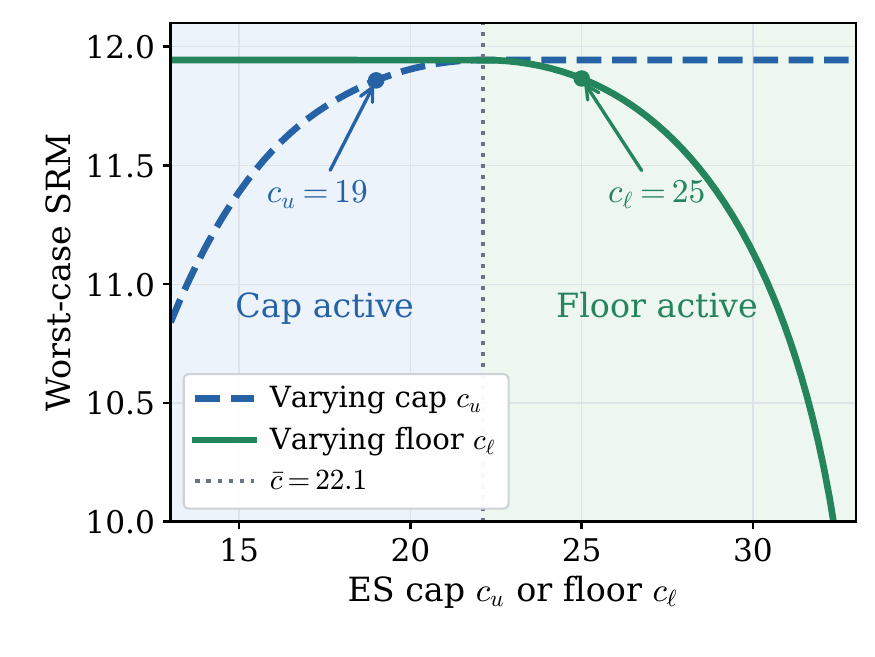}
\caption{Sensitivity to the ES bounds.}
\end{subfigure}\hfill
\begin{subfigure}[t]{0.49\textwidth}
\centering
\includegraphics[width=\linewidth]{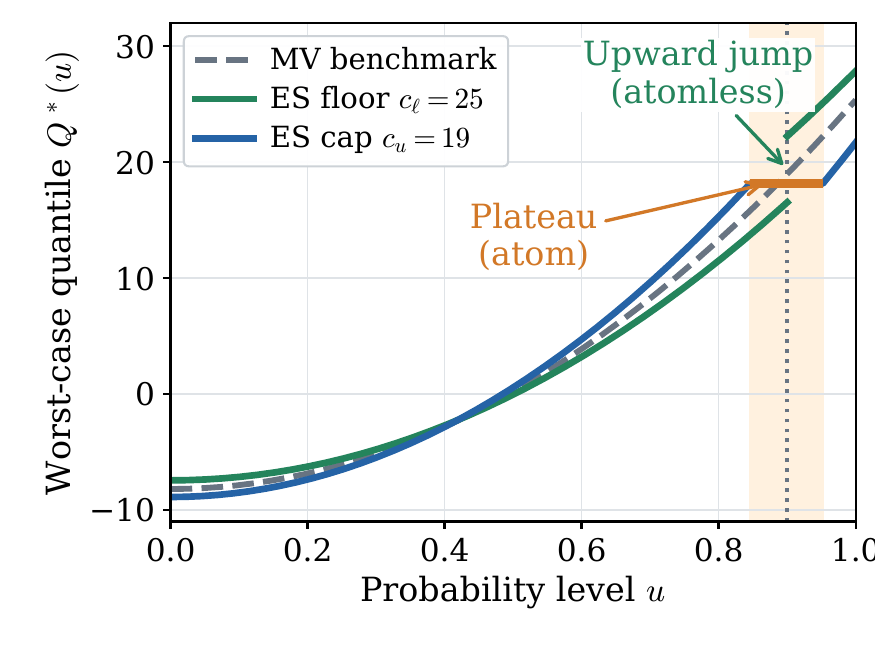}
\caption{Worst-case quantiles.}
\end{subfigure}
\caption{Two-sided ES band with $\mu=3$, $\sigma=10$, $\alpha=0.9$, and $\phi(u)=3u^2$ ($\bar{c}\approx22.1$). \textbf{(a)} With the other bound slack, the worst-case SRM equals its MV benchmark until the cap or floor excludes $\bar{c}$. \textbf{(b)} The upward jump in the plotted floor-active quantile leaves a gap in the loss support; the cap-active plateau corresponds to an atom.}
\label{fig:band-capactive}
\end{figure}

Figure~\ref{fig:band-capactive} illustrates the floor and cap regimes for $\phi(u)=3u^2$. With $\mu=3$, $\sigma=10$, and $\alpha=0.9$, the boundary criterion in Corollary~\ref{cor:band-power} selects the right-boundary case. Solving \eqref{eq:W-power}--\eqref{eq:EA-power} at $u_2=1$, together with the mean and binding variance, gives the transition threshold $c_u^\ddagger\approx16.9$. For $\mu<c_u\leq c_u^\ddagger$, sub-case~\ref{cases:Q-cap}\ref{cases:Q-cap-b} applies. The worst-case quantile changes form at the transition, while the worst-case value remains continuous.

The value follows the same threshold logic. As long as the band contains $\bar{c}$, it adds information without changing the worst case and $\mathcal{R}_\phi(\mathcal{Q}_{12\text{-}\mathrm{ES}}^{\pm})=\mu+\sigma\sigma_\phi$. Once the band excludes $\bar{c}$, the binding tail restriction lowers the worst-case value. Where the value is differentiable, its marginal change is given by the corresponding shadow price: $\partial\mathcal{R}_\phi/\partial c_u=\lambda_u^*$ in the cap-active regime and $\partial\mathcal{R}_\phi/\partial c_\ell=-\lambda_\ell^*$ in the floor-active regime.

\subsection{A General Projection Algorithm}\label{subsec:Q_projection}
For smooth $L^2$ instances of \eqref{prob:Q_WCSRM} without an explicit solution, the normal-cone condition leads to a projected method of multipliers. An unconstrained Lagrangian inner problem need not be bounded below when its multipliers vanish. We therefore use an augmented Lagrangian with an additional proximal term, which makes each inner problem well posed without changing the joint fixed points.

Write
\begin{align*}
\mathcal{M}_i(Q)&=\int_0^1h_i(u)Q(u)\,\mathrm{d}u-c_i, &&i\in\mathcal{I}_1,\\
\mathcal{N}_i(Q)&=\int_0^1H_i(u,Q(u))\,\mathrm{d}u-d_i, &&i\in\mathcal{I}_2.
\end{align*} At outer multiplier iteration $k$, with fixed penalty parameter $\tau>0$ and proximal weight $\omega>0$, define
\begin{align}
\mathcal{J}_k^Q(Q)
&=-\int_0^1\phi(u)Q(u)\,\mathrm{d}u
+\sum_{i\in\mathcal{I}_1}\left(\gamma_i^k\mathcal{M}_i(Q)+\frac{\tau}{2}\mathcal{M}_i(Q)^2\right)\notag\\
&\quad+\frac{1}{2\tau}\sum_{i\in\mathcal{I}_2}\left(\big(\lambda_i^k+\tau\mathcal{N}_i(Q)\big)_+^2-(\lambda_i^k)^2\right)
+\frac{\omega}{2}\|Q-Q^k\|_2^2.
\label{eq:Q_augmented_objective}
\end{align}
The inner problem minimizes $\mathcal{J}_k^Q$ over $\mathcal{Q}_0$. Its penalty terms are convex and bounded below, and its last term dominates the linear spectral objective as $\|Q\|_2\to\infty$. Under continuous convex integral constraints, it therefore has a unique minimizer, even if all initial multipliers are zero.

For $h\in L^2[0,1]$, define
\begin{align*}
\operatorname{Proj}_{\mathcal{Q}_0}(h)\triangleq\arg\min_{V\in\mathcal{Q}_0}\frac{1}{2}\|V-h\|_2^2,
\end{align*}
where $\mathcal{Q}_0$ is the nondecreasing cone in $L^2[0,1]$. This projection exists and is unique because the cone is nonempty, closed, and convex. It enforces monotonicity; the remaining ambiguity constraints are handled by the augmented Lagrangian and multiplier updates. A projected-gradient implementation, with inner iterations indexed by $\ell$, uses
\begin{align}
Q^{k,\ell+1}=\operatorname{Proj}_{\mathcal{Q}_0} \big(Q^{k,\ell}-\delta_\ell\nabla\mathcal{J}_k^Q(Q^{k,\ell})\big),
\label{eq:Q_projected_step}
\end{align}
where $\delta_\ell>0$ is the inner step size and the $L^2$ gradient represents the first variation. On a grid, the projection is weighted isotonic regression. After the inner solve returns $Q^{k+1}$, update
\begin{align*}
\gamma_i^{k+1}&=\gamma_i^k+\tau\mathcal{M}_i(Q^{k+1}), &&i\in\mathcal{I}_1,\\
\lambda_i^{k+1}&=\big(\lambda_i^k+\tau\mathcal{N}_i(Q^{k+1})\big)_+, &&i\in\mathcal{I}_2.
\end{align*}

The original minimization Lagrangian is
\begin{align*}
\mathcal{L}(Q,\gamma,\lambda)
=-\int_0^1\phi(u)Q(u)\,\mathrm{d}u
+\sum_{i\in\mathcal{I}_1}\gamma_i\mathcal{M}_i(Q)
+\sum_{i\in\mathcal{I}_2}\lambda_i\mathcal{N}_i(Q).
\end{align*} Its first variation $\mathrm{D}_Q\mathcal{L}$, with $(\gamma,\lambda)$ fixed, is identified with its $L^2$ gradient. At a joint fixed point, the multiplier updates enforce feasibility and complementary slackness, while the proximal term vanishes and $\nabla\mathcal{J}_k^Q=\mathrm{D}_Q\mathcal{L}$. For any $\delta>0$, the primal fixed-point equation
\begin{align*}
\bar{Q}=\operatorname{Proj}_{\mathcal{Q}_0}\big(\bar{Q}-\delta\mathrm{D}_Q\mathcal{L}(\bar{Q},\gamma,\lambda)\big)
\end{align*}
is equivalent, by the projection characterization, to
\begin{align*}
-\mathrm{D}_Q\mathcal{L}(\bar{Q},\gamma,\lambda)\in N_{\mathcal{Q}_0}(\bar{Q}).
\end{align*} Thus the joint fixed points are precisely the KKT points.

Algorithm~\ref{alg:dual_ascent_Q} in Appendix~\ref{appendix:algorithms} specifies the updates and inner stopping condition. Appendix~\ref{app:fixed-grid-convergence} proves convergence for a fixed smooth convex discretization with a KKT point and summable inner errors. Smoothness here concerns dependence on the decision variables: the spectral objective is linear in $Q$ even when $\phi$ is an ES step spectrum. Nonsmooth constraint functionals require a separately justified inner solver; convergence under grid refinement is not asserted.

\section{Convex Variational Analysis in Distribution Space}\label{sec:cases-F}

Generalized-moment constraints are affine in the density, and KL constraints are convex. After developing the optimality conditions in Section~\ref{subsec:F_optimality}, we study the partial- and geometric-moment models in Section~\ref{subsec:F_endogenous}, followed by an exponential reference under KL ambiguity in Section~\ref{PRRM-case-KL}. Section~\ref{PRRM-case-KL-Gaussian} studies the Gaussian-reference instance of Model~\ref{ex:ambiguity-kl} on the unbounded support $\mathbb{R}$.

\subsection{General Variational Optimality Conditions}\label{subsec:F_optimality}

We first focus on the bounded-support problem \eqref{prob:F_WCSRM}. Fix $a,b\in\mathbb{R}$ with $a<b$, so both endpoints of the admissible interval $[a,b]$ are finite. In this part, we use the CDF classes $\mathcal{F}_0$ and $\mathcal{F}$ defined in Section~\ref{subsec:distribution_formulation}, with $F\in W^{1,r}[a,b]$ and $1<r<\infty$. The following assumption ensures that the maximum is attained.

\begin{assumption}\label{assump_F_regular}
The ambiguity set $\mathcal{F}$ defined in \eqref{eq:F_ambiguity} is nonempty and weakly compact in $W^{1,r}[a,b]$.
\end{assumption}
The objective is weakly upper semicontinuous on $\mathcal{F}_0$, so this assumption ensures attainment. A uniform bound on the $L^r$ norms of the feasible densities, together with weak closedness, ensures the required compactness; moment or KL constraints alone need not do so. Appendix~\ref{proof_prop_P_WCSRM_sol} gives the existence argument. The explicit models below instead establish existence by constructing a feasible optimizer or by separate arguments.

To derive the optimality system at a feasible candidate, we require a constraint qualification for the ambiguity constraints and enough regularity to express their first variations in the dual of $W^{1,r}[a,b]$.

\begin{assumption}\label{assump:F_dual}
The following conditions hold.
\begin{enumerate}[label=(\roman*)]

\item The generalized Slater--Robinson constraint qualification relative to $\mathcal{F}_0$, stated in Appendix~\ref{appendix-functionals}, holds for the affine equalities and convex integral inequalities defining $\mathcal{F}$.

\item The spectrum $\phi$ is continuous on $[0,1]$, and $k_j\in W^{1,r'}[a,b]$ for $j\in\mathcal{I}_1$. For each $j\in\mathcal{I}_2$, the integral functional $I_j(f)=\int_a^bK_j(x,f(x))\,\mathrm{d}x$ is finite, convex, and continuous on the nonnegative cone of $L^r[a,b]$. At the feasible candidate $F^*$ where the optimality system is evaluated, with $f^*=(F^*)'$, there is a supporting derivative $a_j(x)=\left.(\partial K_j(x,y)/\partial y)\right|_{y=f^*(x)}\in W^{1,r'}[a,b]$ such that, for every nonnegative $v\in L^r[a,b]$,
\begin{align*}
\lim_{\varepsilon\downarrow0}\frac{I_j((1-\varepsilon)f^*+\varepsilon v)-I_j(f^*)}{\varepsilon}
=\int_a^ba_j(x)\mathbin{\cdot}(v(x)-f^*(x))\,\mathrm{d}x.
\end{align*}
At $f^*(x)=0$, the derivative with respect to $y$ is understood as a finite right derivative. This is a relative first-variation assumption along nonnegative densities. Pointwise differentiability and membership of a formal derivative in a dual space alone do not imply it.

\end{enumerate}
\end{assumption}

Condition (i) supplies multipliers, while (ii) permits first variations along nonnegative density segments and integration by parts, as detailed in Appendix~\ref{proof_prop_P_WCSRM_sol}. Affine density constraints satisfy the relative derivative condition directly; nonlinear constraints require verification at the candidate. The KL constructions below use direct convexity certificates rather than assuming differentiability at vanishing densities. Appendix~\ref{appendix-functionals} gives the relative KKT principle.

The ambiguity constraints are handled by ordinary Lagrange multipliers. What remains is the set constraint that $F$ be a valid CDF, which cannot be represented by a single scalar multiplier. Let $(W^{1,r}[a,b])^*$ denote the continuous dual space, consisting of continuous linear functionals on $W^{1,r}[a,b]$. For $F\in\mathcal{F}_0$, define its normal cone by
\begin{align*}
N_{\mathcal{F}_0}(F)\triangleq \left\{\pi_F\in (W^{1,r}[a,b])^*~\middle|~\langle \pi_F,V-F\rangle\leq 0,\ \forall V\in\mathcal{F}_0\right\},
\end{align*}
where $\langle\cdot,\cdot\rangle$ is the dual pairing between $(W^{1,r}[a,b])^*$ and $W^{1,r}[a,b]$. Geometrically, the normal cone records the first-order correction imposed by nonnegative density and the CDF boundary conditions. This gives the language needed to state the optimality system.

\begin{proposition}
\label{prop-P-WCSRM-sol}
Under Assumption~\ref{assump_F_regular}, problem \eqref{prob:F_WCSRM} attains its maximum. Let $F^*\in\mathcal{F}$ have density $f^*=(F^*)'$, and suppose Assumption~\ref{assump:F_dual} holds at $F^*$. Then $F^*$ is optimal if and only if there exist multipliers $\eta_j^*\in\mathbb{R}$, $j\in\mathcal{I}_1$, and $\zeta_j^*\geq 0$, $j\in\mathcal{I}_2$, satisfying complementary slackness, $\zeta_j^* \cdot \left(\int_a^b K_j(x,f^*(x))\,\mathrm{d}x-d_j\right)=0$ for $j\in\mathcal{I}_2$, and the normal-cone stationarity condition
\begin{align}
\rho^*\in N_{\mathcal{F}_0}(F^*),
\end{align}
where $\rho^*\in (W^{1,r}[a,b])^*$ is represented, under the regularity conditions in Assumption~\ref{assump:F_dual}, by the strong-form residual
\begin{align}
\rho^*(x)=-\phi(F^*(x))+\sum_{j\in\mathcal{I}_1}\eta_j^* k_j'(x)+\sum_{j\in\mathcal{I}_2}\zeta_j^*\frac{\mathrm{d}}{\mathrm{d}x}\left(\left.\frac{\partial K_j(x,y)}{\partial y}\right|_{y=f^*(x)}\right).
\label{optimal-p-simp}
\end{align}
If $\phi$ is strictly increasing on $[0,1]$, then the objective in \eqref{eq:srm_def_distribution} is strictly concave in $F$, and the optimal CDF is unique.
\end{proposition}

The residual $\rho^*$ measures the first-order effect of changing the CDF after accounting for the ambiguity constraints. For the same bounded-support problem, a direct way to satisfy its normal-cone condition is to impose stationarity where the CDF lies between $0$ and $1$, with suitable signs where it equals either boundary value.

\begin{corollary}\label{corollary-normal-PRRM}
Let $\widehat{F}\in\mathcal{F}$ be feasible, with density $\widehat{f}=\widehat{F}'$, and let $\widehat{\eta}_j\in\mathbb{R}$, $j\in\mathcal{I}_1$, and $\widehat{\zeta}_j\geq0$, $j\in\mathcal{I}_2$, be multipliers. Suppose $k_j\in W^{1,r'}[a,b]$ for $j\in\mathcal{I}_1$ and $a_j(x)=\left.(\partial K_j(x,y)/\partial y)\right|_{y=\widehat{f}(x)}\in W^{1,r'}[a,b]$ whenever $\widehat{\zeta}_j>0$. Assume that, for a.e.\ $x\in[a,b]$, $\phi(\widehat{F}(x))$ supports $\Phi$ at $\widehat{F}(x)$ and each such $a_j(x)$ supports $K_j(x,\cdot)$ at $\widehat{f}(x)$ on $[0,\infty)$. Let $\widehat{\rho}$ be the residual in \eqref{optimal-p-simp} evaluated at these quantities, omitting terms with zero inequality multipliers.

If complementary slackness holds and the residual satisfies (i)~$\widehat{\rho}(x)=0$ when $0<\widehat{F}(x)<1$, (ii)~$\widehat{\rho}(x)\leq0$ when $\widehat{F}(x)=0$, and (iii)~$\widehat{\rho}(x)\geq0$ when $\widehat{F}(x)=1$, for a.e.\ $x\in[a,b]$, then $\widehat{F}$ is optimal for \eqref{prob:F_WCSRM}. If, in addition, $\phi$ is strictly increasing on $[0,1]$, then the optimal CDF is unique.
\end{corollary}

The signs imply $\widehat{\rho}\in N_{\mathcal{F}_0}(\widehat{F})$: every competing CDF lies above $\widehat{F}=0$ and below $\widehat{F}=1$, while the residual vanishes in between. These pointwise conditions are sufficient and need not hold at every optimizer. The corollary uses its own supporting-slope and regularity conditions; neither compactness nor a constraint qualification is needed.

We now turn to unbounded support and allow one or both endpoints to be infinite. Specifically, let $-\infty\leq a<b\leq+\infty$; the result below also permits both endpoints to remain finite. We replace the bounded-interval Sobolev class by absolutely continuous laws with a finite $p$-th moment, where $p\in(1,\infty)$ has conjugate exponent $q$, and assume $\phi\in L^q[0,1]$. The same residual expression applies, but integration by parts now requires explicit control of the boundary products. Define
\begin{align}
\widetilde{\mathcal{F}}_0\triangleq\Big\{F:\mathbb{R}\to[0,1]~\Big|~&
F\text{ is an absolutely continuous CDF},\ f=F'=0\text{ a.e. outside }(a,b),\notag\\*
&{\textstyle\int}|x|^p\,\mathrm{d}F(x)<\infty\Big\}.
\label{eq:F0_unbounded}
\end{align}
H\"older's inequality makes \eqref{eq:F_objective_unbounded} finite on $\widetilde{\mathcal{F}}_0$. Let $\widetilde{\mathcal{F}}$ impose the constraint families in \eqref{eq:F_ambiguity} on this class, with the constraint functions defined on $(a,b)$ and their integrals well defined, and consider $\sup_{F\in\widetilde{\mathcal{F}}}R_\phi(F)$. The following result uses the residual formula \eqref{optimal-p-simp} under local regularity conditions, without extending the Sobolev normal-cone characterization to the whole real line.

\begin{proposition}\label{prop:unbounded-sufficiency}
Let $F^*\in\widetilde{\mathcal{F}}$ have density $f^*$, and let $\eta_j^*\in\mathbb{R}$, $j\in\mathcal{I}_1$, and $\zeta_j^*\geq0$, $j\in\mathcal{I}_2$, satisfy complementary slackness. Assume each $k_j$ is locally absolutely continuous on $(a,b)$ and satisfies $|k_j(x)|\leq C_j(1+|x|^{p_0})$ there for some $C_j<\infty$ and $0\leq p_0<p$. Assume also that $a_j(x)\triangleq\left.(\partial K_j(x,y)/\partial y)\right|_{y=f^*(x)}$ is locally absolutely continuous on $(a,b)$. Suppose the following conditions hold:
\begin{enumerate}[label=(\roman*)]
\item The function $\Phi$ is differentiable at $F^*(x)$ for a.e.\ $x\in(a,b)$, with derivative $\phi(F^*(x))$. For every $F\in\widetilde{\mathcal{F}}$, writing $f=F'$ and $h=f-f^*$,
\begin{align*}
\left.\frac{\mathrm{d}}{\mathrm{d}\varepsilon}\int_a^bK_j\big(x,f^*(x)+\varepsilon h(x)\big)\,\mathrm{d}x\right|_{\varepsilon=0^+}
=\int_a^b\left.\frac{\partial K_j(x,y)}{\partial y}\right|_{y=f^*(x)}h(x)\,\mathrm{d}x
\end{align*}
for every $j\in\mathcal{I}_2$, with a finite Lebesgue integral on the right;
\item The residual $\rho^*$ in \eqref{optimal-p-simp}, evaluated on $(a,b)$, satisfies $\rho^*(x)=0$ for a.e.\ $x\in(a,b)$;
\item for every $F\in\widetilde{\mathcal{F}}$, writing $H=F-F^*$, the product
\begin{align*}
\left(\sum_{j\in\mathcal{I}_1}\eta_j^*k_j(x)+\sum_{j\in\mathcal{I}_2}\zeta_j^*a_j(x)\right)\cdot H(x)
\end{align*}
tends to zero as $x\downarrow a$ and as $x\uparrow b$, whether the endpoint is finite or infinite, and
\begin{align*}
\int_a^b\phi(F^*(x))|H(x)|\,\mathrm{d}x<\infty.
\end{align*}
\end{enumerate}
Then $F^*$ is optimal over $\widetilde{\mathcal{F}}$. It is unique if $\phi$ is strictly increasing on $[0,1]$, or if, for some $j_0\in\mathcal{I}_2$ with $\zeta_{j_0}^*>0$, $K_{j_0}(x,\cdot)$ is strictly convex on $[0,\infty)$ for a.e.\ $x\in(a,b)$.
\end{proposition}

Proposition~\ref{prop:unbounded-sufficiency} is a verification result: existence of a feasible candidate satisfying its conditions must be established separately. Its boundary condition applies at finite as well as infinite endpoints. At a finite endpoint, it holds when the multiplier-weighted factor remains bounded and the competing CDFs have the same endpoint value. At an infinite endpoint, the growth of that factor must be controlled by the tails of the competing laws.

Corollary~\ref{corollary-normal-PRRM} provides the sufficient conditions for the bounded-support models in Sections~\ref{subsec:F_endogenous} and~\ref{PRRM-case-KL}. For moment constraints, its affine specialization in Corollary~\ref{corollary:worst-moment} determines the interior CDF, with boundary signs certifying the support. For the positive exponential-reference KL candidate, the residual vanishes throughout the support interior. Section~\ref{PRRM-case-KL-Gaussian} instead uses Proposition~\ref{prop:unbounded-sufficiency} to verify the Gaussian-reference candidate on $\mathbb{R}$. Smooth bounded-support problems without an explicit solution admit a density-space counterpart of the quantile projection method; see Appendix~\ref{subsec:F_projection}.

\subsection{Moment Constraints and Endogenous Support}\label{subsec:F_endogenous}

Two-sided partial moments record how loss variability is divided above and below the mean without specifying a full distribution. For the partial-moment model (Model~\ref{ex:ambiguity-partial-moments}), we seek a worst-case law whose support is determined as part of the solution. The bounded admissible interval $[a,b]$ is given; the candidate's support endpoints $x_L$ and $x_U$ must be found within it. Corollary~\ref{corollary-normal-PRRM} provides the construction: solve the stationarity equation between these endpoints, then verify the residual signs where the CDF equals $0$ or $1$.

When every ambiguity constraint is affine in the density, the interior equation $\rho^*=0$ no longer contains $f$ explicitly. If the spectrum is strictly increasing, this equation can be solved pointwise for $F$ by inverting $\phi$.

\begin{corollary}\label{corollary:worst-moment}
Suppose $k_j\in W^{1,r'}[a,b]$ for $j\in\mathcal{I}_1$, and each inequality has the form $K_j(x,y)=\ell_j(x)y$ with $\ell_j\in W^{1,r'}[a,b]$ for $j\in\mathcal{I}_2$. Let $\phi$ be continuous and strictly increasing on $[0,1]$. If a feasible $F^*\in\mathcal{F}$ and multipliers $\eta_j^*\in\mathbb{R}$, $j\in\mathcal{I}_1$, and $\zeta_j^*\geq0$, $j\in\mathcal{I}_2$, satisfy complementary slackness and the residual sign conditions in Corollary~\ref{corollary-normal-PRRM}, then $F^*$ is the unique optimal solution of \eqref{prob:F_WCSRM}. For a.e. $x$ such that $0<F^*(x)<1$,
\begin{align}
F^*(x)=\phi^{-1}\left(\sum_{j\in\mathcal{I}_1}\eta_j^*k_j'(x)+\sum_{j\in\mathcal{I}_2}\zeta_j^*\ell_j'(x)\right).
\label{eq:F_inverse_spectrum}
\end{align}
\end{corollary}

Equation~\eqref{eq:F_inverse_spectrum} expresses the interior CDF in terms of finitely many multipliers. The equations $F(x_L)=0$ and $F(x_U)=1$, together with the moment constraints, give a finite system for the candidate's support and multipliers. Optimality still requires a nondecreasing CDF with density in $L^r[a,b]$, support contained in $[a,b]$, and the exterior residual signs. These checks use the corollary's own conditions, without the compactness or constraint qualification assumed for the general optimality system.

\subsubsection{Two-Sided Partial Moments.} For second-order partial moments, let $\mu>0$ and $\sigma>0$ be the prescribed mean and standard deviation. Normalizing the two partial-moment budgets by $\mu^2$, we write the ambiguity set as
\begin{align*}
\mathcal{F}_{12\mathrm{pm}}\triangleq\left\{F\in\mathcal{F}_0~\middle|~
\begin{aligned}
&\int_a^b xf(x)\,\mathrm{d}x=\mu,\\
&\int_a^b(\mu-x)_+^2f(x)\,\mathrm{d}x=s_1\mu^2,\quad
\int_a^b(x-\mu)_+^2f(x)\,\mathrm{d}x=s_2\mu^2
\end{aligned}\right\},
\end{align*}
where $s_1\triangleq (1-t)\sigma^2/(2\mu^2)$ and $s_2\triangleq (1+t)\sigma^2/(2\mu^2)$ are the dimensionless lower- and upper-side budgets. The parameter $t\triangleq \big(\mathbb{E}\big[(X-\mu)_+^2\big]-\mathbb{E}\big[(\mu-X)_+^2\big]\big)/\sigma^2$ measures the difference between the upper and lower partial moments relative to the total variance. The subscript $12\mathrm{pm}$ records the mean and second-order partial moments. We consider nonnegative losses with $0\leq a<\mu<b$. All three constraints are affine in $f$, whereas the corresponding exact partial-moment equalities are generally nonconvex in the quantile variable. Thus Corollary~\ref{corollary:worst-moment} applies in distribution space.

For arbitrary probability laws on $[0,\infty)$, the partial-moment constraints are feasible if and only if  $0<\frac{s_1\mathbin{\cdot}(s_1+s_2)}{s_2}\leq 1$, $s_1>0$ and $s_2>0$ \citep{natarajan2018asymmetry}. Equality forces a two-point law and is excluded by our absolutely continuous formulation. We therefore require $s_1\mathbin{\cdot}(s_1+s_2)/s_2<1$ and $t\in(-1,1)$. For a prescribed interval $[a,b]$, these conditions are only necessary; the construction below establishes feasibility and optimality when its support lies in $[a,b]$ and its density has the stated integrability. We use the power spectrum $\phi(u)=\beta u^{\beta-1}$, $\beta\geq2$, and consider
\begin{align}
\sup_{F\in\mathcal{F}_{12\mathrm{pm}}}\quad \int_a^b x\beta\mathbin{\cdot}(F(x))^{\beta-1}\,\mathrm{d}F(x).
\label{prob_prf}
\end{align}

\begin{theorem}\label{theo-LU}
Consider \eqref{prob_prf} with $0\leq a<\mu<b<\infty$, $\sigma>0$, $t\in(-1,1)$, and power spectrum $\phi(u)=\beta u^{\beta-1}$, $\beta\geq2$. Set $s_1=(1-t)\sigma^2/(2\mu^2)$ and $s_2=(1+t)\sigma^2/(2\mu^2)$, and assume $s_1\mathbin{\cdot}(s_1+s_2)/s_2<1$. The scalar equation
\begin{align}
    &2(s_1+s_2)(\beta-1)^2y^{2\beta}-\left(4s_2\beta\mathbin{\cdot}(\beta-1)+(2\beta-1)\beta s_1\right)y^{2\beta-1}+2s_2\beta^2 y^{2\beta-2} \notag\\
    &\quad~~~~~~~~+\left(4s_2\mathbin{\cdot}(\beta-1)+2(2\beta-1)s_1\right)y^\beta-4s_2\beta y^{\beta-1}-\beta s_1y+2s_2=0
\label{eq:LPM-calibration}
\end{align}
has at least one root in $(0,(1+t)/2]$. Choose any such root $y^*$ and set
\begin{align*}
\eta_1^*&=\beta\mathbin{\cdot}(y^*)^{\beta-1},\\
\eta_2^*&=-\sqrt{\frac{\beta\mathbin{\cdot}(\beta-1)^2}{2s_1\mu^2\mathbin{\cdot}(2\beta-1)}\left(\frac{\eta_1^*}{\beta}\right)^{\frac{2\beta-1}{\beta-1}}},
\end{align*}
and
\begin{align*}
\eta_3^*=\sqrt{\frac{(\beta-1)^2+(\eta_1^*-1)^2(2\beta-1)}{4s_2\mu^2\mathbin{\cdot}(2\beta-1)}-\frac{\beta\mathbin{\cdot}(\beta-1)^2}{2s_2\mu^2\mathbin{\cdot}(2\beta-1)}\left(\frac{\eta_1^*}{\beta}\right)^{\frac{2\beta-1}{\beta-1}}}.
\end{align*}
These expressions satisfy $\eta_2^*<0<\eta_3^*$. Define the endpoints
\begin{align*}
x_L^*\triangleq\frac{2\eta_2^*\mu+\eta_1^*}{2\eta_2^*},
\qquad
x_U^*\triangleq\frac{\beta+2\eta_3^*\mu-\eta_1^*}{2\eta_3^*}.
\end{align*} If $\beta=2$, let $r\in(1,+\infty)$; if $\beta>2$, assume $1<r<(\beta-1)/(\beta-2)$. If $[x_L^*,x_U^*]\subseteq[a,b]$, then the unique worst-case CDF for \eqref{prob_prf} is
\begin{align*}
F^*(x)=
\begin{cases}
0, & a\leq x<x_L^*,\\
\left(\dfrac{-2\eta_2^*}{\beta}x+\dfrac{2\eta_2^*\mu+\eta_1^*}{\beta}\right)^{\frac{1}{\beta-1}}, & x_L^*\leq x<\mu,\\
\left(\dfrac{2\eta_3^*}{\beta}x+\dfrac{\eta_1^*-2\eta_3^*\mu}{\beta}\right)^{\frac{1}{\beta-1}}, & \mu\leq x\leq x_U^*,\\
1, & x_U^*<x\leq b.
\end{cases}
\end{align*}
The corresponding worst-case value is
\begin{align*}
\mathcal{R}_{\phi}(\mathcal{F}_{12\mathrm{pm}})
&=\frac{\beta+2\eta_3^*\mu-\eta_1^*}{2\eta_3^*}
-\frac{\beta\mathbin{\cdot}(\beta-1)}{2(2\beta-1)}
\left(
\frac{1}{-\eta_2^*}\left(\frac{\eta_1^*}{\beta}\right)^{\frac{2\beta-1}{\beta-1}}
+\frac{1}{\eta_3^*}
-\frac{1}{\eta_3^*}\left(\frac{\eta_1^*}{\beta}\right)^{\frac{2\beta-1}{\beta-1}}
\right).
\end{align*}
\end{theorem}

The piecewise-power form in Theorem~\ref{theo-LU} follows from interior stationarity. Let $\eta_1,\eta_2,\eta_3$ be the multipliers for the mean, lower partial moment, and upper partial moment, using the sign convention that $\eta_2$ is the negative of the multiplier in \eqref{eq:F_inverse_spectrum}. Setting the residual to zero where the CDF increases gives
\begin{align*}
F(x)&=\left(\frac{-2\eta_2x+2\eta_2\mu+\eta_1}{\beta}\right)^{1/(\beta-1)}, &&x<\mu,\\
F(x)&=\left(\frac{2\eta_3x+\eta_1-2\eta_3\mu}{\beta}\right)^{1/(\beta-1)}, &&x>\mu.
\end{align*}
Writing $y=F(\mu)$, continuity at the mean gives $\eta_1=\beta y^{\beta-1}$; the lower and upper partial moments determine $\eta_2$ and $\eta_3$, respectively; and the mean constraint reduces the remaining calibration to \eqref{eq:LPM-calibration}. In particular, $y^*=F^*(\mu)$ is the probability accumulated below the mean.

The calibrated signs $\eta_2^*<0$ and $\eta_3^*>0$ make both branches nondecreasing and place the mean between the support endpoints, $x_L^*<\mu<x_U^*$, which are determined by $F^*(x_L^*)=0$ and $F^*(x_U^*)=1$. They also give the two exterior residual signs required by Corollary~\ref{corollary-normal-PRRM}; strict increase of $\phi$ then gives uniqueness. The restriction on $r$ is only a Sobolev regularity condition: when $\beta>2$, the density has a power singularity at $x_L^*$ and belongs to $L^r$ exactly when $r<(\beta-1)/(\beta-2)$. Appendix~\ref{proof_LPM} gives the full calibration.

\begin{figure}[htbp]
\centering
\captionsetup[subfigure]{font=small,skip=2pt,justification=centering}
\begin{subfigure}[t]{0.49\textwidth}
\centering
\includegraphics[width=\linewidth]{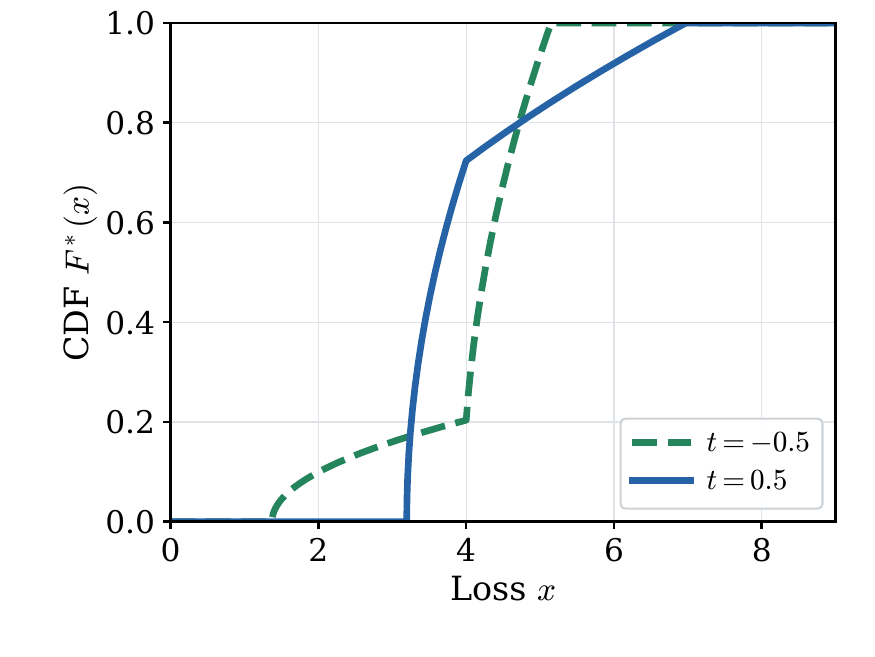}
\caption{Worst-case CDFs.}
\end{subfigure}\hfill
\begin{subfigure}[t]{0.49\textwidth}
\centering
\includegraphics[width=\linewidth]{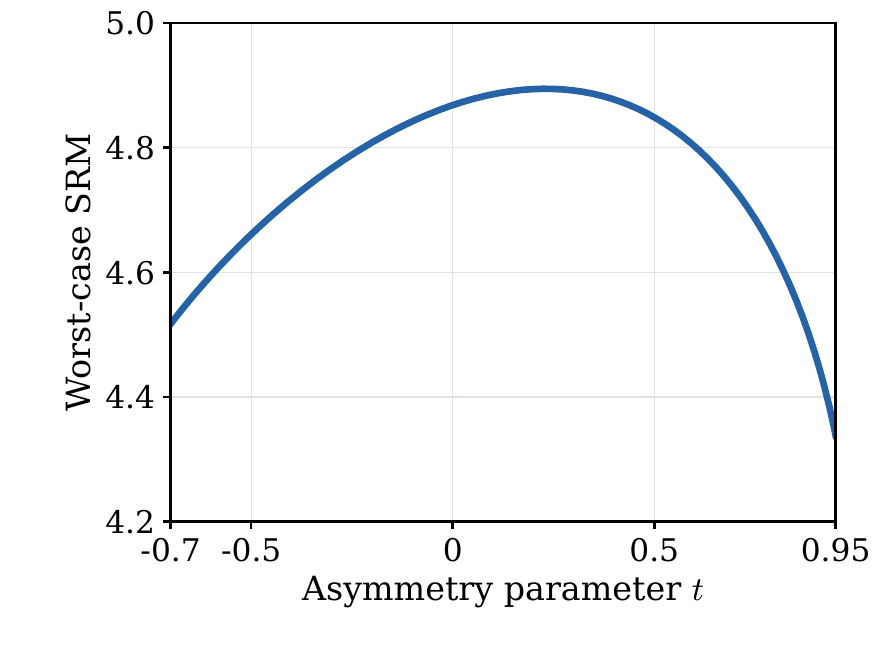}
\caption{Varying the asymmetry parameter.}
\end{subfigure}
\caption{Partial-moment ambiguity with $\mu=4$, $\sigma=1$, $\beta=3$, and admissible support $[0,15]$. \textbf{(a)} Worst-case CDFs at $t=\pm0.5$. \textbf{(b)} Worst-case SRM for $t\in[-0.7,0.95]$; all plotted candidates satisfy the support condition in Theorem~\ref{theo-LU}.}
\label{fig:LPM_Analysis}
\end{figure}

For an illustration, take $\mu=4$, $\sigma=1$, $\beta=3$, and $r=3/2$ on the common admissible interval $[a,b]=[0,15]$. We restrict the plotted asymmetry parameter to $t\in[-0.7,0.95]$, for which the calibrated supports lie within this interval. Figure~\ref{fig:LPM_Analysis}(a) compares $t=-0.5$ and $t=0.5$: a larger lower-side budget extends the support further below the mean, whereas a larger upper-side budget allows a more extended upper tail. In panel (b), these competing effects produce an inverted-U relationship between $t$ and the worst-case SRM: the value initially rises as the upper-side budget relaxes, then falls once the tightening lower-side budget becomes dominant.

\subsubsection{Geometric-Moment Bounds.}\label{subsec:geometric_moment}

The same inverse-spectrum construction applies to the geometric-moment model (Model~\ref{ex:ambiguity-geometric}). Its logarithmic moment measures losses on a multiplicative scale, whereas the partial moments above measure additive deviations from the mean. Let $0<a<\mu<b$, $w_0>0$, and
\begin{align*}
\mathcal{F}_{1\ln}\triangleq\left\{F\in\mathcal{F}_0~\middle|~\int_a^bxf(x)\,\mathrm{d}x=\mu,\quad \int_a^b\ln(w_0x)f(x)\,\mathrm{d}x\geq c\right\}.
\end{align*}
The subscript $1\ln$ records the mean and logarithmic moment. The next lemma records the feasibility boundary.

\begin{lemma}\label{lem:nonempty_geometric}
Let $0<a<\mu<b<\infty$, $w_0>0$, and $c\in\mathbb{R}$. In the absolutely continuous Sobolev class $\mathcal{F}_0$, the set $\mathcal{F}_{1\ln}$ is nonempty if and only if $c<\ln(w_0\mu)$. Whenever it is nonempty, the relative Slater condition holds. Equality at the Jensen bound $\ln(w_0\mu)$ is possible only for the point mass at $\mu$, which lies outside $\mathcal{F}_0$.
\end{lemma}

We next construct a positive-multiplier candidate for which the geometric-moment constraint binds.

For the exponential spectrum $\phi(u)=\nu e^{\nu u}/(e^\nu-1)$, with $\nu>0$, let $\eta_1$ and $\zeta_1\geq0$ be the multipliers for the mean and geometric-moment constraints. Setting the residual to zero where the CDF increases gives
\begin{align}
\phi(F(x))=\eta_1-\frac{\zeta_1}{x},
\qquad
F(x)=\frac{1}{\nu}\ln\left(\frac{e^\nu-1}{\nu}\left(\eta_1-\frac{\zeta_1}{x}\right)\right).
\label{eq:geometric_stationary_shape}
\end{align}
Thus the inverse-spectrum equation again determines the CDF and its support before calibration. Set $\phi_0\triangleq\phi(0)=\nu/(e^\nu-1)$ and $\phi_1\triangleq\phi(1)=\nu e^\nu/(e^\nu-1)$; for $\eta>\phi_1$, define
\begin{align*}
\mathsf{m}(\eta)\triangleq \nu+\ln\left(\frac{\eta-\phi_0}{\eta-\phi_1}\right),
\qquad
\mathrm{Li}_2(t)\triangleq-\int_0^t\frac{\ln(1-y)}{y}\,\mathrm{d}y.
\end{align*}
The mean determines $\zeta_1$ as a function of $\eta_1$, and the active geometric moment leaves one scalar equation.

\begin{theorem}\label{theo-log}
Consider $\mathcal{F}_{1\ln}$ with $0<a<\mu<b<\infty$, $w_0>0$, and $c<\ln(w_0\mu)$, and let $\phi(u)=\nu e^{\nu u}/(e^\nu-1)$ with $\nu>0$. With $\phi_0$, $\phi_1$, and $\mathsf{m}$ as defined above, choose a solution $\eta_1^*>\phi_1$ of
\begin{align}
\ln\left(\frac{w_0\nu\mu}{\mathsf{m}(\eta_1)}\right)
+\frac{1}{\nu}\left(\mathrm{Li}_2\left(\frac{\phi_1}{\eta_1}\right)-\mathrm{Li}_2\left(\frac{\phi_0}{\eta_1}\right)\right)=c,
\label{eq:geo_eta_append}
\end{align}
which has at least one solution in $(\phi_1,\infty)$. Define
\begin{align*}
\zeta_1^*=\frac{\nu\mu\eta_1^*}{\mathsf{m}(\eta_1^*)},\qquad
x_L^*=\frac{\zeta_1^*}{\eta_1^*-\phi_0},\qquad
x_U^*=\frac{\zeta_1^*}{\eta_1^*-\phi_1}.
\end{align*}
If $[x_L^*,x_U^*]\subseteq[a,b]$, then the unique worst-case CDF is
\begin{align}
F^*(x)=
\begin{cases}
0, & a\leq x<x_L^*,\\[2pt]
\dfrac{1}{\nu}\ln\left(\dfrac{e^\nu-1}{\nu}\left(\eta_1^*-\dfrac{\zeta_1^*}{x}\right)\right), & x_L^*\leq x\leq x_U^*,\\[4pt]
1, & x_U^*<x\leq b.
\end{cases}
\end{align}
The geometric-moment bound is attained, with $\zeta_1^*>0$, and
\begin{align*}
\mathcal{R}_\phi(\mathcal{F}_{1\ln})=\frac{\zeta_1^*}{\nu}\ln\left(\frac{x_U^*}{x_L^*}\right).
\end{align*}
\end{theorem}

The support condition in Theorem~\ref{theo-log} is essential: solving the scalar calibration equation alone does not certify feasibility on a prescribed interval. When the calibrated support lies in $[a,b]$, the positive logarithmic-moment multiplier makes the bound active, and the boundary-sign condition certifies optimality. Appendix~\ref{appendix:geometric_moment} proves the feasibility lemma and theorem.

Together with the partial-moment model in Section~\ref{subsec:F_endogenous}, this construction shows how the inverse-spectrum method handles both additive and logarithmic moment information through finite-dimensional calibration.

\subsection{KL Ambiguity with an Exponential Reference}\label{PRRM-case-KL}

When a nominal loss model is available, the mean-anchored KL neighborhood in Model~\ref{ex:ambiguity-kl} limits departures from that model while preserving a prescribed mean. Unlike the moment constraints above, the KL budget introduces the density ratio into the stationarity equation. We derive the resulting tilt and calibrate it for an exponential reference; the same construction will also apply to the Gaussian reference in Section~\ref{PRRM-case-KL-Gaussian}. The prescribed mean need not equal the reference mean.

For probability measures $P$ and $\widehat{P}$, define $\mathrm{KL}(P\|\widehat{P})=\int\ln(\mathrm{d}P/\mathrm{d}\widehat{P})\,\mathrm{d}P$ if $P\ll\widehat{P}$, and $+\infty$ otherwise. Here $P\ll\widehat{P}$ means that $P$ is absolutely continuous with respect to $\widehat{P}$, and $\mathrm{d}P/\mathrm{d}\widehat{P}$ is the corresponding Radon--Nikodym derivative. With a reference density $g>0$ on the admissible support, the constraints are
\begin{align*}
\int x f(x)\,\mathrm{d}x=\mu,
\qquad
\mathrm{KL}(f\|g)=\int f(x)\ln\frac{f(x)}{g(x)}\,\mathrm{d}x\leq\epsilon.
\end{align*}
Here $\mu$ is the prescribed mean, $\epsilon\geq0$ is the KL radius, and we use the convention $0\ln0=0$.
Consider a positive candidate admitting global stationarity multipliers $\eta^*$ for the mean and $\zeta^*>0$ for KL. The entropy derivative is $1+\ln(f/g)$ on positive densities, so setting the residual in \eqref{optimal-p-simp} to zero gives
\begin{align}
-\phi(F^*(x))+\eta^*+\zeta^*\mathbin{\cdot}\left(\ln\frac{f^*(x)}{g(x)}\right)'=0.
\label{eq:KL_generic_stationarity}
\end{align}
Equation~\eqref{eq:KL_generic_stationarity} holds almost everywhere on the interior of the admissible support. A positive density keeps the CDF strictly between $0$ and $1$ there, leaving no exterior regions on which to check residual signs. For the bounded-support exponential model below, this is the zero-residual case of Corollary~\ref{corollary-normal-PRRM}, whose supporting-slope and regularity conditions we verify at the candidate. A binding KL budget alone does not establish the required positive multiplier.

We now specialize the risk measure to ES by setting $\phi(u)=\mathbf{1}_{\{u\geq\alpha\}}/(1-\alpha)$. Let $x_\alpha$ denote the $\alpha$-quantile of $F^*$. Because the ES spectrum is constant on either side of $\alpha$, \eqref{eq:KL_generic_stationarity} gives one slope for $\ln(f^*/g)$ below $x_\alpha$ and another above it. Integration yields
\begin{align}
f^*(x)=
\begin{cases}
C_1g(x)e^{\lambda_1x},&x<x_\alpha,\\[2pt]
C_2g(x)e^{\lambda_2x},&x\geq x_\alpha,
\end{cases}
\qquad
\lambda_1=-\frac{\eta^*}{\zeta^*},\qquad
\lambda_2=\lambda_1+\frac{1}{(1-\alpha)\zeta^*}.
\label{eq:KL_generic_tilt}
\end{align}
The stationary density is therefore a two-piece exponential tilt of the reference, with a larger tilting rate above the ES threshold. The prescribed body and tail masses, continuity at $x_\alpha$, the mean, and the binding KL budget determine the parameters. Whereas moment constraints led to a direct expression for the CDF in Section~\ref{subsec:F_endogenous}, KL ambiguity gives an expression for the density relative to its reference. The composed spectrum $\phi(F^*(x))$ is discontinuous only at $x_\alpha$, which does not affect the almost-everywhere equation.

For an exponential reference, the two tilted pieces in \eqref{eq:KL_generic_tilt} remain exponential, with different rates below and above the ES threshold. Take $g_\theta(x)=\theta e^{-\theta x}$ on $[0,\infty)$ and restrict admissible densities to a fixed bounded interval $[a,b]$ with $0\leq a<\mu<b<\infty$. Define
\begin{align*}
\mathcal{F}_{1\exp}\triangleq\left\{F\in\mathcal{F}_0~\middle|~
\int_a^b xf(x)\,\mathrm{d}x=\mu,\quad \mathrm{KL}(f\|g_\theta)\leq\epsilon\right\}.
\end{align*} The subscript $1\exp$ records the mean constraint and the exponential reference.

Before calibrating the worst case, the following lemma identifies when the mean anchor and KL budget are compatible. It is stated for a general bounded support $[a,b]$ and applied below with $a=0$.

\begin{lemma}\label{lem:nonempty_KL}
Suppose $\theta>0$ and $0\leq a<\mu<b<\infty$. Define the normalizing constant $C(\varsigma)=\varsigma/\left(e^{-\varsigma a}-e^{-\varsigma b}\right)$ for $\varsigma\neq0$, with $C(0)=1/(b-a)$. Let $\varsigma^*\in\mathbb{R}$ be the unique solution of the mean equation
\begin{align*}
\frac{ae^{-\varsigma a}-be^{-\varsigma b}}{e^{-\varsigma a}-e^{-\varsigma b}}+\frac{1}{\varsigma}=\mu,
\end{align*}
where the left-hand side is defined by continuity as $(a+b)/2$ at $\varsigma=0$.

Then the ambiguity set $\mathcal{F}_{1\exp}$ is nonempty if and only if
\begin{align}
\epsilon \geq \ln C(\varsigma^*)-\ln\theta+(\theta-\varsigma^*)\mu.\label{eq:KL_nonempty}
\end{align}
The right-hand side is the minimum KL divergence under the support and mean constraints, attained by the density $C(\varsigma^*)e^{-\varsigma^*x}$ on $[a,b]$. The Slater condition holds if and only if the inequality in \eqref{eq:KL_nonempty} is strict.
\end{lemma}

Let $\eta_1\in\mathbb{R}$ and $\zeta_1\geq0$ denote the multipliers for the mean and KL constraints, respectively. We seek a positive stationary candidate with $\zeta_1>0$ and a binding KL budget. For ES with $\alpha\in(1/2,1)$, substituting $g_\theta$ into \eqref{eq:KL_generic_tilt} gives a two-rate exponential density, $f^*(x)=C_i e^{-\xi_i x}$, where $\xi_1=\theta+\eta_1/\zeta_1$ and $\xi_2=\theta+(\eta_1-(1-\alpha)^{-1})/\zeta_1$.
Define the truncated exponential integrals, for $\xi\neq0$, by
\begin{align*}
I_0(\xi;u,v)&\triangleq\int_u^ve^{-\xi x}\,\mathrm{d}x=\frac{e^{-\xi u}-e^{-\xi v}}{\xi},\\
I_1(\xi;u,v)&\triangleq\int_u^vxe^{-\xi x}\,\mathrm{d}x=\frac{(u+\xi^{-1})e^{-\xi u}-(v+\xi^{-1})e^{-\xi v}}{\xi}.
\end{align*}
For $\xi=0$, use the continuous extensions $I_0(0;u,v)=v-u$ and $I_1(0;u,v)=(v^2-u^2)/2$. Normalization on each side of $x_\alpha$, continuity of the density, the mean anchor, and the active KL budget then reduce calibration to the following system for $(\xi_1,\xi_2,x_\alpha)$:
\begin{gather}
\label{eq:KL_match}
\frac{\alpha e^{-\xi_1x_\alpha}}{I_0(\xi_1;0,x_\alpha)}=
  \frac{(1-\alpha)e^{-\xi_2x_\alpha}}{I_0(\xi_2;x_\alpha,b)},\\
\frac{\alpha I_1(\xi_1;0,x_\alpha)}{I_0(\xi_1;0,x_\alpha)}+\frac{(1-\alpha)I_1(\xi_2;x_\alpha,b)}{I_0(\xi_2;x_\alpha,b)}=\mu,\\
\label{eq:KL_kl}
\begin{aligned}
&\alpha\ln\frac{\alpha}{\theta I_0(\xi_1;0,x_\alpha)}
 +(1-\alpha)\ln\frac{1-\alpha}{\theta I_0(\xi_2;x_\alpha,b)}\\
&\quad+(\theta-\xi_1)\frac{\alpha I_1(\xi_1;0,x_\alpha)}{I_0(\xi_1;0,x_\alpha)}
 +(\theta-\xi_2)\frac{(1-\alpha)\,I_1(\xi_2;x_\alpha,b)}{I_0(\xi_2;x_\alpha,b)}
=\epsilon.
\end{aligned}
\end{gather}
An admissible solution of this calibration system gives the worst-case density in the following theorem.

\begin{theorem}\label{theo-KL}
Consider $\mathcal{F}_{1\exp}$ with $a=0$, $0<\mu<b<\infty$, reference rate $\theta>0$, and the ES spectrum $\phi(u)=(1-\alpha)^{-1}\mathbf{1}_{\{u\geq\alpha\}}$, $\alpha\in(1/2,1)$. Suppose the calibration system \eqref{eq:KL_match}--\eqref{eq:KL_kl} has a solution $(\xi_1^*,\xi_2^*,x_\alpha^*)$ with $\xi_1^*>\xi_2^*$ and $0<x_\alpha^*<b$. Define $C_1^*=\alpha/I_0(\xi_1^*;0,x_\alpha^*)$ and $C_2^*=(1-\alpha)/I_0(\xi_2^*;x_\alpha^*,b)$. Then the unique worst-case density is
\begin{align*}
f^*(x)=\begin{cases}
C_1^*\,e^{-\xi_1^*x},&0\leq x<x_\alpha^*,\\[2pt]
C_2^*\,e^{-\xi_2^*x},&x_\alpha^*\leq x\leq b.
\end{cases}
\end{align*}
The corresponding worst-case ES is
\begin{align*}
\mathcal{R}_\phi(\mathcal{F}_{1\exp})=\frac{I_1(\xi_2^*;\,x_\alpha^*,\,b)}{I_0(\xi_2^*;\,x_\alpha^*,\,b)}=\frac{\left(x_\alpha^*+(\xi_2^*)^{-1}\right)e^{-\xi_2^*x_\alpha^*}
-\left(b+(\xi_2^*)^{-1}\right)e^{-\xi_2^*b}}
{e^{-\xi_2^*x_\alpha^*}-e^{-\xi_2^*b}},
\ \text{when } \xi_2^*\ne0,
\end{align*}
with the continuous extension $\mathcal{R}_\phi(\mathcal{F}_{1\exp})=(x_\alpha^*+b)/2, \ \text{when }\xi_2^*=0.$
\end{theorem}

The theorem is conditional on solvability of the calibration system. Its candidate has a binding KL budget and a positive KL multiplier; conversely, any positive regular optimizer admitting global stationarity multipliers must have a positive KL multiplier and the stated two-piece form. At the minimum feasible radius in Lemma~\ref{lem:nonempty_KL}, the unique feasible law is the mean-anchored entropy projection, and the positive-multiplier calibration need not apply.

\begin{figure}[H]
\centering
\captionsetup[subfigure]{font=small,skip=2pt,justification=centering}
\begin{subfigure}[t]{0.49\textwidth}
\centering
\includegraphics[width=\linewidth]{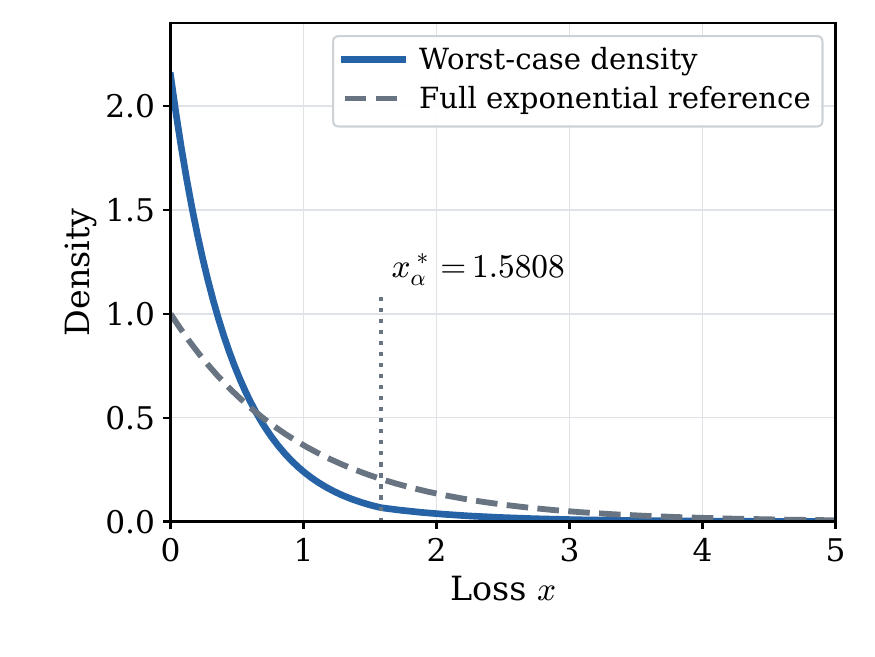}
\caption{Worst-case density at $\epsilon=0.2$.}
\end{subfigure}\hfill
\begin{subfigure}[t]{0.49\textwidth}
\centering
\includegraphics[width=\linewidth]{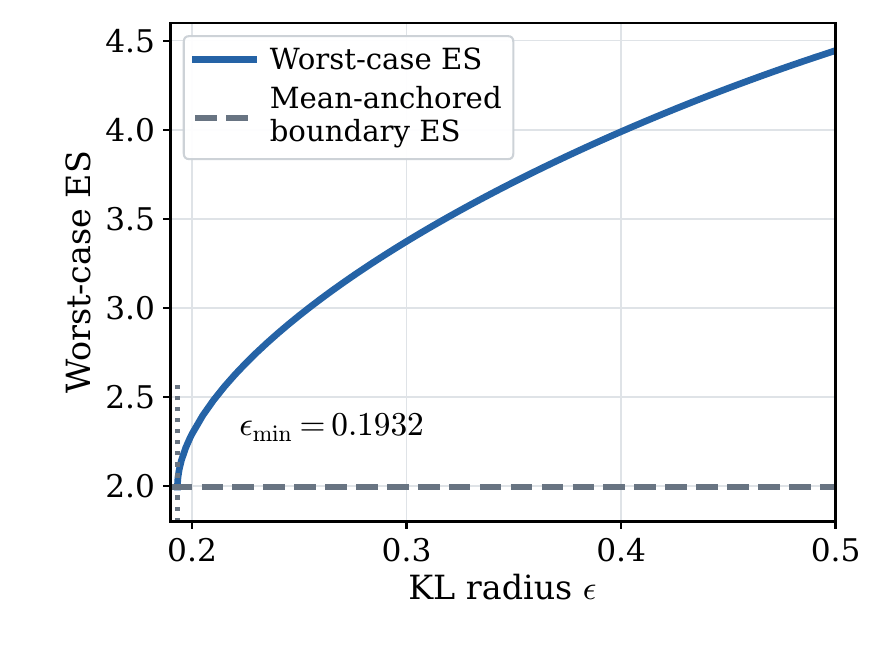}
\caption{Sensitivity to the KL radius.}
\end{subfigure}
\caption{\textbf{(a)} The calibrated density at $\epsilon=0.2$ is continuous at $x_\alpha^*$; its tail decays more slowly than its own body, but faster than the exponential reference. \textbf{(b)} Worst-case ES increases from the mean-anchored entropy-projection value at the minimum feasible radius $\epsilon_{\min}$. The KL reference is the full exponential law, not its renormalized restriction to $[0,5]$.}
\label{fig:KLexp}
\end{figure}

To illustrate the solution numerically, take $g(x)=e^{-x}$, $b=5$, $\alpha=0.95$, and $\mu=0.5$. At $\epsilon=0.2$, calibration gives $\xi_1^*\approx2.1885$, $\xi_2^*\approx1.3360$, and $x_\alpha^*\approx1.5808$. Because the mean anchor differs from the reference mean, the minimum feasible KL radius is $\epsilon_{\min}\approx0.193193$. Figure~\ref{fig:KLexp}(b) uses the ES of the unique mean-anchored entropy projection at this radius, approximately $1.99515$, as its baseline. Worst-case ES approaches this value as $\epsilon\downarrow\epsilon_{\min}$. All KL values use the original exponential reference on $[0,\infty)$, with admissible densities extended by zero outside $[0,5]$.

As an extension of this bounded-support model, the same two-rate family applies on $[0,\infty)$ within the finite-moment class \eqref{eq:F0_unbounded}. If \eqref{eq:KL_match}--\eqref{eq:KL_kl}, with $b=\infty$, has an admissible solution with $\xi_2^*>0$, Proposition~\ref{prop:unbounded-sufficiency} certifies its optimality. For a competing CDF $F$, set $H=F-F^*$. At $0$, the multiplier-weighted factor is bounded and $H(0)=0$; at infinity, its linear growth and the finite $p$-th moments, with $p>1$, give the required boundary limit.

\subsection{KL Ambiguity with a Gaussian Reference}\label{PRRM-case-KL-Gaussian}

For a Gaussian reference, the tilt in \eqref{eq:KL_generic_tilt} produces shifted Gaussian pieces. This unbounded-support model provides the scalar worst-case ES needed for the portfolio problem in Section~\ref{subsec:ext_portfolio}. Let $\varphi_{\mathrm{N}}$ and $\Phi_{\mathrm{N}}$ denote the standard normal density and CDF. For $m\in\mathbb{R}$ and $s>0$, write $\mathsf{N}(m,s^2)$ for the univariate normal law with density $g_{m,s}(x)\triangleq s^{-1}\varphi_{\mathrm{N}}((x-m)/s)$. Taking $(a,b)=\mathbb{R}$ and $p=2$ in \eqref{eq:F0_unbounded}, define the mean-anchored ambiguity set
\begin{align}
\widetilde{\mathcal{F}}_{1\mathrm{N}}(m,s;\mu,\epsilon)\triangleq\Big\{F\in\widetilde{\mathcal{F}}_0~\Big|~\int_{\mathbb{R}} xf(x)\,\mathrm{d}x=\mu,\quad \mathrm{KL}\big(f\,\|\,g_{m,s}\big)\leq\epsilon\Big\}.
\label{eq:F1N_def}
\end{align}
Here the subscript $1\mathrm{N}$ identifies the mean anchor and Gaussian reference. The Gaussian KL budget gives a uniform second-moment bound, so worst-case ES is finite whenever this set is nonempty; Appendix~\ref{proof_theo_KL_normal} establishes the bound. Standardization expresses the problem relative to a standard normal law, with the mean anchor transformed accordingly. The following lemma gives this reduction and identifies the minimum KL budget compatible with the mean anchor.

\begin{lemma}\label{lem:standardize}
Fix $\alpha\in(0,1)$, $m,\mu\in\mathbb{R}$, $s>0$, and $\epsilon\geq0$, and let $\phi$ be the ES spectrum at level $\alpha$. The set $\widetilde{\mathcal{F}}_{1\mathrm{N}}(m,s;\mu,\epsilon)$ is nonempty if and only if $\epsilon\geq(\mu-m)^2/(2s^2)$. Whenever it is nonempty,
\begin{align}
\begin{aligned}
\mathcal{R}_\phi\big(\widetilde{\mathcal{F}}_{1\mathrm{N}}(m,s;\mu,\epsilon)\big)
&=m+s\,\chi(\alpha,\epsilon;\bar{\mu}),\qquad \bar{\mu}\triangleq\frac{\mu-m}{s},\\
\chi(\alpha,\epsilon;\bar{\mu})
&\triangleq\mathcal{R}_\phi\big(\widetilde{\mathcal{F}}_{1\mathrm{N}}(0,1;\bar{\mu},\epsilon)\big).
\end{aligned}
\label{eq:chi_def}
\end{align}
The change of variables $\tilde{f}(u)=s f(m+su)$ is a bijection between the two feasible sets and preserves optimality. In standardized coordinates, the minimum KL divergence under the mean constraint is $\bar{\mu}^2/2$, uniquely attained by $\mathsf{N}(\bar{\mu},1)$. The Slater condition holds if and only if $\epsilon>\bar{\mu}^2/2$. We abbreviate $\chi(\alpha,\epsilon)\triangleq\chi(\alpha,\epsilon;0)$.
\end{lemma}

We henceforth take $\epsilon>\bar{\mu}^2/2$. For the standardized reference, \eqref{eq:KL_generic_tilt} combines with
\begin{align*}
\varphi_{\mathrm{N}}(x)e^{\lambda x}=e^{\lambda^2/2}\varphi_{\mathrm{N}}(x-\lambda).
\end{align*}
Thus the two pieces in \eqref{eq:KL_generic_tilt} are truncated Gaussian densities with shifted means $\lambda_1<\lambda_2$. For $\bar{\mu}=0$, the shifts satisfy $\lambda_1<0<\lambda_2$: the body moves downward and the tail upward while the overall mean remains fixed. The body and tail masses, continuity, the mean anchor, and the active KL budget reduce calibration to $(\lambda_1,\lambda_2,x_\alpha)$. Define the required tilted Gaussian mass and first-moment integrals by
\begin{align}
J_0(\lambda;u,v)&\triangleq\int_u^v \varphi_{\mathrm{N}}(x)e^{\lambda x}\,\mathrm{d}x
= e^{\lambda^2/2}\,\big(\Phi_{\mathrm{N}}(v-\lambda)-\Phi_{\mathrm{N}}(u-\lambda)\big),
\label{eq:J0}\\
J_1(\lambda;u,v)&\triangleq\int_u^v x\,\varphi_{\mathrm{N}}(x)e^{\lambda x}\,\mathrm{d}x\notag\\
&= e^{\lambda^2/2}\,\Big(\lambda\mathbin{\cdot}\big(\Phi_{\mathrm{N}}(v-\lambda)-\Phi_{\mathrm{N}}(u-\lambda)\big)
-\big(\varphi_{\mathrm{N}}(v-\lambda)-\varphi_{\mathrm{N}}(u-\lambda)\big)\Big),
\end{align}
both closed forms following from the shift identity above and the truncated-normal mean formula.

\begin{theorem}\label{theo-KL-normal}
Fix $\alpha\in(0,1)$, $\bar{\mu}\in\mathbb{R}$, and $\epsilon>\bar{\mu}^2/2$, let $\phi(u)=(1-\alpha)^{-1}\mathbf{1}_{\{u\geq\alpha\}}$, and consider $\mathcal{R}_\phi\big(\widetilde{\mathcal{F}}_{1\mathrm{N}}(0,1;\bar{\mu},\epsilon)\big)$ in \eqref{eq:chi_def}. Then the calibration system
\begin{align}
&\frac{\alpha\, e^{\lambda_1 x_\alpha}}{J_0(\lambda_1;-\infty,x_\alpha)}=\frac{(1-\alpha)\, e^{\lambda_2 x_\alpha}}{J_0(\lambda_2;x_\alpha,\infty)},\label{eq:N_match}\\
&\alpha\,\frac{J_1(\lambda_1;-\infty,x_\alpha)}{J_0(\lambda_1;-\infty,x_\alpha)}
+(1-\alpha)\,\frac{J_1(\lambda_2;x_\alpha,\infty)}{J_0(\lambda_2;x_\alpha,\infty)}=\bar{\mu}, \label{eq:N_mean}\\
&\alpha\ln\frac{\alpha}{J_0(\lambda_1;-\infty,x_\alpha)}
+(1-\alpha)\ln\frac{1-\alpha}{J_0(\lambda_2;x_\alpha,\infty)}
\notag\\
&\quad+\lambda_1\,\alpha\,\frac{J_1(\lambda_1;-\infty,x_\alpha)}{J_0(\lambda_1;-\infty,x_\alpha)}
+\lambda_2\mathbin{\cdot}(1-\alpha)\frac{J_1(\lambda_2;x_\alpha,\infty)}{J_0(\lambda_2;x_\alpha,\infty)}=\epsilon.
\label{eq:N_kl}
\end{align}
has exactly one solution $(\lambda_1,\lambda_2,x_\alpha)$ with $\lambda_1<\lambda_2$. Set $C_1=\alpha/J_0(\lambda_1;-\infty,x_\alpha)$ and $C_2=(1-\alpha)/J_0(\lambda_2;x_\alpha,\infty)$. The unique worst-case density is
\begin{align}
\tilde{f}^*(x)=
\begin{cases}
C_1\,\varphi_{\mathrm{N}}(x)\,e^{\lambda_1 x}, & x<x_\alpha,\\[2pt]
C_2\,\varphi_{\mathrm{N}}(x)\,e^{\lambda_2 x}, & x\geq x_\alpha,
\end{cases}
\label{eq:fstar_normal}
\end{align}
which is continuous, places mass $\alpha$ on $(-\infty,x_\alpha)$, and has $\alpha$-quantile $x_\alpha$. The optimal value is the tail mean of $\tilde{f}^*$,
\begin{align}
\chi(\alpha,\epsilon;\bar{\mu})
=\frac{J_1(\lambda_2;x_\alpha,\infty)}{J_0(\lambda_2;x_\alpha,\infty)}
=\lambda_2+\frac{\varphi_{\mathrm{N}}(x_\alpha-\lambda_2)}{1-\Phi_{\mathrm{N}}(x_\alpha-\lambda_2)}.
\label{eq:chi_value}
\end{align}
\end{theorem}

The calibrated density satisfies Proposition~\ref{prop:unbounded-sufficiency} with $\zeta^*=((1-\alpha)(\lambda_2-\lambda_1))^{-1}$ and $\eta^*=-\zeta^*\lambda_1$; strict convexity of the active KL term gives uniqueness. Calibration is available for every $\epsilon>\bar{\mu}^2/2$. Appendix~\ref{proof_theo_KL_normal} establishes this existence claim through a smooth convex dual in the multipliers and the ES threshold: its value equals $\chi(\alpha,\epsilon;\bar{\mu})$, it attains its minimum at a positive entropy multiplier, and its first-order conditions are precisely \eqref{eq:N_match}--\eqref{eq:N_kl}.

The coefficient $\chi$ can therefore be computed either from \eqref{eq:N_match}--\eqref{eq:N_kl} or by minimizing that dual; the two routes give the same density and value, and numerical output should be checked against the matching, mean, and KL equations. For $\bar{\mu}=0$, $\chi(\alpha,\epsilon)$ approaches the classical Gaussian ES multiplier as $\epsilon\downarrow0$ and diverges as $\epsilon\uparrow\infty$. The piecewise exponential tilt is consistent with the entropic-robustness literature \citep{glasserman2014robust}; here the mean anchor, uniqueness certificate, and location--scale reduction support the portfolio application.

\section{Application: KL-Robust Portfolio Selection via Exact Projection}\label{subsec:ext_portfolio}

In this section, we use portfolio optimization to illustrate how the scalar-loss theory in Section~\ref{sec:cases-F} can address high-dimensional uncertainty. Asset losses form a random vector, while each allocation produces a scalar portfolio loss. We consider a joint KL neighborhood of a Gaussian reference with every asset mean fixed at its nominal value.

We establish an exact projection of this joint ambiguity set onto each portfolio-loss direction: every law in the corresponding univariate mean-anchored KL set admits a joint extension preserving all asset means with the same KL cost. Theorem~\ref{theo-KL-normal} then reduces worst-case risk evaluation to one univariate calibration shared by all allocations, leaving an exact SOCP for the portfolio weights. To our knowledge, this reduction for the mean-anchored KL--ES portfolio model has not previously been established.\footnote{\citet{glasserman2014robust} develop the general entropic framework with expectation constraints and study both mean-constrained Gaussian MV portfolios and worst-case CVaR.}

Fix $\alpha\in(0,1)$ and $\epsilon>0$. We consider $d>1$ assets with random loss vector $Y\in\mathbb{R}^d$. Its nominal law is $\widehat{P}=\mathsf{N}(\hat{\mu},\widehat{\Sigma})$, the multivariate normal law with mean vector $\hat{\mu}\in\mathbb{R}^d$ and positive definite covariance matrix $\widehat{\Sigma}\succ0$. Define
\begin{align*}
\mathcal{P}_{\epsilon}\triangleq
\Big\{P\ll\widehat{P}~\Big|~\mathrm{KL}(P\,\|\,\widehat{P})\leq\epsilon,\quad \mathbb{E}_P[Y]=\hat{\mu}\Big\}.
\end{align*} The same joint ambiguity set is used for all portfolios. Let $w\in\mathcal{W}\subseteq\mathbb{R}^d\setminus\{0\}$ be the vector of portfolio weights. Portfolio sets defined by linear constraints and convex quadratic inequalities are second-order-cone representable. We consider
\begin{align}
\min_{w\in\mathcal{W}}\ V(w),\qquad
V(w)\triangleq\sup_{P\in\mathcal{P}_{\epsilon}}\ \mathrm{ES}_{\alpha}\big(w^\top Y;P\big). \label{eq:KL_port_prob}
\end{align}
Write $\mu_w=w^\top\hat{\mu}$, $\sigma_w=(w^\top\widehat{\Sigma} w)^{1/2}$, and let $\mathrm{proj}_w(\mathcal{P}_\epsilon)$ denote the laws of $w^\top Y$ induced by $\mathcal{P}_\epsilon$. By law invariance,
\begin{align}
V(w)=\sup\{\mathrm{ES}_\alpha(F):F\in\mathrm{proj}_w(\mathcal{P}_\epsilon)\}.\label{eq:law_inv_reduction}
\end{align}
The following lemma identifies this projected ambiguity set exactly.

\begin{lemma}\label{lem:KL_proj}
Let $\mathcal{P}_\epsilon$ be the joint mean-anchored KL set above, with $\widehat{P}=\mathsf{N}(\hat{\mu},\widehat{\Sigma})$, $\widehat{\Sigma}\succ0$, and $\epsilon\geq0$. For every $w\in\mathbb{R}^d\setminus\{0\}$,
\begin{align*}
\mathrm{proj}_w(\mathcal{P}_{\epsilon})
=\widetilde{\mathcal{F}}_{1\mathrm{N}}\big(\mu_w,\sigma_w;\,\mu_w,\epsilon\big),
\end{align*}
the Gaussian-reference ambiguity set \eqref{eq:F1N_def} with reference parameters $(\mu_w,\sigma_w)$, mean anchored at the reference mean, and the same radius $\epsilon$.
\end{lemma}

For $P\in\mathcal{P}_\epsilon$, let $f$ be the density of $w^\top Y$ under $P$, and write $P(\cdot\mid t)$ and $\widehat{P}(\cdot\mid t)$ for the conditional laws of $Y$ given $w^\top Y=t$. The entropy chain rule \citep[see, e.g.,][]{cover2006elements,dupuis1997weak} gives
\begin{align}
\mathrm{KL}(P\|\widehat{P})
&=\mathrm{KL}\big(f\,\|\,g_{\mu_w,\sigma_w}\big)+\int_{\mathbb{R}}\mathrm{KL}\big(P(\cdot\mid t)\,\big\|\,\widehat{P}(\cdot\mid t)\big)\,f(t)\,\mathrm{d}t.
\label{eq:KL_chain}
\end{align}
The nonnegative conditional term shows that projection cannot increase KL divergence. Conversely, reweighting $\widehat{P}$ only through $w^\top Y$ preserves its conditional laws, so that term vanishes. Under the Gaussian reference,
\begin{align*}
\mathbb{E}_{\widehat{P}}[Y\mid w^\top Y=t]
=\hat{\mu}+\frac{\widehat{\Sigma}w}{\sigma_w^2}\mathbin{\cdot}(t-\mu_w).
\end{align*} Integrating this affine conditional mean against any reweighted marginal with mean $\mu_w$ returns the full vector $\hat{\mu}$. Thus every law in the univariate set has a feasible joint extension with the same KL cost, establishing equality of the projected sets without rescaling the radius. Convexity of the robust portfolio objective holds more generally, without requiring a Gaussian reference.

\begin{proposition}
\label{prop:port_convex}
For an arbitrary nominal law $\widehat{P}$, suppose the ambiguity set $\mathcal{P}_\epsilon$ is nonempty and independent of $w$. Replace $\mathrm{ES}_\alpha$ in \eqref{eq:KL_port_prob} by a law-invariant coherent risk measure, and assume that all portfolio losses belong to its linear domain under every $P\in\mathcal{P}_\epsilon$. Then $V$ is convex on the convex portfolio set $\mathcal{W}$, with the supremum allowed to take the value $+\infty$.
\end{proposition}

For the Gaussian reference, combining \eqref{eq:law_inv_reduction}, Lemma~\ref{lem:KL_proj}, and the standardization of Lemma~\ref{lem:standardize} with $\bar{\mu}=0$ gives
\begin{align}
V(w)=\mathcal{R}_\phi\big(\widetilde{\mathcal{F}}_{1\mathrm{N}}(\mu_w,\sigma_w;\mu_w,\epsilon)\big)= \mu_w+\sigma_w\,\chi(\alpha,\epsilon)\qquad\text{for every }w\in\mathcal{W}.
\label{eq:V_socp}
\end{align}
Theorem~\ref{theo-KL-normal} gives finiteness of $\chi(\alpha,\epsilon)$, and feasibility of the nominal standard normal law gives $\chi(\alpha,\epsilon)\geq\mathrm{ES}_\alpha(\mathsf{N}(0,1))>0$.

For fixed $(\alpha,\epsilon)$, the same coefficient serves every $w$ and every asset dimension. Theorem~\ref{theo-KL-normal} with $\bar{\mu}=0$ supplies the unique calibration triple $(\lambda_1,\lambda_2,x_\alpha)$ and standardized worst-case density $\tilde{f}^*$; let $f_w^*(x)=\sigma_w^{-1}\tilde{f}^*((x-\mu_w)/\sigma_w)$ and $F_w^*$ denote the corresponding worst-case marginal density and CDF. We write $\widehat{\Sigma}^{1/2}$ for the symmetric positive definite square root of $\widehat{\Sigma}$ and use $\|\cdot\|_2$ for the Euclidean norm of a vector. The following theorem gives an exact SOCP formulation of \eqref{eq:KL_port_prob} and characterizes the worst-case joint law for each portfolio.

\begin{theorem}\label{thm:KL_port}
Fix $\alpha\in(0,1)$ and $\epsilon>0$, and let $\widehat{P}=\mathsf{N}(\hat{\mu},\widehat{\Sigma})$ with $\widehat{\Sigma}\succ0$. Suppose $\mathcal{W}\subseteq\mathbb{R}^d\setminus\{0\}$ is nonempty, compact, and second-order-cone representable. Then problem \eqref{eq:KL_port_prob} is equivalent to
\begin{align}
\min_{w\in\mathcal{W}}\ \Big\{w^\top\hat{\mu}+\chi(\alpha,\epsilon)\,\|\widehat{\Sigma}^{1/2}w\|_2\Big\}.
\label{eq:port_socp}
\end{align}
This problem attains its minimum and admits an SOCP representation. Model ambiguity enters only through the finite positive scalar $\chi(\alpha,\epsilon)$. Moreover:
\begin{enumerate}[label=(\roman*)]
\item For each $w\in\mathcal{W}$, the worst-case joint law $P^*_w\in\mathcal{P}_\epsilon$ is unique and given by the reweighting
\begin{align*}
\mathrm{d}P^*_w(y)=\frac{f^*_w(w^\top y)}{g_{\mu_w,\sigma_w}(w^\top y)}\,\mathrm{d}\widehat{P}(y).
\end{align*}
Its conditional law of $Y$ given $w^\top Y=t$ coincides with the nominal conditional law for $F_w^*$-almost every $t$.
\item As a function on $\mathbb{R}^d\setminus\{0\}$, $V$ is continuously differentiable, with
\begin{align}
\begin{aligned}
\nabla V(w)
&=\hat{\mu}+\chi(\alpha,\epsilon)\,\frac{\widehat{\Sigma} w}{\sqrt{w^\top\widehat{\Sigma} w}}\\
&=\hat{\mu}+\frac{\widehat{\Sigma} w}{\sigma_w^{2}}\big(V(w)-\mu_w\big)\\
&=(1-\alpha)^{-1}\,\mathbb{E}_{P^*_w}\big[Y\,\mathbf{1}_{\{w^\top Y\geq x_\alpha(w)\}}\big],
\end{aligned}
\label{eq:danskin_grad}
\end{align}
where $x_\alpha$ is the $\alpha$-quantile of the standardized worst case $\tilde{f}^*$ of \eqref{eq:fstar_normal}, so that $x_\alpha(w)=\mu_w+\sigma_w x_\alpha$ is the $\alpha$-quantile of $F^*_w$; the last expression is the Danskin gradient evaluated at the unique maximizer $P^*_w$.
\end{enumerate}
\end{theorem}

The conic form in \eqref{eq:port_socp} has precedents in KL-robust expectation and entropic-value-at-risk models \citep{hu2012kullback,ahmadijavid2012entropic,ahmadijavid2019portfolio}. Lemma~\ref{lem:KL_proj} makes this form exact under the joint mean constraint. Optimal allocations are efficient for the nominal mean--standard-deviation trade-off over $\mathcal{W}$. For each $w$, the unique worst marginal exhausts the KL budget, so the worst joint law preserves the nominal conditional law given portfolio loss. It may depend on $w$ and need not preserve the unconditional covariance $\widehat{\Sigma}$; \eqref{eq:danskin_grad} gives the marginal risk contributions.

\section{Conclusion}

We develop a convex variational theory for worst-case spectral risk by choosing the quantile or distribution representation that preserves the ambiguity set's convexity. The resulting optimality conditions characterize extremal laws through stationarity, calibration, and admissibility. They recover known bounds, yield analytical solutions for moment, tail, and divergence models, and reduce a multivariate mean-anchored KL--ES portfolio problem to an exact SOCP. The same conditions support projection algorithms with convergence guarantees for fixed smooth convex discretizations under the stated assumptions; grid-refinement convergence is not established here.

Further work could extend the exact projection beyond Gaussian references and linear losses, incorporate additional dependence constraints, and combine data-based ambiguity with moment and tail information. Reinsurance design and energy procurement offer applications in which decisions transform multivariate uncertainty into a scalar loss.

\noindent\textbf{Code and data availability.} The code, reference data, and README with reproduction instructions are provided in the accompanying ZIP archive.

\clearpage
\appendix
\titleformat{\section}{\large\bfseries}{Appendix \thesection.}{0.6em}{}
\counterwithin{proposition}{section}
\counterwithin{lemma}{section}
\counterwithin{example}{section}
\counterwithin{theorem}{section}
\numberwithin{equation}{section}
\renewcommand{\theequation}{\thesection.\arabic{equation}}

\begin{center}
\large\textbf{Appendices}
\end{center}
The appendices are organized as follows.
\begin{itemize}[leftmargin=*,itemsep=4pt,parsep=0pt,topsep=4pt]
\item \textbf{Appendix~\ref{appendix:preliminaries}} collects the functional-analytic preliminaries, SRM representations, and functional KKT principle.
\item \textbf{Appendix~\ref{appendix:Q_proofs}} provides proofs and supporting calculations for the quantile-space optimality system, Wasserstein--ES model, and two-sided ES band.
\item \textbf{Appendix~\ref{appendix:F_proofs}} provides proofs and supporting calculations for the distribution-space optimality system, partial moments, and KL models.
\item \textbf{Appendix~\ref{appendix:port_proofs}} proves the portfolio results, including exact KL projection.
\item \textbf{Appendix~\ref{appendix:algorithms}} presents the projection algorithms, fixed-grid convergence analysis, and numerical checks.
\item \textbf{Appendix~\ref{appendix:analytical_examples}} proves the central- and geometric-moment results stated in Sections~\ref{subsec:central_moment} and~\ref{subsec:geometric_moment}.
\end{itemize}

\section{Auxiliary Results and Preliminaries}\label{appendix:preliminaries}

\subsection{Functional-Analytic Notation}\label{appendix:functional_notation}

The variational optimality systems in Sections~\ref{sec:cases-Q} and~\ref{sec:cases-F} use dual pairings, first variations, and normal cones. We collect their functional-analytic definitions here for use in the proofs. The notation follows standard Banach-space convex analysis; see, e.g., \citet{luenberger1997optimization,ekeland1999convex}. Let $S$ be a Banach space and let $S^*$ denote its continuous dual space. For $\xi\in S^*$ and $H\in S$, the notation $\langle \xi,H\rangle$ denotes the action of the continuous linear functional $\xi$ on the element $H$, that is, $ \langle \xi,H\rangle=\xi(H)$. Thus, the dual pairing should not be understood as an ordinary product. In concrete $L^p$--$L^q$ settings, the dual pairing is represented by integration. For example, if $S=L^p[0,1]$ and $S^*=L^q[0,1]$, with $1/p+1/q=1$, then
\begin{align*}
\langle \pi,Q\rangle=\int_0^1 \pi(u)Q(u)\,\mathrm{d}u,
\qquad \pi\in L^q[0,1],\ Q\in L^p[0,1].
\end{align*}

In the distribution-based formulation, we take $S=W^{1,r}[a,b]$. Its dual $S^*=(W^{1,r}[a,b])^*$ is the space of continuous linear functionals on $W^{1,r}[a,b]$. In the strong-form calculations below, such dual elements are often represented by integrals after integration by parts. For a functional $\mathcal{U}:S\to\mathbb{R}$, its G\^ateaux derivative at $G\in S$ is denoted by $\mathrm{D}\mathcal{U}(G)\in S^*$ and is defined through
\begin{align*}
\left\langle \mathrm{D}\mathcal{U}(G),H\right\rangle
\triangleq
\lim_{\varepsilon\to0}
\frac{\mathcal{U}(G+\varepsilon H)-\mathcal{U}(G)}{\varepsilon},
\qquad H\in S,
\end{align*}
whenever the limit exists for every direction $H\in S$ and defines a bounded linear functional of $H$. If a functional depends on several variables, such as a Lagrangian $\mathcal{L}(G,\gamma,\lambda)$ (with Lagrange multipliers $\gamma,\lambda$ as introduced in Appendix~\ref{appendix-functionals}), then $\mathrm{D}_G\mathcal{L}(G,\gamma,\lambda)$
denotes the G\^ateaux derivative of $\mathcal{L}$ with respect to the primal variable $G$ only, with the multipliers $(\gamma,\lambda)$ held fixed.

In later proofs, the abstract primal variable $G$ is specialized according to the formulation under consideration. In the quantile-based formulation, $G$ is replaced by the quantile function $Q$, and we write $\mathrm{D}_Q\mathcal{L}(Q,\gamma,\lambda)$.
In the distribution-based formulation, $G$ is replaced by the distribution function $F$, and we write
$\mathrm{D}_F\mathcal{L}(F,\eta,\zeta)$. When only admissible one-sided directions are available, the same notation denotes a bounded linear functional representing the relative first variation specified in Appendix~\ref{appendix-functionals}; ambient G\^ateaux differentiability is not implied.

For a nonempty closed convex set $\mathcal{G}_0\subseteq S$, the normal cone at $G\in\mathcal{G}_0$ is defined by $ N_{\mathcal{G}_0}(G)
\triangleq
\left\{\xi\in S^*~\middle|~\langle \xi,V-G\rangle\leq0,\ \forall V\in\mathcal{G}_0\right\}.$
Thus, an inclusion such as $-\mathrm{D}_G\mathcal{L}(G^*,\gamma^*,\lambda^*)\in N_{\mathcal{G}_0}(G^*)$ means that the negative first-variation functional of the Lagrangian, viewed as an element of $S^*$, supports the convex set $\mathcal{G}_0$ at $G^*$. Equivalently, it means that $\left\langle \mathrm{D}_G\mathcal{L}(G^*,\gamma^*,\lambda^*),V-G^*\right\rangle\geq0,
\qquad \forall V\in\mathcal{G}_0.$
This is the infinite-dimensional analogue of the first-order variational inequality for constrained convex optimization.

\subsection{Distribution-Based Representation of SRMs}\label{append:obj2prob}

For the bounded-support formulation in Section~\ref{subsec:distribution_formulation}, we derive the CDF objective \eqref{eq:srm_def_distribution} from the quantile representation and explain its concavity. The general identity allowing unbounded support is treated in Appendix~\ref{app:min-rep}. For a loss random variable with quantile function $Q$, the SRM induced by the spectrum $\phi$ is $R_{\phi}(Q)=\int_0^1 Q(u)\phi(u)\,\mathrm{d}u$, where $\phi$ is nonnegative and nondecreasing, and satisfies $\int_0^1\phi(u)\,\mathrm{d}u=1$. Define $\Phi(t)\triangleq\int_0^t\phi(u)\,\mathrm{d}u$, $t\in[0,1]$. Then $\Phi(0)=0$, $\Phi(1)=1$, and $\Phi$ is convex. Hence $\Phi$ is a distortion function, and the SRM above can be viewed as the distortion risk measure generated by $\Phi$; see, e.g., \citet{wang1995choquet,wang2000choquet,dhaene2012LS}.

Let $F$ be the distribution function of the loss, supported on $[a,b]$, and let $Q$ be its quantile function. At the level of Stieltjes integrals, the corresponding distribution-based representation is $\int_a^b x\,\mathrm{d}\Phi(F(x)).$
Since $\Phi'(u)=\phi(u)$ a.e., the Stieltjes change of variables gives $\int_a^b x\,\mathrm{d}\Phi(F(x))=\int_0^1 Q(u)\,\mathrm{d}\Phi(u)=R_{\phi}(Q)$.
Therefore, the quantile-based SRM representation and the distribution-based Stieltjes representation are equivalent whenever the above integrals are well defined.

For the bounded-support variational analysis, we further assume that $F$ is absolutely continuous on $[a,b]$, with density $f=F'$. Then
\begin{align*}
R_{\phi}(F)=\int_a^b x\phi(F(x))\,\mathrm{d}F(x)=\int_a^b x\phi(F(x))f(x)\,\mathrm{d}x.
\end{align*}
For variational analysis, it is more convenient to remove the derivative from the objective. Using the definition of $\Phi$ and integration by parts, we obtain $\int_a^b x\,\mathrm{d}\Phi(F(x))=x\Phi(F(x))\big|_a^b-\int_a^b\Phi(F(x))\,\mathrm{d}x$.
Since $F(a)=0$, $F(b)=1$, $\Phi(0)=0$, and $\Phi(1)=1$, this reduces to $R_{\phi}(F)=b-\int_a^b\Phi(F(x))\,\mathrm{d}x.$
This is the objective form used in the bounded-support distribution-based WCSRM analysis. Since $\phi$ is nondecreasing, $\Phi$ is convex, and hence $F\mapsto b-\int_a^b\Phi(F(x))\,\mathrm{d}x$ is concave in $F$. This concavity is the basis for treating the distribution-based WCSRM problem as a concave maximization problem over a convex ambiguity set.

\subsection{A Functional KKT Principle}\label{appendix-functionals}

The quantile and bounded-support distribution optimality systems in Propositions~\ref{prop-Q-WCSRM-sol} and~\ref{prop-P-WCSRM-sol} follow from the same functional KKT principle. We state it here, distinguishing existence of an optimizer from the constraint qualification needed for multipliers. Let $S$ be a reflexive Banach space and let $\mathcal{G}_0\subseteq S$ be a nonempty closed convex set. Consider the constrained convex functional optimization problem
\begin{align}
\inf_{G\in\mathcal{G}_0} &\quad \mathcal{T}(G) \label{prob:con-fun} \\
\text{s.t.} &~ \mathcal{M}_i(G)=0,\ i=1,\ldots,m, \notag \\
            &~ \mathcal{N}_j(G)\leq0,\ j=1,\ldots,n,\notag
\end{align}
where $\mathcal{T}:S\to\overline{\mathbb{R}}$ is proper, convex, and lower semicontinuous, each $\mathcal{M}_i$ is continuous affine, and each $\mathcal{N}_j$ is convex and lower semicontinuous. Assume $\mathcal{T}$ and the $\mathcal{N}_j$ are finite on $\mathcal{G}_0$; they may be extended by $+\infty$ outside their natural domains.
At the feasible point $G^*$ under consideration, assume that the one-sided variations of $\mathcal{T}$ and the constraint functionals along every segment from $G^*$ to $V\in\mathcal{G}_0$ are represented by bounded linear functionals in $S^*$. For such a functional $\mathcal{U}$, write
\begin{align*}
\lim_{\varepsilon\downarrow0}\frac{\mathcal{U}(G^*+\varepsilon\mathbin{\cdot}(V-G^*))-\mathcal{U}(G^*)}{\varepsilon}
=\langle\mathrm{D}\mathcal{U}(G^*),V-G^*\rangle.
\end{align*}
An ordinary G\^ateaux derivative is sufficient, but this relative version does not require inadmissible perturbations, such as negative densities. The representing functional need not be unique outside the span of $\mathcal{G}_0-G^*$; all normal-cone statements below depend only on its pairings with admissible differences. For the sufficiency direction alone, a supporting functional on $\mathcal{G}_0$ suffices, even if the relative derivative does not exist.
The feasible set is denoted by
\begin{align*}
\mathcal{G}\triangleq\left\{G\in\mathcal{G}_0~\middle|~
\begin{aligned}
&\mathcal{M}_i(G)=0, &&i=1,\ldots,m,\\
&\mathcal{N}_j(G)\leq0, &&j=1,\ldots,n
\end{aligned}\right\}.
\end{align*}
If $\mathcal{G}$ is nonempty and weakly compact, then \eqref{prob:con-fun} admits an optimal solution. The same conclusion holds when weak compactness follows from a coercive constraint. If $\mathcal{T}$ is strictly convex on $\mathcal{G}$, the optimal solution is unique.

For a multiplier characterization at a feasible $G^*$, impose the following constraint qualification: there exists $\bar{G}\in\mathcal{G}_0$ with $\mathcal{N}_j(\bar{G})<0$ for all $j=1,\ldots,n$, and the equality constraints are qualified in the sense that
\begin{align*}
0\in\operatorname{int}\Big\{\big(\mathcal{M}_i(G)\big)_{i=1}^m~:~G\in\mathcal{G}_0,\ \mathcal{N}_j(G)<0\ \text{for all }j=1,\ldots,n\Big\},
\end{align*}
the interior taken in $\mathbb{R}^m$. This is the generalized Slater, or Robinson, constraint qualification for a convex program over an abstract convex set $\mathcal{G}_0$; the interior condition on the attainable equality values, and not merely the existence of a strictly feasible point, is what excludes abnormal multipliers when $\mathcal{G}_0$ has empty interior---as does the monotone cone $\mathcal{Q}_0$ in the quantile formulation \citep{bonnans2000perturbation}.
In the quantile models where necessity is used, the sole equality is a mean; small constant shifts of a strictly feasible quantile vary that mean in a neighborhood while preserving strict inequalities. The explicit distribution-space constructions instead use sufficiency and do not presume a constraint qualification for every moment model.
Define the Lagrangian functional
\begin{align*}
\mathcal{L}(G,\gamma,\lambda)\triangleq\mathcal{T}(G)+\sum_{i=1}^m\gamma_i\mathcal{M}_i(G)+\sum_{j=1}^n\lambda_j\mathcal{N}_j(G),
\end{align*}
where $\gamma_i\in\mathbb{R}$ and $\lambda_j\geq0$.

By the notation introduced in Appendix~\ref{appendix:functional_notation}, the linear functional representing the relative first variation of the Lagrangian at $G^*$ is
\begin{align*}
\mathrm{D}_G\mathcal{L}(G^*,\gamma,\lambda) =\mathrm{D}\mathcal{T}(G^*)
+\sum_{i=1}^m\gamma_i\mathrm{D}\mathcal{M}_i(G^*)
+\sum_{j=1}^n\lambda_j\mathrm{D}\mathcal{N}_j(G^*)
\in S^*.
\end{align*}
For a continuous affine functional $\mathcal{M}_i$, its derivative $\mathrm{D}\mathcal{M}_i(G)$ is the continuous linear part of $\mathcal{M}_i$ and is independent of $G$.

Under this Slater-type condition, a feasible $G^*$ is optimal for \eqref{prob:con-fun} if and only if there exist multipliers $\gamma^*\in\mathbb{R}^m$ and $\lambda^*\in\mathbb{R}^n_+$ such that
\begin{align}
-\mathrm{D}_G\mathcal{L}(G^*,\gamma^*,\lambda^*)\in N_{\mathcal{G}_0}(G^*),
\label{eq:functional_kkt_stationarity}
\end{align}
and
\begin{align*}
\mathcal{M}_i(G^*)=0,\quad i=1,\ldots,m,\qquad \mathcal{N}_j(G^*)\leq0,\quad \lambda_j^*\geq0,\quad \lambda_j^*\mathcal{N}_j(G^*)=0,\quad j=1,\ldots,n.
\end{align*}
Here $\mathrm{D}_G\mathcal{L}$ represents the relative first variation at $G^*$, and $N_{\mathcal{G}_0}$ is the normal cone defined in Appendix~\ref{appendix:functional_notation}. Equivalently, $\langle\mathrm{D}_G\mathcal{L},V-G^*\rangle\geq0$ for every $V\in\mathcal{G}_0$. This is the infinite-dimensional analogue of the finite-dimensional KKT conditions.

The sufficiency direction (if the multipliers and the stationarity and complementary-slackness conditions hold, then $G^*$ is optimal) is elementary convex analysis and needs only convexity and the supporting inequalities at $G^*$; see, e.g., \citet{luenberger1997optimization,ekeland1999convex}. The necessity direction (an optimal $G^*$ admits such multipliers) uses the constraint qualification above and follows from the generalized Karush--Kuhn--Tucker theory for convex programs over convex sets \citep{bonnans2000perturbation}. In the sequel, we apply \eqref{eq:functional_kkt_stationarity} to the minimization form of the WCSRM problems, obtained by changing the sign of the maximization objective. Specifically, we use $S=L^p[0,1]$ for the quantile-based formulation and $S=W^{1,r}[a,b]$ for the distribution-based formulation.

\subsection{Minimization Representations and Duality with the Distribution-Based Formulation}
\label{app:min-rep}

Section~\ref{sec:formulation} uses a direct CDF representation of the SRM. We justify that identity and its bounded-support specialization, then relate the latter to the classical positive-part minimization formulas and the corresponding worst-case minimax exchange.

For the general identity, the image under $Q$ of the probability measure with density $\phi$ on $(0,1)$ has CDF $\int_0^1\mathbf{1}_{\{Q(u)\leq x\}}\phi(u)\,\mathrm{d}u=\Phi(F(x))$. Its first absolute moment is $\int_0^1|Q(u)|\phi(u)\,\mathrm{d}u$, which is finite under the integrability condition in Section~\ref{sec:formulation}. Its expectation is therefore both $R_\phi(F)=\int_{\mathbb{R}}x\,\mathrm{d}\big(\Phi\circ F\big)(x)$ and the difference of the positive and negative tail integrals in \eqref{eq:F_objective_unbounded}. This proves the identity without a bounded-support or absolute-continuity assumption.

To obtain \eqref{eq:srm_def_distribution} from \eqref{eq:F_objective_unbounded}, suppose the law is supported on a finite interval $[a,b]$. Since $\Phi(0)=0$ and $\Phi(1)=1$, we have $\Phi(F(x))=0$ for $x<a$ and $\Phi(F(x))=1$ for $x\geq b$. Set $a_0=\min\{a,0\}$ and $b_0=\max\{b,0\}$. The two tail integrals then give
\begin{align*}
R_\phi(F)&=\int_0^{b_0}\big(1-\Phi(F(x))\big)\,\mathrm{d}x-\int_{a_0}^0\Phi(F(x))\,\mathrm{d}x\\
&=b_0-(b_0-b)-\int_a^b\Phi(F(x))\,\mathrm{d}x
=b-\int_a^b\Phi(F(x))\,\mathrm{d}x.
\end{align*}
Here the integral over $(a_0,a)$ is zero, and the integral over $(b,b_0)$ equals $b_0-b$. The calculation applies whether or not $[a,b]$ contains zero, and also allows atoms at either endpoint, since individual points do not affect these Lebesgue integrals. If $F$ is absolutely continuous, the chain rule additionally yields the density form used in Sections~\ref{subsec:distribution_formulation} and~\ref{PRRM-case-KL-Gaussian}.

For the minimization representations below, take a bounded, nondecreasing, right-continuous spectrum on $[0,1)$, with its finite left limit used at $1$, and bounded loss support $[a,b]$. These conditions include the ES and smooth spectra used in the analytical models. We relate the bounded-support objective \eqref{eq:srm_def_distribution} to the classical minimization representations of ES and SRMs through pointwise Fenchel conjugacy. Recall the representation of \citet{rockafellar2000optimization,rockafellar2002conditional}:
\begin{align}
\mathrm{ES}_{\alpha}(F)=\min_{t\in\mathbb{R}}\left\{t+\frac{1}{1-\alpha}\int_a^b(x-t)_+\,\mathrm{d}F(x)\right\}.
\label{eq:RU_ES}
\end{align}
For $0<\alpha<1$, the minimizers are precisely the $t\in[a,b]$ satisfying $F(t^-)\leq\alpha\leq F(t)$; $Q(\alpha)$ is one canonical selection. A flat segment of $F$ at level $\alpha$ can give an interval of minimizers. At $\alpha=0$, use the endpoint convention $Q(0)=a$ and choose $t(0)=Q(0)$.
 By the Kusuoka representation \citep{kusuoka2001law}, every SRM with nondecreasing spectrum $\phi$ normalized by $\int_0^1\phi(u)\,\mathrm{d}u=1$ is a mixture of ES levels, $R_\phi(F)=\int_{[0,1)}\mathrm{ES}_{\alpha}(F)\,\mu_\phi(\mathrm{d}\alpha)$, where $\mu_\phi(\mathrm{d}\alpha)=\phi(0)\,\delta_0(\mathrm{d}\alpha)+(1-\alpha)\,\mathrm{d}\phi(\alpha)$ is a probability measure on $[0,1)$, the second term understood as a Lebesgue--Stieltjes measure; for the ES spectrum $\phi=(1-\alpha)^{-1}\mathbf{1}_{[\alpha,1]}$ it reduces to $\delta_\alpha$. Substituting \eqref{eq:RU_ES} level by level and interchanging the minimization with the mixture integral --- legitimate because the minimization is pointwise in $\alpha$ and admits the measurable selection $t(\alpha)=Q(\alpha)$ --- yields the minimization representation
\begin{align}
R_\phi(F)=\min_{t(\cdot)}\int_{[0,1)}\left(t(\alpha)+\frac{1}{1-\alpha}\int_a^b\big(x-t(\alpha)\big)_+\,\mathrm{d}F(x)\right)\mu_\phi(\mathrm{d}\alpha),
\label{eq:min_rep_SRM}
\end{align}
over measurable $t:[0,1)\to[a,b]$, attained at $t^*(\alpha)=Q(\alpha)$. Up to sign and orientation conventions, \eqref{eq:min_rep_SRM} is the representation derived by \citet{acerbi2002portfolio} for general spectra; its finite-mixture form is the mixed-quantile quadrangle of \citet{rockafellar2013fundamental}, and its single-level analogue for general convex risk measures is the optimized certainty equivalent of \citet{2010bental}.

The link with \eqref{eq:srm_def_distribution} is the following. Since $\Phi$ is convex and finite on $[0,1]$, with $\phi$ a measurable selection of its supporting slopes, it agrees with its biconjugate, $\Phi(v)=\sup_{y}\{yv-\Phi^*(y)\}$, where $\Phi^*(y)=\sup_{v\in[0,1]}\{yv-\Phi(v)\}$ is the convex conjugate. Applying this identity pointwise under the integral in \eqref{eq:srm_def_distribution}, and again selecting the maximizer measurably, gives
\begin{align}
R_\phi(F)=b-\int_a^b\Phi(F(x))\,\mathrm{d}x
=\min_{y(\cdot)}\left\{b+\int_a^b\Big(\Phi^*\big(y(x)\big)-y(x)F(x)\Big)\,\mathrm{d}x\right\},
\end{align}
over measurable $y:[a,b]\to[\phi(0),\phi(1)]$. Pointwise minimizers satisfy $y(x)\in\partial\Phi(F(x))$, where the subdifferential uses the extension of $\Phi$ by $+\infty$ outside $[0,1]$ and is intersected with the displayed range of $y$. The choice $y(x)=\phi(F(x))$ is one such selection. Thus the two representations have the same value and canonical selections $t(\alpha)=Q(\alpha)$ and $y(x)=\phi(F(x))$; there is no asserted one-to-one correspondence between all minimizing auxiliary functions. In particular, the ES kink makes the second minimizer set-valued when $F(x)=\alpha$.

Both formulas express the SRM as an infimum of affine functionals of the same law and therefore show its concavity in $F$. At an optimizer $F^*$ of a given distribution-based ambiguity problem, one may select its own quantile $Q_{F^*}$ in \eqref{eq:min_rep_SRM}. This does not identify an optimizer of a quantile-based problem with a different ambiguity set.

For the minimax exchange, assume in addition that $\mathcal{F}$ is nonempty, convex, and weakly compact in $W^{1,r}[a,b]$, as in Assumption~\ref{assump_F_regular}. Let $\mathcal{H}(F,t)$ denote the integral in \eqref{eq:min_rep_SRM} and let $t$ range over the convex subset $\{t\in L^\infty(\mu_\phi):a\leq t\leq b\}$ with its norm topology. Boundedness of the spectrum gives $\int_{[0,1)}\frac{\mu_\phi(\mathrm{d}\alpha)}{1-\alpha}=\phi(1)<\infty.$
For fixed $t$, its payoff integrand is bounded on $[a,b]$, so $\mathcal{H}(\cdot,t)$ is affine and weakly continuous in $F$. For fixed $F$, $\mathcal{H}(F,\cdot)$ is convex and norm-continuous, with a Lipschitz bound uniform in $F$. These are the compactness and semicontinuity conditions needed by Sion's minimax theorem, which yields
\begin{align*}
\sup_{F\in\mathcal{F}}\inf_t\mathcal{H}(F,t)
=\inf_t\sup_{F\in\mathcal{F}}\mathcal{H}(F,t).
\end{align*}
Without these or alternative verified minimax conditions, affinity and convexity alone do not justify the exchange. Under the stated conditions, this is the classical route through worst-case expectations of piecewise-linear losses \citep{zhu2009worst,ben2013robust}. The direct variational treatment additionally constructs the adverse laws for the models in Sections~\ref{sec:cases-Q} and~\ref{sec:cases-F}.

\section{Proofs and Calculations for Quantile-Based WCSRM}\label{appendix:Q_proofs}

The proofs below follow the quantile-space results in Section~\ref{sec:cases-Q}, with feasibility results placed before the corresponding model solutions. The central-moment result in Section~\ref{subsec:central_moment} is proved separately in Appendix~\ref{proof_theo_1_k}.

\subsection{Proof of Proposition~\ref{prop-Q-WCSRM-sol}}\label{proof_prop_Q_WCSRM_sol}

We prove the general optimality conditions for the quantile-based WCSRM problem. In this proof, the abstract primal variable $G$ in Appendix~\ref{appendix-functionals} is specialized to the quantile function $Q$. Accordingly, $\mathrm{D}_Q\mathcal{L}$ denotes the G\^ateaux derivative of the Lagrangian with respect to $Q$.

Let $1<p<\infty$ and let $q$ be the conjugate exponent of $p$, i.e., $1/p+1/q=1$. The decision space is $L^p[0,1]$, whose dual is identified with $L^q[0,1]$. The dual pairing is $\langle \pi,Q\rangle=\int_0^1\pi(u)Q(u)\,\mathrm{d}u$, $\pi\in L^q[0,1]$, $Q\in L^p[0,1].$

The maximization problem \eqref{prob:Q_WCSRM} is equivalent to minimizing the negative SRM over the same feasible set. Define
\begin{align*}
\mathcal{T}(Q)&\triangleq-\int_0^1\phi(u)Q(u)\,\mathrm{d}u,\\
\mathcal{M}_i(Q)&\triangleq\int_0^1h_i(u)Q(u)\,\mathrm{d}u-c_i,\qquad i\in\mathcal{I}_1,\\
\mathcal{N}_i(Q)&\triangleq\int_0^1H_i(u,Q(u))\,\mathrm{d}u-d_i,\qquad i\in\mathcal{I}_2.
\end{align*}
Then \eqref{prob:Q_WCSRM} can be written as
\begin{align*}
\inf_{Q\in\mathcal{Q}_0}\quad \mathcal{T}(Q)\quad
\text{s.t.}\quad \mathcal{M}_i(Q)=0,\ i\in\mathcal{I}_1,\qquad \mathcal{N}_i(Q)\leq0,\ i\in\mathcal{I}_2.
\end{align*}
The objective $\mathcal{T}$ is a continuous linear functional on $L^p[0,1]$, and the equality-constraint residuals $\mathcal{M}_i$ are continuous affine functionals.
Each $\mathcal{N}_i$ is finite, continuous, and convex on $L^p$ under Assumption~\ref{assump:Q_dual}(ii). For $Q,v\in L^p$ and $|t|\leq1$, the mean-value formula bounds the absolute difference quotient by
\begin{align*}
\big(b_i(u)+C_i\mathbin{\cdot}(|Q(u)|^{p-1}+|v(u)|^{p-1})\big)|v(u)|,
\end{align*}
which is integrable by H\"older's inequality. Dominated convergence therefore gives the derivative in \eqref{Q-gateaux}. The same growth bounds give continuity of the integral functional under $L^p$ convergence; convexity then gives weak lower semicontinuity.
 Thus, the problem is a constrained convex functional optimization problem in the sense of Appendix~\ref{appendix-functionals}, and its constraint qualification is Assumption~\ref{assump:Q_dual}(i).

The Lagrangian associated with the minimization form is
\begin{align}
\mathcal{L}(Q,\gamma,\lambda)
&=-\int_0^1\phi(u)Q(u)\,\mathrm{d}u+\sum_{i\in\mathcal{I}_1}\gamma_i\mathbin{\cdot}\left(\int_0^1h_i(u)Q(u)\,\mathrm{d}u-c_i\right) \notag\\
&\quad+\sum_{i\in\mathcal{I}_2}\lambda_i\mathbin{\cdot}\left(\int_0^1H_i(u,Q(u))\,\mathrm{d}u-d_i\right),
\end{align}
where $\gamma_i\in\mathbb{R}$ and $\lambda_i\geq0$. For any direction $v\in L^p[0,1]$, the G\^ateaux derivative of $\mathcal{L}$ with respect to $Q$ is
\begin{align}
\left\langle \mathrm{D}_Q\mathcal{L}(Q,\gamma,\lambda),v\right\rangle
=\int_0^1\left(-\phi(u)+\sum_{i\in\mathcal{I}_1}\gamma_i h_i(u)+\sum_{i\in\mathcal{I}_2}\lambda_i\left.\frac{\partial H_i(u,y)}{\partial y}\right|_{y=Q(u)}\right)v(u)\,\mathrm{d}u.
\label{Q-gateaux}
\end{align}
The bracketed term belongs to $L^q[0,1]$ by Assumption~\ref{assump:Q_dual}(ii); hence $\mathrm{D}_Q\mathcal{L}(Q,\gamma,\lambda)$ is a bounded linear functional on $L^p[0,1]$.

By the functional KKT principle in Appendix~\ref{appendix-functionals}, a feasible $Q^*\in\mathcal{Q}$ is optimal if and only if there exist multipliers $\gamma_i^*\in\mathbb{R}$, $i\in\mathcal{I}_1$, and $\lambda_i^*\geq0$, $i\in\mathcal{I}_2$, such that
$-\mathrm{D}_Q\mathcal{L}(Q^*,\gamma^*,\lambda^*)\in N_{\mathcal{Q}_0}(Q^*)$
and $\lambda_i^* \cdot \left(\int_0^1H_i(u,Q^*(u))\,\mathrm{d}u-d_i\right)=0$, for $i\in\mathcal{I}_2.$
Using \eqref{Q-gateaux}, the stationarity condition becomes
\begin{align}
g^*\in N_{\mathcal{Q}_0}(Q^*),
\label{optimal-q-norm-append}
\end{align}
where $g^*(\cdot)\in L^q[0,1]$ is defined by
\begin{align*}
g^*(u)=\phi(u)-\sum_{i\in\mathcal{I}_1}\gamma_i^*h_i(u)-\sum_{i\in\mathcal{I}_2}\lambda_i^*\left.\frac{\partial H_i(u,y)}{\partial y}\right|_{y=Q^*(u)},\qquad \text{for a.e. }u\in[0,1].
\end{align*}
This proves Proposition~\ref{prop-Q-WCSRM-sol}.
\Halmos

Assumption~\ref{assump:Q_regular} ensures attainment because the objective is weakly continuous. At an existing optimizer, the necessity of multipliers follows from Assumption~\ref{assump:Q_dual}(i), with the derivatives justified by part~(ii); weak compactness is not needed for this step. A standard way to verify weak compactness is through a coercive constraint. If $H_{i_0}(u,y)\geq a_0|y|^p-b_0$ for a.e. $u\in[0,1]$ and all $y\in\mathbb{R}$, with constants $a_0>0$ and $b_0\geq0$, then every feasible $Q\in\mathcal{Q}$ satisfies $a_0\|Q\|_p^p-b_0\leq\int_0^1H_{i_0}(u,Q(u))\,\mathrm{d}u\leq d_{i_0}$.

Thus $\mathcal{Q}$ is bounded in $L^p[0,1]$. Each inequality sublevel set $\{Q:\int_0^1 H_i(u,Q(u))\,\mathrm{d}u\leq d_i\}$ is convex, since $H_i(u,\cdot)$ is convex, and weakly closed, by the norm continuity and convexity just established under Assumption~\ref{assump:Q_dual}(ii); the equality constraints are affine and continuous, and $\mathcal{Q}_0$ is a closed convex cone and hence weakly closed. Therefore $\mathcal{Q}$ is weakly closed, and since $L^p[0,1]$ is reflexive for $1<p<\infty$, the feasible set is weakly compact. The objective $Q\mapsto\int_0^1\phi(u)Q(u)\,\mathrm{d}u$ is weakly continuous, and hence the supremum in \eqref{prob:Q_WCSRM} is attained.

\subsection{Proof of Corollary~\ref{corollary-normal-QRRM}}

We prove Corollary~\ref{corollary-normal-QRRM}. Let $\widehat{Q}\in\mathcal{Q}$ be feasible and let multipliers $\hat{\gamma}_i\in\mathbb{R}$, $i\in\mathcal{I}_1$, and $\hat{\lambda}_i\geq0$, $i\in\mathcal{I}_2$, satisfy complementary slackness and the pointwise stationarity condition
\begin{align*}
\phi(u)-\sum_{i\in\mathcal{I}_1}\hat{\gamma}_i h_i(u)-\sum_{i\in\mathcal{I}_2}\hat{\lambda}_i\left.\frac{\partial H_i(u,y)}{\partial y}\right|_{y=\widehat{Q}(u)}=0,\qquad \text{for a.e. }u\in[0,1].
\end{align*}
Under Assumption~\ref{assump:Q_dual}(ii), the convex constraint functionals have the derivatives used in this condition. The residual in \eqref{optimal-q-norm-append} is zero a.e., and feasibility includes $\widehat{Q}\in\mathcal{Q}_0$. Since $0\in N_{\mathcal{Q}_0}(\widehat{Q})$, the full KKT system holds. The sufficiency argument in Appendix~\ref{appendix-functionals} therefore proves optimality without weak compactness or a constraint qualification.
\Halmos

\subsection{Feasibility of Wasserstein Ambiguity with ES Stress}

The analysis in Section~\ref{sub-sec-quantile-direct-solution} requires the ES floor to be compatible with the Wasserstein radius. The following lemma establishes the feasibility and Slater conditions used there before Theorem~\ref{theo-cw}.

\begin{lemma}\label{lem:nonempty_Qcw}
Let $\epsilon>0$, $\alpha\in(1/2,1)$, $c\in\mathbb{R}$, and $Q_G\in\mathcal{Q}_0\subset L^2[0,1]$. With $c_G=\mathrm{ES}_\alpha(Q_G)$, the ambiguity set $\mathcal{Q}_{\mathrm{W}\text{-}\mathrm{ES}}$ in \eqref{eq:Q_cw} is nonempty if and only if $c\leq c_G+\epsilon/\sqrt{1-\alpha}$. Moreover, the Slater condition holds if the inequality is strict.
\end{lemma}

\textbf{\emph{Proof.}} Recall that $c_G=\mathrm{ES}_\alpha(Q_G)=(1-\alpha)^{-1}\int_\alpha^1Q_G(u)\,\mathrm{d}u$. The constraints are $\mathrm{ES}_\alpha(Q)\geq c$ and $\|Q-Q_G\|_2\leq\epsilon$.

We first prove necessity. Define $h(u)\triangleq(1-\alpha)^{-1}\mathbf{1}_{\{u\geq\alpha\}}$. For any $Q$ satisfying $\|Q-Q_G\|_2\leq\epsilon$, H\"older's inequality gives
\begin{align*}
\frac{1}{1-\alpha}\int_\alpha^1 Q(u)\,\mathrm{d}u-c_G
=\int_0^1h(u)\mathbin{\cdot}(Q(u)-Q_G(u))\,\mathrm{d}u
\leq \|h\|_2\|Q-Q_G\|_2
\leq \frac{\epsilon}{\sqrt{1-\alpha}}.
\end{align*}
Thus nonemptiness requires $c\leq c_G+\epsilon/\sqrt{1-\alpha}$.

Conversely, suppose $c\leq c_G+\epsilon/\sqrt{1-\alpha}$. Choose $\delta\in[0,\epsilon]$ such that $c\leq c_G+\delta/\sqrt{1-\alpha}$ and define $\bar{Q}(u)=Q_G(u)+\delta\mathbf{1}_{\{u\geq\alpha\}}/\sqrt{1-\alpha}$. Then, since $Q_G\in\mathcal{Q}_0$ and the added step function is nondecreasing, the function $\bar{Q}$ remains nondecreasing and belongs to $\mathcal{Q}_0$. Moreover,
\begin{align*}
\|\bar{Q}-Q_G\|_2^2=\int_\alpha^1\frac{\delta^2}{1-\alpha}\,\mathrm{d}u=\delta^2\leq\epsilon^2,\qquad
\frac{1}{1-\alpha}\int_\alpha^1\bar{Q}(u)\,\mathrm{d}u=c_G+\frac{\delta}{\sqrt{1-\alpha}}\geq c.
\end{align*}
Hence $\bar{Q}\in\mathcal{Q}_{\mathrm{W}\text{-}\mathrm{ES}}$, proving sufficiency. If $c<c_G+\epsilon/\sqrt{1-\alpha}$, one can choose $\delta<\epsilon$ and $c<c_G+\delta/\sqrt{1-\alpha}$, so both the Wasserstein constraint and the ES lower-bound constraint are strictly satisfied. Therefore, the Slater-type condition holds.
\Halmos

For later use, the remaining regularity conditions are straightforward. The ES lower bound can be represented by $H_1(u,y)=-(1-\alpha)^{-1}\mathbf{1}_{\{u\geq\alpha\}}y$ with threshold $d_1=-c$, and the Wasserstein constraint by $H_2(u,y)=(y-Q_G(u))^2$ with threshold $d_2=\epsilon^2$. Their derivatives with respect to $y$ are $\partial H_1(u,y)/\partial y=-\mathbf{1}_{\{u\geq\alpha\}}/(1-\alpha)$ and $\partial H_2(u,y)/\partial y=2(y-Q_G(u))$.
If $Q_G\in L^2[0,1]$ and $Q\in L^2[0,1]$, both derivatives belong to $L^2[0,1]$.

Moreover, the Wasserstein ball gives $\|Q\|_2\leq\|Q_G\|_2+\epsilon$, and hence the feasible set is bounded in $L^2[0,1]$. Since $\mathcal{Q}_0$ is a closed convex cone, the Wasserstein ball is closed and convex, and the ES lower bound is affine, the feasible set is weakly closed. Therefore, by reflexivity of $L^2[0,1]$, the compactness and differentiability conditions required by the quantile-based optimality system are satisfied.

\subsection{Proof of Theorem~\ref{theo-cw}}

We derive the solution for the ambiguity set $\mathcal{Q}_{\mathrm{W}\text{-}\mathrm{ES}}$ in \eqref{eq:Q_cw}. Let $\lambda_c\geq0$ be the multiplier for the ES lower-bound constraint and $\lambda_w\geq0$ be the multiplier for the squared Wasserstein constraint $\|Q-Q_G\|_2^2\leq\epsilon^2$. The unconstrained stationarity condition is 
\begin{align*}
\phi(u)+\lambda_c\mathbf{1}_{\{u\geq\alpha\}}/(1-\alpha)-2\lambda_w\mathbin{\cdot}\big(Q(u)-Q_G(u)\big)=0,
\end{align*}
 and hence, in the nondegenerate case $\lambda_w>0$,
\begin{align*}
Q(u)=Q_G(u)+\frac{\phi(u)+\dfrac{\lambda_c}{1-\alpha}\mathbf{1}_{\{u\geq\alpha\}}}{2\lambda_w}.
\end{align*}
The Wasserstein constraint is active at the worst case; otherwise the linear objective could be locally increased along the direction of the spectrum while remaining feasible. Hence $\|Q-Q_G\|_2=\epsilon$, and stationarity forces $\lambda_w>0$. The candidate is admissible, $Q\in\mathcal{Q}_0$: since $Q_G$ and $\phi$ are nondecreasing, $\lambda_w>0$, and the added term $(\lambda_c/(1-\alpha))\mathbf{1}_{\{u\geq\alpha\}}$ has an upward (never downward) jump at $\alpha$ as $\lambda_c\geq0$, the sum is nondecreasing; the zero-residual certificate in Corollary~\ref{corollary-normal-QRRM} applies.

For convenience, define
\begin{align*}
A\triangleq\int_\alpha^1\phi(u)\,\mathrm{d}u,\qquad B\triangleq\int_0^1\phi(u)^2\,\mathrm{d}u,\qquad D\triangleq\frac{(1-\alpha)(c-c_G)}{\epsilon}.
\end{align*}
If the ES lower-bound constraint is active, substituting the candidate $Q$ into this constraint gives $c-c_G=(A+\lambda_c)/(2\lambda_w\mathbin{\cdot}(1-\alpha))$, or equivalently, $\lambda_w=(A+\lambda_c)/(2D\epsilon)$.
The active Wasserstein constraint gives
\begin{align*}
\epsilon^2
=\frac{1}{4\lambda_w^2}\int_0^1\left(\phi(u)+\frac{\lambda_c}{1-\alpha}\mathbf{1}_{\{u\geq\alpha\}}\right)^2\,\mathrm{d}u
=\frac{1}{4\lambda_w^2}\left(B+\frac{2A\lambda_c}{1-\alpha}+\frac{\lambda_c^2}{1-\alpha}\right).
\end{align*}

Substituting $\lambda_w=(A+\lambda_c)/(2D\epsilon)$ into the last equation yields
\begin{align*}
(A+\lambda_c)^2= D^2\mathbin{\cdot}\left(B+\frac{2A\lambda_c}{1-\alpha}+\frac{\lambda_c^2}{1-\alpha}\right).
\end{align*}
Solving this quadratic equation for the nonnegative root gives
\begin{align*}
\lambda_c^*=-A+\sqrt{A^2-(1-\alpha)\frac{A^2-D^2B}{1-\alpha-D^2}},
\end{align*}
whenever the ES lower-bound constraint is active. The associated Wasserstein multiplier is $\lambda_w^*=(A+\lambda_c^*)/(2D\epsilon)$.

It remains to identify the active regime. If $\lambda_c=0$, the unconstrained Wasserstein solution is $Q(u)=Q_G(u)+\epsilon\phi(u)/\sqrt{B}$, with multiplier $\lambda_w^*=\sqrt{B}/(2\epsilon)$. The corresponding ES is $c_G+\epsilon A/((1-\alpha)\sqrt{B})$; denote this threshold by $c^{\dagger}$.
If $c<c^{\dagger}$, the ES lower bound is slack; at $c=c^{\dagger}$ it is met at equality with zero multiplier. In both cases, $\lambda_c^*=0$ and $\lambda_w^*=\sqrt{B}/(2\epsilon)$. If $c^{\dagger}<c<c_G+\epsilon/\sqrt{1-\alpha}$, the ES lower bound is active with a positive multiplier, and the expression above for $\lambda_c^*$ applies. The upper endpoint is the nonemptiness boundary from Lemma~\ref{lem:nonempty_Qcw}.

Finally, substituting the optimal quantile function into the SRM objective gives
\begin{align*}
\mathcal{R}_{\phi}(\mathcal{Q}_{\mathrm{W}\text{-}\mathrm{ES}})
&=\int_0^1\phi(u)Q_G(u)\,\mathrm{d}u+\frac{1}{2\lambda_w^*}\int_0^1\phi(u)\mathbin{\cdot}\left(\phi(u)+\frac{\lambda_c^*}{1-\alpha}\mathbf{1}_{\{u\geq\alpha\}}\right)\,\mathrm{d}u\\
&=\int_0^1\phi(u)Q_G(u)\,\mathrm{d}u+\frac{1}{2\lambda_w^*}\left(B+\frac{\lambda_c^*}{1-\alpha}A\right).
\end{align*}
Since $\lambda_w^*>0$, the minimization Lagrangian is strictly convex in $Q$. The displayed candidate is its unique minimizer, and every primal optimizer must minimize this same Lagrangian by feasibility, weak duality, and complementary slackness. Hence the optimal quantile is unique. This proves Theorem~\ref{theo-cw}.
\Halmos

\subsection{Proof of Proposition~\ref{prop:ncone}}

We now prove the concrete characterization of $N_{\mathcal{Q}_0}$ stated in Proposition~\ref{prop:ncone}, in the case where no additional support or quantile-band constraints are imposed, so that $\mathcal{Q}_0$ is the cone of nondecreasing functions in $L^p[0,1]$. A monotone $L^p$ function is bounded on compact subintervals of $(0,1)$ but may diverge at either endpoint. Its distributional derivative is a locally finite nonnegative Lebesgue--Stieltjes measure on $(0,1)$, independent of the choice of monotone representative at jump points. We assign no masses to $0$ or $1$. For any nonnegative continuous $\Pi$, the integral $\int_{(0,1)}\Pi(u)\,\mathrm{d}Q(u)$ is defined as the increasing limit of its integrals over compact subintervals exhausting $(0,1)$. For $\pi\in L^q[0,1]$, define $\Pi(u)\triangleq\int_0^u\pi(v)\,\mathrm{d}v$ for $u\in[0,1]$.
The polar cone of the monotone cone is $\mathcal{Q}_0^\circ=\{\pi\in L^q[0,1]:\Pi(u)\geq0\text{ for all }u\in[0,1],\ \Pi(1)=0\}$.
Indeed, testing the polar inequality against the constant functions $1$ and $-1$ gives $\Pi(1)=0$, while testing it against $\mathbf{1}_{[u,1]}$ gives $\int_u^1\pi(v)\,\mathrm{d}v\leq0$, and hence $\Pi(u)\geq0$.

Conversely, suppose that $\Pi(u)\geq0$ for all $u\in[0,1]$ and $\Pi(1)=0$. Let $V\in\mathcal{Q}_0$ and choose a nondecreasing representative of $V$. If $V$ is bounded, then the Stieltjes integration-by-parts formula gives
\begin{align*}
\int_0^1\pi(u)V(u)\,\mathrm{d}u =\int_0^1 V(u)\,\mathrm{d}\Pi(u)= -\int_{(0,1)}\Pi(u)\,\mathrm{d}V(u)\leq0,
\end{align*}
because $\Pi(u)\geq 0$ and $\mathrm{d}V$ is a nonnegative measure. For a general $V\in\mathcal{Q}_0\subset L^p[0,1]$, use the monotone truncation $V_m(u)\triangleq\min\{m,\max\{-m,V(u)\}\}$.
Then $V_m$ is bounded and nondecreasing, $V_m\to V$ in $L^p[0,1]$, and $\int_0^1\pi(u)V(u)\,\mathrm{d}u=\lim_{m\to\infty}\int_0^1\pi(u)V_m(u)\,\mathrm{d}u\leq0$.
Thus $\pi\in\mathcal{Q}_0^\circ$.

Since the monotone cone is a convex cone, its normal cone at $Q$ satisfies $N_{\mathcal{Q}_0}(Q)=\mathcal{Q}_0^\circ\cap\{Q\}^{\perp}$.
Equivalently,
\begin{align*}
N_{\mathcal{Q}_0}(Q)=\left\{\pi\in L^q[0,1]~\middle|~\Pi(u)\geq0,\ \Pi(1)=0,\ \int_{(0,1)}\Pi(u)\,\mathrm{d}Q(u)=0\right\}.
\end{align*}
To justify the last equivalence for an unbounded $Q$, take $Q_m=\min\{m,\max\{-m,Q\}\}$. The measures $\mathrm{d}Q_m$ increase to $\mathrm{d}Q$ on $(0,1)$: for $n\geq m$, the function $Q_n-Q_m$ is nondecreasing, and on each compact subinterval the truncation eventually leaves $Q$ unchanged. For $\pi\in\mathcal{Q}_0^\circ$, monotone convergence and the bounded integration-by-parts identity therefore give
\begin{align*}
\int_{(0,1)}\Pi(u)\,\mathrm{d}Q(u)
=\lim_{m\to\infty}\int_{(0,1)}\Pi(u)\,\mathrm{d}Q_m(u)
=-\lim_{m\to\infty}\int_0^1\pi(u)Q_m(u)\,\mathrm{d}u
=-\int_0^1\pi(u)Q(u)\,\mathrm{d}u.
\end{align*}
The final limit follows from $L^p$ convergence and H\"older's inequality, so this integral is finite for every polar element. Boundary terms are taken only for bounded truncations, where $\Pi(0)=\Pi(1)=0$ eliminates them; no product of a vanishing primitive and an infinite endpoint value is used. The identity proves the asserted equivalence with $\{Q\}^{\perp}$. Thus the cumulative residual may be nonzero only across flat portions of $Q$. If direct stationarity already produces a valid quantile, the zero element gives the certificate in Corollary~\ref{corollary-normal-QRRM}.
\Halmos

\subsection{Feasibility of the Two-Sided ES Band}

For the mean--variance model in Section~\ref{sub-sec-quantile-normal-cone}, feasibility requires the ES band to intersect the range permitted by the moment information. We establish the exact feasibility criterion and the strict-feasibility condition used in Theorem~\ref{thm:band-general}.

\begin{lemma}\label{lem:nonempty_Q12c_band}
Let $\mu\in\mathbb{R}$, $\sigma\geq0$, $\alpha\in(1/2,1)$, and $c_\ell<c_u$. The ambiguity set $\mathcal{Q}_{12\text{-}\mathrm{ES}}^{\pm}$ in \eqref{eq:Q_12c_band} is nonempty if and only if
\begin{align}
\mu\leq c_u,\qquad
c_\ell\leq\mu+\sigma\sqrt{\frac{\alpha}{1-\alpha}}.
\label{eq:nonempty_Q12c_band}
\end{align}
If $\sigma>0$, the Slater condition holds whenever both inequalities in \eqref{eq:nonempty_Q12c_band} are strict; in particular, it holds when $c_\ell<\mu<c_u$.
\end{lemma}

\textbf{\emph{Proof.}} Recall the sharp ES bound $M_\alpha$ introduced after \eqref{eq:Q_12c_band} and $\mathrm{ES}_{\alpha}(X)=\left(1/(1-\alpha)\right)\int_\alpha^1 Q(u)\,\mathrm{d}u$. Assume $\sigma>0$; the case $\sigma=0$ forces the point mass at $\mu$, for which \eqref{eq:nonempty_Q12c_band} reduces to $c_\ell\leq\mu\leq c_u$ and the claim is immediate.

We first prove necessity. For any $Q\in\mathcal{Q}_0$ with $\int_0^1 Q(u)\,\mathrm{d}u=\mu$, the upper-tail average of a nondecreasing function is at least its global average,
$$\frac{1}{1-\alpha}\int_\alpha^1 Q(u)\,\mathrm{d}u \geq\int_0^1 Q(u)\,\mathrm{d}u=\mu.$$
Thus, the cap can hold only if $c_u\geq\mu$. To obtain the upper endpoint, set $Z=Q-\mu$ and $I=\int_\alpha^1Z(u)\,\mathrm{d}u=-\int_0^\alpha Z(u)\,\mathrm{d}u$. Cauchy--Schwarz gives
\begin{align*}
\int_0^1Z(u)^2\,\mathrm{d}u
\geq \frac{I^2}{1-\alpha}+\frac{I^2}{\alpha}
=\frac{I^2}{\alpha\mathbin{\cdot}(1-\alpha)}.
\end{align*}
Hence $I\leq\sigma\sqrt{\alpha\mathbin{\cdot}(1-\alpha)}$ and $\mathrm{ES}_\alpha(Q)\leq M_\alpha$, so $c_\ell\leq M_\alpha$ is also necessary.

We next prove sufficiency. Let $X$ take $x_1=\mu-\sigma\sqrt{(1-\alpha)/\alpha}$ with probability $\alpha$ and $x_2=\mu+\sigma\sqrt{\alpha/(1-\alpha)}$ with probability $1-\alpha$; then $X$ has mean $\mu$, variance $\sigma^2$, and $\mathrm{ES}_{\alpha}(X)=x_2=M_\alpha$. For $t\in[0,1]$ set $X_t=\mu+t\mathbin{\cdot}(X-\mu)$, whose quantile function is nondecreasing, with mean $\mu$, variance $t^2\sigma^2\leq\sigma^2$, and $\mathrm{ES}_{\alpha}(X_t)=\mu+t\mathbin{\cdot}(M_\alpha-\mu)$; as $t$ runs over $[0,1]$ this sweeps the whole interval $[\mu,M_\alpha]$. Choose $c^*\triangleq\max\{c_\ell,\mu\}$. Conditions \eqref{eq:nonempty_Q12c_band} together with $c_\ell<c_u$ give $\mu\leq c^*\leq M_\alpha$ and $c_\ell\leq c^*\leq c_u$, so $t^*=(c^*-\mu)/(M_\alpha-\mu)\in[0,1]$ is well defined and $X_{t^*}$ is feasible with $\mathrm{ES}_{\alpha}=c^*\in[c_\ell,c_u]$; hence $\mathcal{Q}_{12\text{-}\mathrm{ES}}^{\pm}\neq\emptyset$. When the inequalities in \eqref{eq:nonempty_Q12c_band} are strict (i.e., $c_\ell<M_\alpha$ and $\mu<c_u$), the intersection $(c_\ell,c_u)\cap(\mu,M_\alpha)$ is nonempty, since $c_\ell<\min\{c_u,M_\alpha\}$ and $\mu<\min\{c_u,M_\alpha\}$; picking any $v$ in it and setting $t=(v-\mu)/(M_\alpha-\mu)\in(0,1)$ gives a feasible $X_t$ with strict variance $t^2\sigma^2<\sigma^2$ and $\mathrm{ES}_{\alpha}(X_t)=v\in(c_\ell,c_u)$, so all inequality constraints hold strictly. Small constant shifts of this quantile vary its mean in a neighborhood of $\mu$ while preserving strictness of the variance and ES inequalities. This verifies the equality qualification as well, and hence the relative Slater condition. \Halmos

\subsection{Proof of Theorem~\ref{thm:band-general}}

In $L^2[0,1]$, the feasible set $\mathcal{Q}_{12\text{-}\mathrm{ES}}^{\pm}$ is weakly compact: the variance budget $\int_0^1(Q(u)-\mu)^2\,\mathrm{d}u\leq\sigma^2$ bounds $\|Q\|_2\leq|\mu|+\sigma$, the mean and ES constraints are affine and weakly closed, the variance sublevel set is convex and weakly closed, and $\mathcal{Q}_0$ is a weakly closed cone. Thus Assumption~\ref{assump:Q_regular} holds, and Assumption~\ref{assump:Q_dual}(ii) is immediate for the affine and quadratic constraint integrands.

We first dispose of the two feasibility boundaries, where Slater's condition fails. If $c_u=\mu$, the upper-tail average of a nondecreasing function equals its global average only when that function is constant a.e.; hence $Q^*=\mu$. If $c_\ell=M_\alpha$, equality must hold in the sharp MV bound for $\mathrm{ES}_\alpha$. The equality conditions in the two Cauchy--Schwarz inequalities used to obtain that bound force $Q-\mu$ to be constant on $[0,\alpha)$ and on $[\alpha,1]$, giving the two-point quantile stated in the theorem. Both boundary solutions are therefore unique.

For the remainder of the proof, $c_u>\mu$ and $c_\ell<M_\alpha$. Lemma~\ref{lem:nonempty_Q12c_band} then supplies Slater's condition, so Proposition~\ref{prop-Q-WCSRM-sol} is necessary and sufficient. It is therefore enough in each regime to exhibit a feasible $Q^*$ and multipliers $\gamma_1^*\in\mathbb{R}$ and $\lambda_1^*,\lambda_\ell^*,\lambda_u^*\geq0$ satisfying complementary slackness and $g^*\in N_{\mathcal{Q}_0}(Q^*)$.

In regime (i) the benchmark remains feasible: take $\lambda_\ell^*=\lambda_u^*=0$, $\gamma_1^*=1$, and $\lambda_1^*=\sigma_\phi/(2\sigma)>0$, so that \eqref{eq:cand-gen} becomes $Q^\circ$ defined before \eqref{eq:cbar-gen}. Then $Q^\circ$ is nondecreasing (as $\phi$ is), satisfies the mean and the binding variance by construction, and its tail average equals $\bar{c}\in[c_\ell,c_u]$. The zero ES multipliers satisfy complementary slackness, including when $\bar{c}$ equals an endpoint of the band. Since $g^*\equiv0\in N_{\mathcal{Q}_0}(Q^\circ)$, Corollary~\ref{corollary-normal-QRRM} applies, and the worst-case value is $R_{\phi}(Q^\circ)=\mu+\sigma\sigma_\phi$.

In regime (ii) the floor is active: for $\lambda_\ell\geq0$ set $\Lambda=\lambda_\ell$ and
\begin{align*}
\psi_{\lambda_\ell}(u)\triangleq\phi(u)-1+\lambda_\ell\mathbin{\cdot}\Big(\tfrac{1}{1-\alpha}\mathbf{1}_{\{u\geq\alpha\}}-1\Big),
\end{align*}
which integrates to zero. The mean equation gives $\gamma_1=1+\lambda_\ell$ and the binding variance gives $2\sigma\lambda_1=\|\psi_{\lambda_\ell}\|_{2}>0$, so the candidate \eqref{eq:cand-gen} is $\widehat{Q}=\mu+\sigma\,\psi_{\lambda_\ell}/\|\psi_{\lambda_\ell}\|_{2}$, which is nondecreasing because its only jump, at $\alpha$, is upward. Its tail average is
\begin{align*}
T(\lambda_\ell)=\mu+\sigma\,\frac{A_0+\lambda_\ell\frac{\alpha}{1-\alpha}}{\sqrt{\sigma_\phi^2+2\lambda_\ell A_0+\lambda_\ell^2\frac{\alpha}{1-\alpha}}},
\qquad
A_0\triangleq\frac{1}{1-\alpha}\int_\alpha^1\big(\phi(u)-1\big)\,\mathrm{d}u>0,
\end{align*}
where $A_0>0$ because the tail average of a nondecreasing, nonconstant spectrum exceeds its overall mean. A direct computation gives $\operatorname{sgn}T'(\lambda_\ell)=\operatorname{sgn}\big(\alpha\sigma_\phi^2/(1-\alpha)-A_0^2\big)$, and the Cauchy--Schwarz inequality yields $A_0^2\leq\alpha\sigma_\phi^2/(1-\alpha)$, with equality only if $\phi-1$ is proportional to $\mathbf{1}_{\{u\geq\alpha\}}/(1-\alpha)-1$, which is excluded by continuity of $\phi$. Hence $T$ is continuous and strictly increasing, with $T(0)=\bar{c}$ and $T(\lambda_\ell)\uparrow\mu+\sigma\sqrt{\alpha/(1-\alpha)}$ as $\lambda_\ell\to\infty$; in particular $\bar{c}<\mu+\sigma\sqrt{\alpha/(1-\alpha)}$. For every $c_\ell\in\big(\bar{c},\ \mu+\sigma\sqrt{\alpha/(1-\alpha)}\big)$ there is thus a unique $\lambda_\ell^*>0$ with $T(\lambda_\ell^*)=c_\ell$; the floor binds, the cap is slack because $c_u>c_\ell$, and Corollary~\ref{corollary-normal-QRRM} applies since $\widehat{Q}\in\mathcal{Q}_0$ and $g^*\equiv0$. (At the boundary $c_\ell=\mu+\sigma\sqrt{\alpha/(1-\alpha)}$ the feasible set reduces to the two-point extremal quantile function used in the proof of Lemma~\ref{lem:nonempty_Q12c_band}, and the conclusion is immediate.)

In regime (iii), $\mu<c_u<\bar{c}$. The cap must bind: otherwise a sufficiently small convex combination of an optimizer with $Q^\circ$ preserves the band and the variance bound and strictly improves the objective, since $Q^\circ$ is the unique mean--variance optimizer and is infeasible here. Thus the floor is slack. Moreover, $\lambda_u^*>0$; if it were zero, the same KKT certificate would make $Q^*$ optimal for the MV problem without the band, again forcing $Q^*=Q^\circ$.

We next prove $\lambda_1^*>0$. Suppose instead that $\lambda_1^*=0$. The residual is then $g^*(u)=\phi(u)-\gamma_1^*-(\lambda_u^*/(1-\alpha))\mathbf{1}_{\{u\geq\alpha\}}$, and its primitive $\Pi^*(u)\triangleq\int_0^u g^*(v)\,\mathrm{d}v$ is strictly convex on each side of $\alpha$. By Proposition~\ref{prop:ncone}, $\Pi^*\geq0$ and $\Pi^*(0)=\Pi^*(1)=0$. Strict convexity rules out another zero on $(0,\alpha]$, because the chord from $0$ to that zero would force $\Pi^*<0$ between them. The chord to $1$ rules out any zero on $[\alpha,1)$ in the same way. Hence $\Pi^*>0$ on $(0,1)$, so complementarity forces $Q^*$ to be constant. Its mean would then give $\mathrm{ES}_\alpha(Q^*)=\mu$, contradicting the binding cap $c_u>\mu$. Thus $\lambda_1^*>0$ and the variance binds.

For these multipliers, $Q^*$ minimizes a strictly convex quadratic Lagrangian over $\mathcal{Q}_0$, equivalently it is the $L^2$ projection of the bounded zero-residual candidate \eqref{eq:cand-gen}. It is bounded, since truncation to the range of that candidate preserves monotonicity and decreases the squared distance. On every connected component $J$ of $\{\Pi^*>0\}$, complementarity makes $Q^*$ constant. If $J$ did not contain $\alpha$ in its interior, $g^*$ would be strictly increasing on $J$; the resulting strict convexity of $\Pi^*$, together with its zero endpoint values, would contradict $\Pi^*>0$. This argument also applies when an endpoint of $J$ is $0$ or $1$. At least one such component exists, because $\Pi^*\equiv0$ would give the inadmissible downward jump in \eqref{eq:cand-gen}. Therefore there is exactly one component $(u_1,u_2)$, with $u_1<\alpha<u_2$, and $Q^*=\bar{q}$ on it. Outside its closure, $\Pi^*=g^*=0$ and the stationary branches apply.

At an interior endpoint, the stationary branch must meet $\bar{q}$ continuously. To see this, note that $\Pi^*\geq0$ has a local minimum of zero there, so its left derivative is nonpositive and its right derivative nonnegative. Any upward jump of $Q^*$ at that point would create a downward jump of $g^*$, since $\phi$ is continuous and the endpoint differs from $\alpha$; this is incompatible with those derivative signs. The remaining possibility $u_1=0,u_2=1$ would make $Q^*$ constant and has already been excluded. This proves the asserted three-case classification, including the matching of each surviving stationary branch.

For an interior plateau, write $k\triangleq\lambda_u^*/(1-\alpha)$. Matching gives $\gamma_1^*+2\lambda_1^*\mathbin{\cdot}(\bar{q}-\mu)=\phi(u_1)=\phi(u_2)-k$.
The first endpoint identity yields \eqref{eq:W-gen}; integrating the plateau residual $g^*(u)=\phi(u)-\phi(u_1)-k\mathbf{1}_{\{u\geq\alpha\}}$ over $[u_1,u_2]$ yields \eqref{eq:EA-gen}. Conversely, these two equations make $g^*\geq0$ before $\alpha$ and $g^*\leq0$ after $\alpha$, with zero total integral over the plateau. Thus $\Pi^*\geq0$, $\Pi^*(1)=0$, and $\Pi^*\,\mathrm{d}Q^*=0$, which verifies the normal-cone certificate.

For a right-boundary plateau, the cap gives $\bar{q}=c_u$ and matching gives $\gamma_1^*+2\lambda_1^*\mathbin{\cdot}(c_u-\mu)=\phi(u_1)$. Closure $\Pi^*(1)=0$ yields \eqref{eq:lamu-boundary}. Necessity of the sign condition follows from the left derivative at $1$: $g^*(1^-)\leq0$. Conversely, that condition makes $g^*\leq0$ throughout $[\alpha,1]$, whereas $g^*\geq0$ on $[u_1,\alpha)$; closure again guarantees $\Pi^*\geq0$ and complementarity.

For a left-boundary plateau, matching at $u_2$ gives
\begin{align*}
\gamma_1^*+2\lambda_1^*\mathbin{\cdot}(\bar{q}-\mu)=\phi(u_2)-k,
\qquad
g^*(u)=\phi(u)-\phi(u_2)+k\mathbf{1}_{\{u<\alpha\}},\quad u\in[0,u_2].
\end{align*}
Closure $\Pi^*(u_2)=0$ yields \eqref{eq:lamu-left}. Since $\Pi^*(0)=0$ and $\Pi^*\geq0$, its right derivative gives $g^*(0^+)\geq0$, exactly the stated sign condition. Conversely, that condition makes the primitive increase on $[0,\alpha]$ and decrease back to zero on $[\alpha,u_2]$; it vanishes on $[u_2,1]$. Hence the normal-cone certificate holds here as well.

Existence and the KKT conditions have supplied an optimizer and multipliers, and the exhaustive classification shows that they solve one of the three calibration systems. Conversely, any solution of a stated system with the required endpoint ranges, positive multipliers, mean, variance, and cap has the certificate just verified and is globally optimal. The atom masses are respectively $u_2-u_1$, $1-u_1$, and $u_2$.

In every nonboundary regime the certificate has $\lambda_1^*>0$, making the minimization Lagrangian strictly convex in $Q$. Any primal optimizer must minimize this same Lagrangian by weak duality and complementary slackness. It is therefore unique, and its variance equals $\sigma^2$. The two feasibility-boundary cases were treated separately above.
\Halmos

\subsection{Proof of Corollary~\ref{cor:band-power}}

Substitution of $\phi(u)=\beta u^{\beta-1}$ gives the benchmark formulas, \eqref{eq:W-power}, and \eqref{eq:EA-power}. Rearrangement of the equal-area equation yields \eqref{eq:power-endpoint-mean}. Its right-hand side is the weighted mean of $u$ on $[u_1,u_2]$ with positive weight $u^{\beta-2}$. It strictly increases in either endpoint, so along the interior solution curve $u_2$ strictly increases as $u_1$ decreases. At $u_1=0$, it requires $u_2=\beta\alpha/(\beta-1)$. This endpoint is below $1$ exactly when $\beta\mathbin{\cdot}(1-\alpha)>1$; when the inequality is reversed, the curve reaches $u_2=1$ at a positive $u_1$ instead. Substitution into \eqref{eq:lamu-left} gives \eqref{eq:power-left}. At equality, the two geometric endpoints meet only at $(0,1)$, which cannot occur for $c_u>\mu$.

For completeness, widening corresponds to tightening the cap. For two cap levels in $(\mu,\bar{c})$, both optimizers satisfy the mean, variance, and cap constraints with equality, so adding the two Lagrangian optimality inequalities gives $(\lambda_u^{(1)}-\lambda_u^{(2)})(c_u^{(1)}-c_u^{(2)})\leq0$. Within the interior case, \eqref{eq:W-power} then makes the plateau widen as the cap decreases.

To justify the endpoint limit when $\beta\mathbin{\cdot}(1-\alpha)=1$, write $u_2=b(u_1)$ along \eqref{eq:power-endpoint-mean}. Strict monotonicity of the weighted mean gives a continuous curve with $b(0)=1$ and $b(\alpha)=\alpha$, where the latter denotes the zero-width limit. For $a=u_1\in(0,\alpha]$, define the stationary profile
\begin{align*}
w_a(u)=\begin{cases}
\phi(u),&0\leq u<a,\\
\phi(a),&a\leq u\leq b(a),\\
\phi(u)-\phi(b(a))+\phi(a),&b(a)<u\leq1,
\end{cases}
\qquad
\overline{w}_a=\int_0^1w_a(u)\,\mathrm{d}u,\quad
z_a=\frac{w_a-\overline{w}_a}{\|w_a-\overline{w}_a\|_2}.
\end{align*}
The mean and binding variance give $Q_a=\mu+\sigma z_a$. On any compact parameter interval $[\delta,\alpha]$, with $\delta>0$, the map $a\mapsto w_a$ is continuous in $L^2$ by dominated convergence, and its centered norm is positive because $w_a$ increases strictly on $(0,a)$. That norm therefore has a positive minimum, so $a\mapsto z_a$ is also $L^2$-continuous. Every $z_a$ is nonconstant, nondecreasing, and has mean zero; hence its upper-tail average is strictly positive. Consequently, the continuous function $a\mapsto(1-\alpha)^{-1}\int_\alpha^1z_a(u)\,\mathrm{d}u$ has a positive minimum on $[\delta,\alpha]$. If the binding cap tends to $\mu$, the identity $c_u-\mu=\sigma(1-\alpha)^{-1}\int_\alpha^1z_a(u)\,\mathrm{d}u$ rules out every such compact interval. Thus $u_1\to0$, and continuity of $b$ gives $u_2\to1$.

For $\beta=2$ in the cap-active regime $\mu<c_u<\bar{c}$, an interior plateau satisfies $u_1+u_2=2\alpha$ by \eqref{eq:EA-power}, so
$u_2=2\alpha-u_1$ decreases as $u_1$ increases.
Sub-case~\ref{cases:Q-cap}\ref{cases:Q-cap-a}
requires $u_2<1$, equivalently $u_1>2\alpha-1$; the
critical value is therefore $u_1^\ddagger=2\alpha-1$,
at which $u_2=1$ and the plateau reaches the right
endpoint. The same monotonicity result holds for $\beta=3$:
the positive root $u_2(u_1)$ of \eqref{eq:EA-power}
is strictly decreasing in $u_1$, with
$u_1^\ddagger\approx0.792$ when $\alpha=0.9$.

It remains to derive $c_u^\ddagger$ for $\beta=2$.
Setting $u_2=1$ and $\phi(u)=2u$, the two-piece
quantile \eqref{eq:Q-cap-boundary} has
$\bar{q}=c_u$ and $\delta\triangleq c_u-\mu$. The
left-branch matching gives
$\gamma_1=\phi(u_1)-2\lambda_1\delta
=2u_1-2\lambda_1\delta$. Substituting into the mean
constraint:
\[
\frac{u_1^2-(2u_1-2\lambda_1\delta)\,u_1}
     {2\lambda_1}+\delta\mathbin{\cdot}(1-u_1)
=\frac{-u_1^2}{2\lambda_1}+\delta=0,
\]
giving $2\lambda_1\delta=u_1^2$ and
$\gamma_1=u_1\mathbin{\cdot}(2-u_1)$. For the variance, the
left-branch deviation is
$\phi(u)-\gamma_1=2(u-u_1)+u_1^2$, and
\[
\frac{1}{4\lambda_1^2}\int_0^{u_1}
  \!\big[2(u-u_1)+u_1^2\big]^2\mathrm{d}u
+\delta^2\mathbin{\cdot}(1-u_1)=\sigma^2.
\]
The substitution $v=u/u_1$ evaluates the integral as
$u_1^3\mathbin{\cdot}(4/3-2u_1+u_1^2)$, and the atom term contributes
$\delta^2\mathbin{\cdot}(1-u_1)=u_1^4\mathbin{\cdot}(1-u_1)/(4\lambda_1^2)$. Combining with
$\delta=u_1^2/(2\lambda_1)$ and simplifying:
\[
\frac{u_1^3\mathbin{\cdot}(4-3u_1)}{12\lambda_1^2}=\sigma^2,
\qquad
\delta=\sigma\sqrt{\frac{3u_1}{4-3u_1}}\,.
\]
Hence $c_u=\mu+\delta$ evaluated at
$u_1^\ddagger=2\alpha-1$ gives
\[
c_u^\ddagger
=\mu+\sigma\sqrt{\frac{3(2\alpha-1)}{7-6\alpha}}\,.
\]
For $c_u^\ddagger<c_u<\bar{c}$,
\eqref{eq:W-power} gives the symmetric endpoints
$u_{1,2}=\alpha\mp\lambda_u^*/(4(1-\alpha))$ with
$\lambda_u^*<4(1-\alpha)^2$; for
$\mu<c_u\le c_u^\ddagger$,
sub-case~\ref{cases:Q-cap}\ref{cases:Q-cap-b} applies.
\Halmos

\section{Proofs and Calculations for Distribution-Based WCSRM}\label{appendix:F_proofs}

The proofs below follow Section~\ref{sec:cases-F}, from the general optimality conditions to the partial-moment and KL models. The geometric-moment results in Section~\ref{subsec:geometric_moment} are proved separately in Appendix~\ref{appendix:geometric_moment}.

\subsection{Proof of Proposition~\ref{prop-P-WCSRM-sol}}\label{proof_prop_P_WCSRM_sol}

We prove the general optimality conditions for the distribution-based WCSRM problem. Let $1<r<\infty$ and let $r'$ be the conjugate exponent of $r$. The decision variable is the distribution function $F\in W^{1,r}[a,b]$, with density $f=F'$. The boundary conditions are $F(a)=0$ and $F(b)=1$, while the monotonicity and nonnegativity conditions are encoded by the admissible set $\mathcal{F}_0$. In this proof, the abstract primal variable $G$ in Appendix~\ref{appendix-functionals} is specialized to the distribution function $F$. Accordingly, $\mathrm{D}_F\mathcal{L}$ denotes the bounded linear functional representing the relative first variation at $F^*$ along segments in $\mathcal{F}_0$, as in Appendix~\ref{appendix-functionals}.

By Appendix~\ref{append:obj2prob}, the distribution-based objective is $R_\phi(F)=b-\int_a^b\Phi(F(x))\,\mathrm{d}x$, where $\Phi(t)=\int_0^t\phi(u)\,\mathrm{d}u$. Maximizing $R_\phi(F)$ is therefore equivalent to minimizing $\mathcal{T}(F)\triangleq\int_a^b\Phi(F(x))\,\mathrm{d}x-b$.
For the equality and inequality constraints, define
\begin{align*}
\mathcal{M}_j(F)&\triangleq\int_a^b k_j(x)f(x)\,\mathrm{d}x-c_j,\qquad j\in\mathcal{I}_1,\\
\mathcal{N}_j(F)&\triangleq\int_a^b K_j(x,f(x))\,\mathrm{d}x-d_j,\qquad j\in\mathcal{I}_2.
\end{align*}
Then \eqref{prob:F_WCSRM} can be written in minimization form as
\begin{align*}
\inf_{F\in\mathcal{F}_0}\quad \mathcal{T}(F)\quad
\text{s.t.}\quad \mathcal{M}_j(F)=0,\ j\in\mathcal{I}_1,\qquad \mathcal{N}_j(F)\leq0,\ j\in\mathcal{I}_2.
\end{align*}
The functional $\mathcal{T}$ is convex because $\Phi$ is convex. The moment constraints $\mathcal{M}_j$ are continuous affine functionals of $F$, and the functionals $\mathcal{N}_j$ are convex under Assumption~\ref{assump:F_dual}(ii). Therefore, the distribution-based WCSRM problem is a constrained convex functional optimization problem after changing the sign of the objective.

The Lagrangian associated with the minimization form is
\begin{align*}
\mathcal{L}(F,\eta,\zeta)
&=\int_a^b\Phi(F(x))\,\mathrm{d}x-b+\sum_{j\in\mathcal{I}_1}\eta_j\mathbin{\cdot}\left(\int_a^b k_j(x)f(x)\,\mathrm{d}x-c_j\right)\\
&\quad+\sum_{j\in\mathcal{I}_2}\zeta_j\mathbin{\cdot}\left(\int_a^b K_j(x,f(x))\,\mathrm{d}x-d_j\right),
\end{align*}
where $\eta_j\in\mathbb{R}$ and $\zeta_j\geq0$.
Evaluate the following formulas at $F=F^*$ and $f=f^*$. For $V\in\mathcal{F}_0$, let $H=V-F^*$. Then $H(a)=H(b)=0$ and $f^*+\varepsilon H'$ is nonnegative for $0\leq\varepsilon\leq1$. All derivatives below are right derivatives at $\varepsilon=0$. Continuity of $\phi$ bounds the objective difference quotient by $\|\phi\|_\infty|H|$, and Assumption~\ref{assump:F_dual}(ii) supplies the relative derivatives of the density constraints.
 The first variation of the objective is
\begin{align*}
\left.\frac{\mathrm{d}}{\mathrm{d}\varepsilon}\int_a^b\Phi(F(x)+\varepsilon H(x))\,\mathrm{d}x\right|_{\varepsilon=0}
=\int_a^b\phi(F(x))H(x)\,\mathrm{d}x.
\end{align*}
For the linear moment constraints, integration by parts gives
\begin{align*}
\left.\frac{\mathrm{d}}{\mathrm{d}\varepsilon}\int_a^b k_j(x)\mathbin{\cdot}(f(x)+\varepsilon H'(x))\,\mathrm{d}x\right|_{\varepsilon=0}
=\int_a^b k_j(x)H'(x)\,\mathrm{d}x=-\int_a^b k_j'(x)H(x)\,\mathrm{d}x,
\end{align*}
where the boundary term vanishes because $H(a)=H(b)=0$. Similarly, for the density-dependent inequality constraints,
\begin{align*}
\left.\frac{\mathrm{d}}{\mathrm{d}\varepsilon}\int_a^b K_j(x,f(x)+\varepsilon H'(x))\,\mathrm{d}x\right|_{\varepsilon=0}
&=\int_a^b\left.\frac{\partial K_j(x,y)}{\partial y}\right|_{y=f(x)}H'(x)\,\mathrm{d}x\\
&=-\int_a^b\frac{\mathrm{d}}{\mathrm{d}x}\left(\left.\frac{\partial K_j(x,y)}{\partial y}\right|_{y=f(x)}\right)H(x)\,\mathrm{d}x.
\end{align*}
The integration-by-parts step is justified by Assumption~\ref{assump:F_dual}(ii), which ensures that
\begin{align*}
x\mapsto\left.\frac{\partial K_j(x,y)}{\partial y}\right|_{y=f(x)}
\end{align*}
has the required Sobolev regularity.

Combining the preceding variations, we obtain
\begin{align}
\left\langle \mathrm{D}_F\mathcal{L}(F,\eta,\zeta),H\right\rangle
=\int_a^b\left\{\phi(F(x))-\sum_{j\in\mathcal{I}_1}\eta_j k_j'(x)-\sum_{j\in\mathcal{I}_2}\zeta_j\frac{\mathrm{d}}{\mathrm{d}x}\left(\left.\frac{\partial K_j(x,y)}{\partial y}\right|_{y=f(x)}\right)\right\}H(x)\,\mathrm{d}x.
\label{F-gateaux}
\end{align}
By the functional KKT principle in Appendix~\ref{appendix-functionals}, a feasible $F^*\in\mathcal{F}$ is optimal if and only if there exist multipliers $\eta_j^*\in\mathbb{R}$, $j\in\mathcal{I}_1$, and $\zeta_j^*\geq0$, $j\in\mathcal{I}_2$, satisfying $\zeta_j^*\mathbin{\cdot}(\int_a^b K_j(x,f^*(x))\,\mathrm{d}x-d_j)=0$ for all $j\in\mathcal{I}_2$ and $-\mathrm{D}_F\mathcal{L}(F^*,\eta^*,\zeta^*)\in N_{\mathcal{F}_0}(F^*)$.
Using \eqref{F-gateaux}, this normal-cone stationarity condition can be written as
\begin{align}
\rho^*\in N_{\mathcal{F}_0}(F^*),
\end{align}
where, in strong form,
\begin{align}
\rho^*(x)=-\phi(F^*(x))+\sum_{j\in\mathcal{I}_1}\eta_j^* k_j'(x)+\sum_{j\in\mathcal{I}_2}\zeta_j^*\frac{\mathrm{d}}{\mathrm{d}x}\left(\left.\frac{\partial K_j(x,y)}{\partial y}\right|_{y=f^*(x)}\right).
\end{align}
This proves Proposition~\ref{prop-P-WCSRM-sol}.
For continuous $\phi$, the relative objective variation used here is legitimate because admissible CDF segments remain in $[0,1]$. The discontinuous ES spectrum and the entropy integrand are handled in the KL models by direct supporting inequalities: the constructed densities are positive, their $\alpha$-level sets are singletons, and their entropy supporting functions satisfy the required integrability. These arguments do not assume ambient G\^ateaux differentiability at zero density.

If $\phi$ is strictly increasing on $[0,1]$, then $\Phi$ is strictly convex, so $\mathcal{T}(F)$ is strictly convex. Equivalently, the original maximization objective $R_\phi(F)$ is strictly concave. Hence the optimal distribution function is unique whenever it exists.
\Halmos

A standard existence argument follows from Assumption~\ref{assump_F_regular}. The functional $\mathcal{T}$ is convex and norm-continuous on $\mathcal{F}_0$, hence weakly lower semicontinuous there. Nonemptiness and weak compactness of $\mathcal{F}$ therefore ensure attainment. In applications, this compactness can be obtained from coercive density-dependent constraints. For example, if one constraint implies a bound of the form $\int_a^b |f(x)|^r\,\mathrm{d}x\leq C$, then the feasible densities are bounded in $L^r[a,b]$, and the corresponding CDFs are bounded in $W^{1,r}[a,b]$. Since $W^{1,r}[a,b]$ is reflexive for $1<r<\infty$, weak compactness follows after imposing closedness of the feasible set.

Here weak compactness supplies existence. At an existing optimizer, Assumption~\ref{assump:F_dual}(i) supplies multipliers, while part~(ii) justifies the first variations and integration by parts; weak compactness is not needed for this multiplier step. The sufficient conditions in Corollaries~\ref{corollary-normal-PRRM} and~\ref{corollary:worst-moment} require neither weak compactness nor a constraint qualification. Once a feasible candidate and multipliers satisfy the stated supporting-slope and regularity conditions, complementary slackness, and stationarity, the convex supporting inequalities imply global optimality. If the objective, or the relevant active Lagrangian, is strictly convex, the optimizer is unique. This is the argument used to obtain the explicit model solutions, where weak compactness of each concrete moment-defined ambiguity set need not be verified separately.

\subsection{Proof of Corollary~\ref{corollary-normal-PRRM}}

We next prove Corollary~\ref{corollary-normal-PRRM}. Let $V\in\mathcal{F}_0$ be
arbitrary. Since $V$ is a CDF on $[a,b]$, we have $0\le V(x)\le 1$ for
all $x\in[a,b]$. On the set $\{\widehat{F}=0\}$, this gives
$V-\widehat{F}=V\ge 0$, and
$\widehat{\rho}\le 0$ implies
$\widehat{\rho}(x)\mathbin{\cdot}(V(x)-\widehat{F}(x))\le 0.$
On the set $\{0<\widehat{F}<1\}$, the residual vanishes, so the contribution
is zero. On the set $\{\widehat{F}=1\}$, we have
$V-\widehat{F}=V-1\le 0$, and the assumed inequality
$\widehat{\rho}\ge 0$ again implies
$\widehat{\rho}(x)\mathbin{\cdot}(V(x)-\widehat{F}(x))\le 0.$
Combining the three regions yields
\[
\langle \widehat{\rho},V-\widehat{F}\rangle
=
\int_a^b \widehat{\rho}(x)\mathbin{\cdot}(V(x)-\widehat{F}(x))\,\mathrm{d}x
\le 0,
\qquad \forall V\in\mathcal{F}_0.
\]
Hence $\widehat{\rho}\in N_{\mathcal{F}_0}(\widehat{F})$.
To conclude without compactness or a constraint qualification, use the supporting inequalities for $\Phi$ and $K_j$. For the minimization Lagrangian they give, after integration by parts,
\begin{align*}
\mathcal{L}(V,\widehat{\eta},\widehat{\zeta})-\mathcal{L}(\widehat{F},\widehat{\eta},\widehat{\zeta})
\geq-\int_a^b\widehat{\rho}(x)\mathbin{\cdot}(V(x)-\widehat{F}(x))\,\mathrm{d}x\geq0.
\end{align*}
For feasible $V$, primal feasibility and complementary slackness give 
\begin{align*}
\mathcal{T}(\widehat{F})=\mathcal{L}(\widehat{F},\widehat{\eta},\widehat{\zeta})\leq\mathcal{L}(V,\widehat{\eta},\widehat{\zeta})\leq\mathcal{T}(V).
\end{align*}
 This is the sufficiency argument of Appendix~\ref{appendix-functionals}, and proves global optimality.
 If $\phi$ is
strictly increasing, the objective in \eqref{eq:srm_def_distribution} is
strictly concave in $F$, and the optimal CDF is unique. This proves Corollary~\ref{corollary-normal-PRRM}.
\Halmos

Moreover, suppose that $\widehat{F}$, with density $\widehat{f}=\widehat{F}'$, and multipliers $\widehat{\eta}_j$, $\widehat{\zeta}_j$ satisfy primal feasibility, complementary slackness, and the boundary conditions. If $\widehat{\rho}(x)=0
\  \text{on } \{0<\widehat{F}<1\}$, then the stationarity equation on this region turns into
\begin{align*}
-\phi(\widehat{F}(x))+\sum_{j\in\mathcal{I}_1}\widehat{\eta}_j k_j'(x)+\sum_{j\in\mathcal{I}_2}\widehat{\zeta}_j\frac{\mathrm{d}}{\mathrm{d}x}\left(\left.\frac{\partial K_j(x,y)}{\partial y}\right|_{y=\widehat{f}(x)}\right)=0.
\end{align*}
In concrete cases, if we can further verify the two sign conditions on the flat
regions, and $\widehat{F}\in\mathcal{F}_0$, then $\widehat{F}$ is optimal for \eqref{prob:F_WCSRM}.

\subsection{Proof of Proposition~\ref{prop:unbounded-sufficiency}}

We prove Proposition~\ref{prop:unbounded-sufficiency}. For $F\in\widetilde{\mathcal{F}}_0$ with density $f$ and finite constraint integrals, define the Lagrangian
\begin{align*}
L(F)\triangleq R_\phi(F)-\sum_{j\in\mathcal{I}_1}\eta_j^*\mathbin{\cdot}\Big(\int k_j(x)f(x)\,\mathrm{d}x-c_j\Big)-\sum_{j\in\mathcal{I}_2}\zeta_j^*\mathbin{\cdot}\Big(\int K_j(x,f(x))\,\mathrm{d}x-d_j\Big),
\end{align*}
where $R_\phi$ is understood in the two-sided Choquet form \eqref{eq:F_objective_unbounded}, which is finite on $\widetilde{\mathcal{F}}_0$ by H\"older's inequality.

\emph{Step 1 (reduction to Lagrangian maximality).}\;
For any $F\in\widetilde{\mathcal{F}}$ the equality terms vanish, and $\zeta_j^*\geq0$ together with $\int K_j(x,f(x))\,\mathrm{d}x\leq d_j$ gives $L(F)\geq R_\phi(F)$; complementary slackness gives $L(F^*)=R_\phi(F^*)$. Hence it suffices to prove $L(F)\leq L(F^*)$ for every $F\in\widetilde{\mathcal{F}}$.

\emph{Step 2 (concavity and the directional derivative).}\;
Fix $F\in\widetilde{\mathcal{F}}$ and set $H\triangleq F-F^*$, $h\triangleq H'$, and $F_{\varepsilon}\triangleq F^*+\varepsilon H$ for $\varepsilon\in[0,1]$; mixtures preserve absolute continuity, the support restriction, and the finite $p$-th moment, so $F_{\varepsilon}\in\widetilde{\mathcal{F}}_0$. The map $\psi(\varepsilon)\triangleq L(F_{\varepsilon})$ is concave on $[0,1]$: pointwise in $x$, $\varepsilon\mapsto-\Phi(F_{\varepsilon}(x))$ is concave because $\Phi$ is convex, $\varepsilon\mapsto k_j(x)f_{\varepsilon}(x)$ is affine, and $\varepsilon\mapsto-K_j(x,f_{\varepsilon}(x))$ is concave because $K_j(x,\cdot)$ is convex. By concavity, $\psi(1)\leq\psi(0)+\psi'(0^+)$, so it suffices to show $\psi'(0^+)=0$.

For the objective term, the difference quotients $\varepsilon^{-1}\mathbin{\cdot}\big(\Phi(F_{\varepsilon}(x))-\Phi(F^*(x))\big)$ are nondecreasing in $\varepsilon$ by convexity of $\Phi$ and converge pointwise to $\phi(F^*(x))H(x)$ as $\varepsilon\downarrow0$. The quotient at $\varepsilon=1$ is integrable, because the two-sided integrals in \eqref{eq:F_objective_unbounded} are finite at both $F$ and $F^*$, and the limit is integrable by condition (iii). Every intermediate quotient lies between these two integrable functions, so dominated convergence yields
\begin{align*}
\frac{\mathrm{d}}{\mathrm{d}\varepsilon}R_\phi(F_{\varepsilon})\Big|_{\varepsilon=0^+}=-\int_a^b\phi(F^*(x))H(x)\,\mathrm{d}x .
\end{align*}
The equality-constraint terms are affine in $\varepsilon$ with derivative $-\sum_{j\in\mathcal{I}_1}\eta_j^*\int k_j(x)h(x)\,\mathrm{d}x$; the integrals are finite by the growth condition on $k_j$ and the $p$-th moments. For the inequality-constraint terms, condition (i) gives the derivative $-\sum_{j\in\mathcal{I}_2}\zeta_j^*\int\left.(\partial K_j(x,y)/\partial y)\right|_{y=f^*(x)}\,h(x)\,\mathrm{d}x$.

\emph{Step 3 (integration by parts).}\;
Write 
\begin{align*}
w(x)\triangleq\sum_{j\in\mathcal{I}_1}\eta_j^*k_j(x)+\sum_{j\in\mathcal{I}_2}\zeta_j^*\left.(\partial K_j(x,y)/\partial y)\right|_{y=f^*(x)}
\end{align*}
 for the multiplier-weighted constraint term, so that the constraint contribution to $\psi'(0^+)$ found in Step~2 is $-\int_a^b w(x)h(x)\,\mathrm{d}x$. The functions $w$ and $H$ are absolutely continuous on every compact subinterval of $(a,b)$. The product $wh$ is integrable by the growth bound and the finite first variations in (i). By stationarity, $w'H=\phi(F^*)H$ is integrable by (iii). We may therefore integrate by parts on compact subintervals and pass to both endpoint limits using (iii), obtaining
\begin{align*}
\int_a^b w(x)h(x)\,\mathrm{d}x=\big(w(x)H(x)\big)_a^b-\int_a^b w'(x)H(x)\,\mathrm{d}x=-\int_a^b w'(x)H(x)\,\mathrm{d}x.
\end{align*}
The boundary term vanishes at both finite and infinite endpoints by (iii). This condition controls the multiplier-weighted product; separate endpoint conditions for the individual constraints are unnecessary. Since 
\begin{align*}
w'(x)=\sum_{j\in\mathcal{I}_1}\eta_j^*k_j'(x)+\sum_{j\in\mathcal{I}_2}\zeta_j^*(\mathrm{d}/\mathrm{d}x)\big(\left.(\partial K_j(x,y)/\partial y)\right|_{y=f^*(x)}\big),
\end{align*}
 collecting the three contributions gives
\begin{align*}
\psi'(0^+)
&=\int_a^b\Big(-\phi(F^*(x))+\sum_{j\in\mathcal{I}_1}\eta_j^*k_j'(x)
+\sum_{j\in\mathcal{I}_2}\zeta_j^*\frac{\mathrm{d}}{\mathrm{d}x}
\big(\left.\frac{\partial K_j(x,y)}{\partial y}\right|_{y=f^*(x)}\big)\Big)H(x)\,\mathrm{d}x\\
&=\int_a^b\rho^*(x)H(x)\,\mathrm{d}x=0
\end{align*}
by condition (ii). Hence $L(F)\leq L(F^*)$ for every $F\in\widetilde{\mathcal{F}}$, and $F^*$ is optimal by Step 1.

For uniqueness, suppose $F_1^*\neq F_2^*$ are both optimal, and let $F_{1/2}$ be their midpoint, which is feasible by convexity of the constraints. Steps 2--3 show $L(F)\leq L(F^*)$ for \emph{every} $F\in\widetilde{\mathcal{F}}$. If $\phi$ is strictly increasing, $\Phi$ is strictly convex, and since $F_1^*\neq F_2^*$ on a set of positive Lebesgue measure, $L(F_{1/2})>(1/2)L(F_1^*)+(1/2)L(F_2^*)=L(F^*)$, a contradiction. If instead $K_{j_0}(x,\cdot)$ is strictly convex with $\zeta_{j_0}^*>0$, the same strict inequality holds unless $f_1^*=f_2^*$ a.e., which forces $F_1^*=F_2^*$.
\Halmos

\subsection{Proof of Corollary~\ref{corollary:worst-moment}}

Suppose that every density-dependent inequality constraint is linear in the density, namely $K_j(x,y)=\ell_j(x)y$ for $j\in\mathcal{I}_2$. Then
\begin{align*}
\left.\frac{\partial K_j(x,y)}{\partial y}\right|_{y=f(x)}=\ell_j(x),\qquad
\frac{\mathrm{d}}{\mathrm{d}x}\left(\left.\frac{\partial K_j(x,y)}{\partial y}\right|_{y=f(x)}\right)=\ell_j'(x).
\end{align*}
If the three residual sign conditions in Corollary~\ref{corollary-normal-PRRM} hold and $F^*\in\mathcal{F}_0$, then $\phi(F^*(x))=\sum_{j\in\mathcal{I}_1}\eta_j^*k_j'(x)+\sum_{j\in\mathcal{I}_2}\zeta_j^*\ell_j'(x)$ wherever $0<F^*(x)<1$.
If $\phi$ is strictly increasing on $[0,1]$, it admits an inverse on its range. Hence
\begin{align*}
F^*(x)=\phi^{-1}\left(\sum_{j\in\mathcal{I}_1}\eta_j^*k_j'(x)+\sum_{j\in\mathcal{I}_2}\zeta_j^*\ell_j'(x)\right).
\end{align*}
Under the feasibility and complementary slackness conditions stated in the corollary, this $F^*$ satisfies Corollary~\ref{corollary-normal-PRRM}, and hence is optimal. Strict monotonicity of $\phi$ gives strict concavity of the original objective, so the optimizer is unique. This proves Corollary~\ref{corollary:worst-moment}.
\Halmos

\subsection{Proof of Theorem~\ref{theo-LU}}
\label{proof_LPM}

We first distinguish feasibility among arbitrary nonnegative laws from feasibility in $\mathcal{F}_{12\mathrm{pm}}$. Normalize $Y=X/\mu$ and write $d=\mathbb{E}[(1-Y)_+]=\mathbb{E}[(Y-1)_+]$, $p_-=\mathbb{P}(Y<1)$, and $p_+=\mathbb{P}(Y>1)$. Since $0\leq(1-Y)_+\leq1$, and by Cauchy--Schwarz on the two sides of the mean, $s_1\leq d$, $d^2/s_1\leq p_-$, $d^2/s_2\leq p_+$, and $p_-+p_+\leq1$.
Thus $s_1\mathbin{\cdot}(s_1+s_2)\leq s_2$. Conversely, when this inequality holds, placing probabilities $s_2/(s_1+s_2)$ and $s_1/(s_1+s_2)$ at $1-\sqrt{s_1\mathbin{\cdot}(s_1+s_2)/s_2}$ and $1+\sqrt{s_2\mathbin{\cdot}(s_1+s_2)/s_1}$, respectively, gives a nonnegative law for $Y$ with the required moments. At equality, all the preceding inequalities must be equalities: the only possible law is supported on $\{0,1+s_2/s_1\}$ with those probabilities. It is not absolutely continuous. More directly, absolute continuity and $s_1>0$ imply $s_1<d$, so the feasibility inequality must be strict in our class. Strictness alone does not ensure feasibility on an arbitrary fixed $[a,b]$; the support and $L^r$ checks below supply a sufficient construction.

In the present parametrization, $s_1=(1-t)\sigma^2/(2\mu^2)$ and $s_2=(1+t)\sigma^2/(2\mu^2)$, so $t\in(-1,1)$ ensures $s_1,s_2>0$. All constraints in $\mathcal{F}_{12\mathrm{pm}}$ are linear equalities in the density, so no complementary slackness condition is needed.

The three moment kernels are $k_1(x)=x$, $k_2(x)=(\mu-x)_+^2$, and $k_3(x)=(x-\mu)_+^2$, with weak derivatives $k_1'(x)=1$, $k_2'(x)=-2(\mu-x)\mathbf{1}_{\{x<\mu\}}$, and $k_3'(x)=2(x-\mu)\mathbf{1}_{\{x>\mu\}}$.
These kernels belong to $W^{1,r'}[a,b]$. Corollary~\ref{corollary:worst-moment} therefore applies once the candidate below satisfies the moment constraints, support inclusion, density integrability, and exterior residual signs.

We now reduce the optimality system to the scalar calibration equation \eqref{eq:LPM-calibration}. We use the power spectrum $\phi(u)=\beta u^{\beta-1}$ with $\beta\geq2$. Since all three constraints are equality constraints, their multipliers are unrestricted. Write the three equality multipliers in \eqref{eq:F_inverse_spectrum} as $\eta_1$, $-\eta_2$, and $\eta_3$, respectively, matching the sign convention in Theorem~\ref{theo-LU}. The stationarity residual is $\rho(x)=-\beta F(x)^{\beta-1}+\eta_1+2\eta_2\mathbin{\cdot}(\mu-x)$ for $x<\mu$, and $\rho(x)=-\beta F(x)^{\beta-1}+\eta_1+2\eta_3\mathbin{\cdot}(x-\mu)$ for $x>\mu$.
Setting $\rho=0$ where $0<F<1$ and solving pointwise gives $F(x)=[(-2\eta_2x+2\eta_2\mu+\eta_1)/\beta]^{1/(\beta-1)}$ for $x<\mu$, and $F(x)=[(2\eta_3x+\eta_1-2\eta_3\mu)/\beta]^{1/(\beta-1)}$ for $x>\mu$.
Let $y\triangleq F(\mu)$. Continuity at $x=\mu$ gives $\eta_1=\beta y^{\beta-1}$. The lower and upper support endpoints, determined by $F=0$ and $F=1$ on the respective branches, are $x_L=(2\eta_2\mu+\eta_1)/(2\eta_2)$ and $x_U=(\beta+2\eta_3\mu-\eta_1)/(2\eta_3)$.
The inverse quantile function generated by this CDF is
\begin{align*}
Q(u)=\mu+\frac{\eta_1-\beta u^{\beta-1}}{2\eta_2},\qquad 0\leq u<y,\qquad
Q(u)=\mu+\frac{\beta u^{\beta-1}-\eta_1}{2\eta_3},\qquad y\leq u\leq1.
\end{align*}

The lower partial moment constraint gives
\begin{align*}
s_1\mu^2=\int_0^y(\mu-Q(u))^2\,\mathrm{d}u=\frac{1}{4\eta_2^2}\int_0^y\left(\eta_1-\beta u^{\beta-1}\right)^2\,\mathrm{d}u=\frac{\beta\mathbin{\cdot}(\beta-1)^2}{2\eta_2^2\mathbin{\cdot}(2\beta-1)}y^{2\beta-1}.
\end{align*}
Similarly, the upper partial moment constraint gives
\begin{align*}
s_2\mu^2&=\frac{1}{4\eta_3^2}\int_y^1\left(\beta u^{\beta-1}-\eta_1\right)^2\,\mathrm{d}u
=\frac{(\beta-1)^2+(\eta_1-1)^2(2\beta-1)-2\beta\mathbin{\cdot}(\beta-1)^2y^{2\beta-1}}{4\eta_3^2\mathbin{\cdot}(2\beta-1)}.
\end{align*}
Choosing $\eta_2<0<\eta_3$ makes both CDF branches increasing. Solving these two moment equations gives the multiplier expressions in Theorem~\ref{theo-LU}, with $\eta_1=\beta y^{\beta-1}$.

It remains to impose the mean constraint. Using the quantile representation above,
\begin{align*}
0=\int_0^1(Q(u)-\mu)\,\mathrm{d}u=\frac{(\beta-1)y^\beta}{2\eta_2}+\frac{1-\beta y^{\beta-1}+(\beta-1)y^\beta}{2\eta_3}.
\end{align*}
Equivalently, $(\beta-1)y^\beta/|\eta_2|=[1-\beta y^{\beta-1}+(\beta-1)y^\beta]/\eta_3$.
Both sides are strictly positive for $0<y<1$: the numerator on the right equals $\int_y^1\beta\mathbin{\cdot}(u^{\beta-1}-y^{\beta-1})\,\mathrm{d}u>0$, and the moment equations give $|\eta_2|,\eta_3>0$. Substituting the multipliers, squaring, and clearing positive denominators therefore gives \eqref{eq:LPM-calibration} without introducing extraneous roots in this interval.

To establish a root in $(0,(1+t)/2]$, observe that \eqref{eq:LPM-calibration} is homogeneous of degree one in $(s_1,s_2)$. For this root-existence argument only, normalize $s_1+s_2=1$, so $s_2=(1+t)/2$, and let $A(y)$ denote the left-hand side of \eqref{eq:LPM-calibration}. Then $A(0)=2s_2>0$, and $A(s_2)=0$ when $\beta=2$. For $\beta>2$, factorization gives
\begin{align*}
A(s_2)=s_2\mathbin{\cdot}(s_2^{\beta-1}-1)A_1(s_2),\qquad
A_1(s)=(2-\beta)s^\beta+\beta s^{\beta-1}-\beta s+\beta-2.
\end{align*}
Since $A_1(1)=A_1'(1)=0$ and $A_1''(s)=\beta\mathbin{\cdot}(\beta-1)(\beta-2)s^{\beta-3}(1-s)>0$ for $0<s<1$, we have $A_1'(s)<0$ and $A_1(s)>0$ there. Thus $A(s_2)<0$, and continuity gives a root in $(0,s_2)$.

Return to the original moment budgets, choose such a root $y^*$, and define the multipliers, endpoints, and CDF $F^*$ as in Theorem~\ref{theo-LU}. The preceding calculations give $\eta_2^*<0<\eta_3^*$ and a nondecreasing CDF, continuous at $\mu$ with $F^*(\mu)=y^*$, satisfying the mean and both partial-moment constraints. On $(x_L^*,\mu)$, its density is a positive constant times $(x-x_L^*)^{-(\beta-2)/(\beta-1)}$; it is bounded on every $[x_L^*+\delta,b]$ with $\delta>0$. Hence, for $\beta>2$, the density belongs to $L^r[a,b]$ exactly when $r\mathbin{\cdot}(\beta-2)/(\beta-1)<1$. For $\beta=2$, it is piecewise constant and belongs to every finite $L^r[a,b]$. Together with the assumed support inclusion, these checks establish $F^*\in\mathcal{F}_{12\mathrm{pm}}$.

On the left flat region, $\rho(x)=\eta_1^*+2\eta_2^*(\mu-x)$ is increasing and vanishes at $x_L^*$, so $\rho(x)\leq0$ for $x\leq x_L^*$. On the right flat region, $\rho(x)=-\beta+\eta_1^*+2\eta_3^*(x-\mu)$ is increasing and vanishes at $x_U^*$, so $\rho(x)\geq0$ for $x\geq x_U^*$. Interior stationarity and these exterior signs satisfy Corollary~\ref{corollary-normal-PRRM}. It gives optimality without a compactness assumption on the moment-defined set, and strict increase of $\phi$ gives uniqueness.

Finally, insert the calibrated multipliers into the quantile derived above. Evaluating
\begin{align*}
\mathcal{R}_{\phi}(\mathcal{F}_{12\mathrm{pm}})=\int_0^{y^*}\beta u^{\beta-1}Q^*(u)\,\mathrm{d}u+\int_{y^*}^1\beta u^{\beta-1}Q^*(u)\,\mathrm{d}u
\end{align*}
gives the worst-case value stated in Theorem~\ref{theo-LU}.
\Halmos

\subsection{Proof of Lemma~\ref{lem:nonempty_KL}}

We prove Lemma~\ref{lem:nonempty_KL}. For a density $f$ supported on $[a,b]$ and the exponential reference density $g_\theta(x)=\theta e^{-\theta x}$, the KL divergence is $\mathrm{KL}(f\|g_\theta)=\int_a^bf(x)\ln\big(f(x)/g_\theta(x)\big)\,\mathrm{d}x$.
By the classical minimum-discrimination-information principle, or equivalently
the information projection of $g_\theta$ onto the affine mean constraint
(see, e.g., \citealp{cover2006elements}, Chapter 11; \citealp{dupuis1997weak}),
the KL-minimizing density has the exponential-tilting form
\begin{align*}
f_{\varsigma}(x)=C(\varsigma)e^{-\varsigma x},\qquad
C(\varsigma)=\frac{\varsigma}{e^{-\varsigma a}-e^{-\varsigma b}},\qquad \varsigma\neq0.
\end{align*}
At $\varsigma=0$, the family is understood by continuity, with $f_0(x)=C(0)=1/(b-a)$.
The parameter $\varsigma$ is chosen to satisfy the mean constraint
\begin{align*}
\frac{a e^{-\varsigma a}-b e^{-\varsigma b}}{e^{-\varsigma a}-e^{-\varsigma b}}+\frac{1}{\varsigma}=\mu.
\end{align*}
At $\varsigma=0$, the uniform density has mean $(a+b)/2$, which gives the continuous extension in the lemma. The induced mean $m(\varsigma)\triangleq\int_a^b x f_{\varsigma}(x)\,\mathrm{d}x$
is continuous and strictly decreasing in $\varsigma$, with
$m(\varsigma)\to a \ \text{as } \varsigma\to+\infty,\ m(\varsigma)\to b \  \text{as } \varsigma\to-\infty.
$ Hence, for any $\mu \in (a,b)$, the mean equation admits a unique solution $\varsigma^*$. Then
\begin{align*}
\mathrm{KL}(f_{\varsigma^*}\|g_\theta)=\int_a^b f_{\varsigma^*}(x)\mathbin{\cdot}\left(\ln C(\varsigma^*)-\varsigma^*x-\ln\theta+\theta x\right)\,\mathrm{d}x
=\ln C(\varsigma^*)-\ln\theta+(\theta-\varsigma^*)\mu.
\end{align*}
Thus the ambiguity set is nonempty exactly when $\epsilon\geq\ln C(\varsigma^*)-\ln\theta+(\theta-\varsigma^*)\mu$.

If the inequality is strict, then $f_{\varsigma^*}$ itself satisfies the KL constraint strictly while preserving the mean. Small changes in the tilt parameter vary the mean on both sides of $\mu$ while preserving the strict KL inequality, so the Slater-type condition holds. At equality, every feasible density attains the minimum KL divergence, so strict feasibility is impossible. Below this minimum, the ambiguity set is empty.
\Halmos

\subsection{Proof of Theorem~\ref{theo-KL}}

We derive the semi-closed-form solution in Theorem~\ref{theo-KL}. For the ES spectrum $\phi(u)=(1-\alpha)^{-1}\mathbf{1}_{\{u\geq\alpha\}}$, the objective is the upper-tail expectation. In the explicit construction below, we take the lower support endpoint to be $0$ and assume that the upper support $b<\infty$ is fixed and given.

The ES spectrum has a jump at $\alpha$, so the continuous-spectrum assumption in Proposition~\ref{prop-P-WCSRM-sol} does not apply verbatim. We instead use the same first-variation argument directly in density space. For a positive candidate, the level set $F(x)=\alpha$ is a singleton and hence negligible, so the strong-form stationarity equation remains valid almost everywhere. For a positive stationary candidate with $\zeta_1>0$, it is
\begin{align}
\label{eq:KL-stats-proof}
-\phi(F(x))+\eta_1+\zeta_1\mathbin{\cdot}\left(\frac{f^{\prime}(x)}{f(x)}+\theta\right)=0.
\end{align}
Integrating \eqref{eq:KL-stats-proof} on either side of $x_\alpha$ gives $f(x)=C_1e^{-\xi_1x}$ on $(0,x_\alpha)$ and $f(x)=C_2e^{-\xi_2x}$ on $(x_\alpha,b)$, where $\xi_1=\theta+\eta_1/\zeta_1$ and $\xi_2=\xi_1-1/[(1-\alpha)\zeta_1]$.

Global stationarity also determines the matching condition at $x_\alpha$. Write $A(x)=\eta_1x+\zeta_1\mathbin{\cdot}(1+\ln(f(x)/g_\theta(x)))$. For a smooth CDF variation $H$ supported in $(0,b)$, integrating $\int_0^b A(x)H'(x)\,\mathrm{d}x$ by parts on the two subintervals produces the internal boundary term $(A(x_\alpha^-)-A(x_\alpha^+))H(x_\alpha)$. Since $H(x_\alpha)$ is arbitrary, $A$ must be continuous there. Continuity of $g_\theta$ and $\zeta_1>0$ then give $C_1e^{-\xi_1x_\alpha}=C_2e^{-\xi_2x_\alpha}$.

The body and tail masses determine $C_1=\alpha/I_0(\xi_1;0,x_\alpha)$ and $C_2=(1-\alpha)/I_0(\xi_2;x_\alpha,b)$, giving \eqref{eq:KL_match}. Substitution into the mean constraint gives the next equation in the calibration system. On each piece, $\ln(f(x)/g_\theta(x))=\ln(C_i/\theta)+(\theta-\xi_i)x$; integrating against $f$ gives \eqref{eq:KL_kl}. Thus the three conditions are exactly \eqref{eq:KL_match}--\eqref{eq:KL_kl}. An admissible solution gives a strictly positive continuous density with bounded entropy supporting function on $[0,b]$; positivity follows from this construction, not from complementary slackness alone.

The following compactness argument establishes existence and the minimum-radius limit in the larger $L^1$ density class.

Since $g_\theta(x)=\theta e^{-\theta x}$ is strictly positive and bounded on the compact interval $[0,b]$, there exist constants $m_g, M_g$
such that $0<m_g\le g_\theta(x)\le M_g<\infty, x\in[0,b]$. Thus a KL constraint bound implies a
bound on $\int_0^b f(x)\ln f(x)\,\mathrm{d}x$. By the de la
Vallée--Poussin criterion, the feasible densities are uniformly integrable,
and hence relatively weakly compact in $L^1[0,b]$ by the Dunford--Pettis
theorem. The probability and mean constraints are weakly closed because they are tested against the bounded functions $1$ and $x$ on $[0,b]$. The nonnegativity constraint is weakly closed, and the KL sublevel set is weakly closed by the weak lower semicontinuity of relative entropy. Here the feasible set is the larger class of all $L^1$ densities with the probability, mean, and KL restrictions. It is weakly closed and, together with relative weak compactness, is weakly compact in $L^1[0,b]$; this does not establish weak compactness of the original $W^{1,r}$ class. Moreover, suppose $f_n\rightharpoonup f$ in
$L^1[0,b]$. By the definition of weak convergence in $L^1$, tested against $\mathbf{1}_{[0,x]}\in L^{\infty}[0,b]=(L^1[0,b])^*$, we have
\[
F_n(x)=\int_0^x f_n(y)\,\mathrm{d}y
\longrightarrow
F(x)=\int_0^x f(y)\,\mathrm{d}y
\quad\text{for every }x\in[0,b].
\]
Since
$\Phi(u)=(u-\alpha)_+/(1-\alpha)$ is continuous and bounded on $[0,1]$,
the dominated convergence theorem gives
\[
\int_0^b \Phi(F_n(x))\,\mathrm{d}x
\longrightarrow
\int_0^b \Phi(F(x))\,\mathrm{d}x .
\]
Thus the ES objective is continuous under this weak convergence. Because the feasible set is weakly compact, an optimizer exists in the density formulation.

This compactness argument also justifies the boundary benchmark in Figure~\ref{fig:KLexp}(b). Let $\epsilon_{\min}$ be the minimum KL radius in Lemma~\ref{lem:nonempty_KL}, and let $f_{\min}$ be its unique mean-anchored entropy projection. For any sequence $\epsilon_n\downarrow\epsilon_{\min}$, choose maximizers $f_n$ in the larger $L^1$ class. They lie in a common weakly compact KL sublevel set. Lower semicontinuity of relative entropy implies that every weak subsequential limit has KL divergence at most $\epsilon_{\min}$ and hence equals $f_{\min}$. The ES continuity proved above therefore gives convergence of the optimal values to the ES of $f_{\min}$. Since $f_{\min}$ is bounded on $[0,b]$, its CDF belongs to the original Sobolev class and is feasible at every $\epsilon_n$. The worst-case value in that class lies between this boundary value and the larger $L^1$ optimum, so it has the same limit.

The $L^1$ existence argument does not by itself provide smooth stationarity multipliers or a solution of the calibration equations. Conditional on a positive regular optimizer admitting the global density stationarity system, $\zeta_1^*=0$ is impossible because the two ES regions would require two different values of the same mean multiplier. Thus $\zeta_1^*>0$ in that conditional argument, and integration of stationarity gives the two exponential pieces and their continuous matching. The candidate constructed from a solution of the system is bounded and belongs to $L^r[0,b]$ for every finite $r$, so its CDF belongs to the original Sobolev class. We now verify its sufficiency directly, without claiming calibration solvability for every feasible radius.

Let $(\xi_1^*,\xi_2^*,x_\alpha^*)$ solve \eqref{eq:KL_match}--\eqref{eq:KL_kl} with $\xi_1^*>\xi_2^*$ and $0<x_\alpha^*<b$, and let $f^*$ be the density defined in Theorem~\ref{theo-KL}. The calibration equations give its prescribed mass and mean and its binding KL budget.

The matching equation makes $f^*$ continuous and strictly positive on $[0,b]$. Recover $\zeta_1^*=((1-\alpha)(\xi_1^*-\xi_2^*))^{-1}>0$ and $\eta_1^*=\zeta_1^*\mathbin{\cdot}(\xi_1^*-\theta)$. The entropy supporting slope $1+\ln(f^*/g_\theta)$ is continuous and piecewise affine, hence belongs to $W^{1,r'}[0,b]$, and \eqref{eq:KL-stats-proof} gives zero residual a.e. Thus Corollary~\ref{corollary-normal-PRRM} certifies optimality in the stated Sobolev class. The direct argument below also gives uniqueness through strict convexity of entropy. The bounded supporting function
\begin{align*}
B(y)=\int_y^b\phi(F^*(x))\,\mathrm{d}x+\eta_1^*y
+\zeta_1^*\mathbin{\cdot}\left(1+\ln\frac{f^*(y)}{g_\theta(y)}\right)
\end{align*}
is Lipschitz on $[0,b]$: the integral term has a bounded integrand, and $\ln(f^*/g_\theta)$ is continuous and piecewise affine. It is therefore absolutely continuous. Equation~\eqref{eq:KL-stats-proof} gives $B'=0$ a.e., so $B$ is constant. Let $\mathcal{T}(F)=\int_0^b\Phi(F(x))\,\mathrm{d}x-b$ and use its minimization Lagrangian. For any competing density $f$ with finite KL, the supporting inequalities for $\Phi$ and relative entropy give
\begin{align*}
\mathcal{L}(F,\eta_1^*,\zeta_1^*)-\mathcal{L}(F^*,\eta_1^*,\zeta_1^*)
\geq\int_0^b B(y)\mathbin{\cdot}(f(y)-f^*(y))\,\mathrm{d}y=0.
\end{align*}
Fubini is valid because the spectrum and $B$ are bounded and both densities are integrable. Strict convexity of entropy makes the first inequality strict unless $f=f^*$ a.e. For feasible $F$, calibration and complementary slackness therefore give
\begin{align*}
\mathcal{T}(F^*)=\mathcal{L}(F^*,\eta_1^*,\zeta_1^*)
\leq\mathcal{L}(F,\eta_1^*,\zeta_1^*)\leq\mathcal{T}(F),
\end{align*}
with a strict first inequality unless $F=F^*$. This proves uniqueness and global optimality in the larger finite-KL density class, and hence in the stated Sobolev class. No extension of entropy to negative densities is used.

The worst-case ES is the conditional mean of the exponential tail,
\begin{align*}
\mathcal{R}_{\phi}(\mathcal{F}_{1\exp})
=\frac{1}{1-\alpha}\int_{x_\alpha^*}^{b}xf^*(x)\,\mathrm{d}x
=\frac{I_1(\xi_2^*;x_\alpha^*,b)}{I_0(\xi_2^*;x_\alpha^*,b)}.
\end{align*}
The definitions of $I_0$ and $I_1$ give the explicit value in Theorem~\ref{theo-KL}. If $\xi_2^*=0$, the tail density is constant and the ratio equals $(x_\alpha^*+b)/2$, its continuous extension.
\Halmos

\subsection{Proof of Lemma~\ref{lem:standardize}}

We first prove the standardization identity.
The affine map $x=m+su$ induces a bijection between $\widetilde{\mathcal{F}}_{1\mathrm{N}}(m,s;\mu,\epsilon)$ and $\widetilde{\mathcal{F}}_{1\mathrm{N}}(0,1;\bar{\mu},\epsilon)$: if $F$ has density $f$, the standardized density is $\tilde{f}(u)=s\,f(m+su)$, and the substitution $x=m+su$ gives
\begin{align*}
\int_{\mathbb{R}}\tilde{f}(u)\ln\frac{\tilde{f}(u)}{\varphi_{\mathrm{N}}(u)}\,\mathrm{d}u
=\int_{\mathbb{R}} f(x)\ln\frac{s\,f(x)}{\varphi_{\mathrm{N}}((x-m)/s)}\,\mathrm{d}x
=\mathrm{KL}\big(f\,\|\,g_{m,s}\big),
\end{align*}
since $g_{m,s}(x)=s^{-1}\varphi_{\mathrm{N}}((x-m)/s)$; likewise $\int u\tilde{f}(u)\,\mathrm{d}u=(\mu-m)/s=\bar{\mu}$ and the second moment stays finite, so membership in $\widetilde{\mathcal{F}}_0$ is preserved. Finally, if $\widetilde{X}$ has density $\tilde{f}$, then $m+s\widetilde{X}$ has density $f$, and translation equivariance together with positive homogeneity of ES give $\mathrm{ES}_\alpha(F)=m+s\,\mathrm{ES}_\alpha(\widetilde{F})$. Taking suprema over the bijection yields \eqref{eq:chi_def}, and optimizers correspond through the same change of variables.

We next prove the feasibility claim.
For any density $f$ with mean $\bar{\mu}$ and $\mathrm{KL}(f\|\varphi_{\mathrm{N}})<\infty$, using $\ln\big(g_{\bar{\mu},1}(x)/\varphi_{\mathrm{N}}(x)\big)=\bar{\mu} x-\bar{\mu}^2/2$,
\begin{align*}
\mathrm{KL}\big(f\,\|\,\varphi_{\mathrm{N}}\big)
=\mathrm{KL}\big(f\,\|\,g_{\bar{\mu},1}\big)+\int_{\mathbb{R}} f(x)\mathbin{\cdot}\Big(\bar{\mu} x-\frac{\bar{\mu}^2}{2}\Big)\mathrm{d}x
=\mathrm{KL}\big(f\,\|\,g_{\bar{\mu},1}\big)+\frac{\bar{\mu}^2}{2}
\geq\frac{\bar{\mu}^2}{2},
\end{align*}
with equality if and only if $f=g_{\bar{\mu},1}$ a.e.; conversely, $g_{\bar{\mu},1}$ is feasible whenever $\epsilon\geq\bar{\mu}^2/2$, and infeasibility for $\epsilon<\bar{\mu}^2/2$ follows from the same identity. If $\epsilon>\bar{\mu}^2/2$, the law $\mathsf{N}(\bar{\mu},1)$ has strict KL slack, and nearby normal means retain that slack while varying the mean on either side of $\bar{\mu}$. This verifies the relative Slater condition. At equality, the feasible set is the singleton $\mathsf{N}(\bar{\mu},1)$, so strict feasibility is impossible.
\Halmos

\subsection{Proof of Theorem~\ref{theo-KL-normal}}\label{proof_theo_KL_normal}

Let $(\lambda_1,\lambda_2,x_\alpha)$ be any solution of \eqref{eq:N_match}--\eqref{eq:N_kl} with $\lambda_1<\lambda_2$; existence and uniqueness of such a solution are proved at the end. We verify the conditions of Proposition~\ref{prop:unbounded-sufficiency} for the problem $\mathcal{R}_\phi\big(\widetilde{\mathcal{F}}_{1\mathrm{N}}(0,1;\bar{\mu},\epsilon)\big)$ of \eqref{eq:chi_def}, with $a=-\infty$, $b=+\infty$, $p=2$, reference density $\varphi_{\mathrm{N}}$, and constraint families $k_1(x)=x$ (mean anchor, multiplier $\eta^*$) and $K_1(x,y)=y\ln\big(y/\varphi_{\mathrm{N}}(x)\big)$ (KL, multiplier $\zeta^*$).

By \eqref{eq:N_match} and the choice of $C_1,C_2$, the density $\tilde{f}^*$ in \eqref{eq:fstar_normal} is continuous at $x_\alpha$, strictly positive, integrates to $\alpha+(1-\alpha)=1$, and places mass $\alpha$ on $(-\infty,x_\alpha)$, so $x_\alpha$ is the $\alpha$-quantile of the corresponding CDF $\widetilde{F}^*$. Completing the square as in \eqref{eq:J0}, each piece is a multiple of the shifted normal density $\varphi_{\mathrm{N}}(\cdot-\lambda_i)$, so all moments are finite. Equation \eqref{eq:N_mean} states $\int x\tilde{f}^*(x)\,\mathrm{d}x=\bar{\mu}$, and since $\ln\big(\tilde{f}^*(x)/\varphi_{\mathrm{N}}(x)\big)=\ln C_i+\lambda_ix$ on the respective pieces, the left side of \eqref{eq:N_kl} equals $\mathrm{KL}(\tilde{f}^*\|\varphi_{\mathrm{N}})$. Hence $\widetilde{F}^*$ is feasible and the KL constraint is active.

Next, set $\zeta^*\triangleq((1-\alpha)(\lambda_2-\lambda_1))^{-1}>0$ and $\eta^*\triangleq-\zeta^*\lambda_1$. Since $\partial K_1(x,y)/\partial y=\ln(y/\varphi_{\mathrm{N}}(x))+1$, we have $(\mathrm{d}/\mathrm{d}x)\left(\left.(\partial K_1(x,y)/\partial y)\right|_{y=\tilde{f}^*(x)}\right)=\lambda_i$ on piece $i$, so the residual is
\begin{align*}
\rho^*(x)=-\phi(\widetilde{F}^*(x))+\eta^*+\zeta^*\lambda_i
=\begin{cases}
\eta^*+\zeta^*\lambda_1=0, & x<x_\alpha\quad(\phi=0),\\[2pt]
-(1-\alpha)^{-1}+\eta^*+\zeta^*\lambda_2=0, & x>x_\alpha\quad(\phi=(1-\alpha)^{-1}),
\end{cases}
\end{align*}
by the definition of $\zeta^*,\eta^*$; the single point $x_\alpha$ is Lebesgue-null, and the discontinuity of the ES spectrum is immaterial because $\widetilde{F}^*$ is continuous and strictly increasing, so $\{x:\widetilde{F}^*(x)=\alpha\}=\{x_\alpha\}$. Complementary slackness holds because the KL constraint is active.

For the decay conditions, every feasible $F$ satisfies $\mathrm{KL}(f\|\varphi_{\mathrm{N}})\leq\epsilon$, so by the Donsker--Varadhan inequality $\mathbb{E}_F[X^2/4]\leq\epsilon+\ln\mathbb{E}_{\mathsf{N}(0,1)}[e^{X^2/4}]=\epsilon+\ln\sqrt{2}$, and second moments are uniformly bounded over the feasible set. For $H=F-\widetilde{F}^*$, Markov's inequality then gives $|H(x)|\leq C/x^2$ for $|x|$ large, so $|x|\,|H(x)|\to0$ as $|x|\to\infty$; the bracket of condition (iii) in Proposition~\ref{prop:unbounded-sufficiency} equals $\eta^*x+\zeta^*\mathbin{\cdot}(\ln C_i+\lambda_ix+1)$, affine in $x$ on each piece, so the boundary products vanish. Also $\phi\leq(1-\alpha)^{-1}$ and $\int|H|\,\mathrm{d}x<\infty$, giving $\int\phi(\widetilde{F}^*)|H|\,\mathrm{d}x<\infty$; the growth condition holds for $k_1(x)=x$ with $p_0=1<p=2$; and $x\mapsto\left.(\partial K_1(x,y)/\partial y)\right|_{y=\tilde{f}^*(x)}$ is continuous and piecewise affine, hence locally absolutely continuous. To verify (i), write $a_1(x)=1+\ln(\tilde{f}^*(x)/\varphi_{\mathrm{N}}(x))$. Its at-most-linear growth makes $a_1(x)\mathbin{\cdot}(f(x)-\tilde{f}^*(x))$ integrable. Convexity bounds each directional difference quotient of $K_1$ between this function and $K_1(x,f(x))-K_1(x,\tilde{f}^*(x))$. The latter is integrable because both KL divergences are finite and the negative part of $K_1(x,f(x))$ is bounded by $\varphi_{\mathrm{N}}(x)/e$. Dominated convergence now gives the finite first variation required in (i).

Proposition~\ref{prop:unbounded-sufficiency} therefore applies: $\widetilde{F}^*$ is optimal, and since $K_1(x,\cdot)$ is strictly convex with $\zeta^*>0$, it is the unique optimizer.

For the value formula, since $\widetilde{F}^*$ is continuous, $\mathrm{ES}_\alpha(\widetilde{F}^*)$ is its conditional mean on $[x_\alpha,\infty)$, i.e., $J_1(\lambda_2;x_\alpha,\infty)/J_0(\lambda_2;x_\alpha,\infty)$. The normalized tail piece is the $\mathsf{N}(\lambda_2,1)$ density truncated below at $x_\alpha$, whose mean is $\lambda_2+\varphi_{\mathrm{N}}(x_\alpha-\lambda_2)/(1-\Phi_{\mathrm{N}}(x_\alpha-\lambda_2))$ by the standard truncated-normal formula; this is \eqref{eq:chi_value}.

We next justify the numerical dual on the unbounded support. Write $G=\mathsf{N}(0,1)$, $h_t(x)=(x-t)_+/(1-\alpha)$, and let $\mathcal{P}$ be the set of all probability laws $P$ with $\mathbb{E}_P[X]=\bar{\mu}$ and $\mathrm{KL}(P\|G)\leq\epsilon$. Finite KL implies absolute continuity with respect to $G$ and hence with respect to Lebesgue measure. The entropy inequality above, applied first to truncated $X^2/4$, gives $\sup_{P\in\mathcal{P}}\mathbb{E}_P[X^2]\leq4\epsilon+2\ln2$. Thus $\mathcal{P}$ is exactly the law class in \eqref{eq:chi_def}, with no extra smoothness restriction. The second-moment bound gives tightness and uniform integrability of $|X|$. Lower semicontinuity of KL and continuity of the mean under $W_1$ convergence then make $\mathcal{P}$ compact in $W_1$; here this means weak convergence together with convergence of first absolute moments.

For each $t$, $P\mapsto t+\mathbb{E}_P[h_t(X)]$ is affine and $W_1$-continuous because $h_t$ is Lipschitz. For each $P$, it is continuous and convex in $t$. The ES minimization representation therefore extends to these integrable laws, and Sion's minimax theorem yields
\begin{align*}
\chi(\alpha,\epsilon;\bar{\mu})
=\inf_{t\in\mathbb{R}}\left\{t+\sup_{P\in\mathcal{P}}\mathbb{E}_P[h_t(X)]\right\}.
\end{align*}
This verifies the minimax exchange separately from the bounded-support argument in Appendix~\ref{app:min-rep}.

For fixed $t$, entropy duality can be established directly from the moment perturbation function $v_t(m,e)\triangleq\sup\{\mathbb{E}_P[h_t(X)]:\mathbb{E}_P[X]=m,\ \mathrm{KL}(P\|G)\leq e\}$. This function is concave, nondecreasing in $e$, and finite on its domain by the second-moment bound. The laws $\mathsf{N}(m,1)$ show that it is finite on a neighborhood of $(\bar{\mu},\epsilon)$, since $\epsilon>\bar{\mu}^2/2$. It consequently has a supporting affine majorant there, with slopes $\eta\in\mathbb{R}$ and $\zeta\geq0$. Applying this majorant to the mean and KL of any law of finite entropy, and using the reverse weak-duality inequality, gives
\begin{align*}
v_t(\bar{\mu},\epsilon)
=\eta\bar{\mu}+\zeta\epsilon+
\sup_P\left\{\mathbb{E}_P[h_t(X)-\eta X]-\zeta\,\mathrm{KL}(P\|G)\right\}.
\end{align*}
Here the supremum ranges over laws of finite entropy. The supporting $\zeta$ is positive: if it were zero, shifted unit-variance Gaussian laws would make the supremum infinite for every $\eta$, because $h_t(x)-\eta x$ is unbounded above on at least one tail. For $\zeta>0$, let $Z=\mathbb{E}_G[\exp((h_t(X)-\eta X)/\zeta)]<\infty$ and let $P_f$ have density $f_{\zeta,\eta,t}(x)=Z^{-1}\varphi_{\mathrm{N}}(x)\exp((h_t(x)-\eta x)/\zeta)$. The identity $\mathbb{E}_P[h_t(X)-\eta X]-\zeta\,\mathrm{KL}(P\|G)=\zeta\ln Z-\zeta\,\mathrm{KL}(P\|P_f)$ shows that the last supremum is $\zeta\ln Z$, attained at $P_f$. This proves strong duality for each fixed threshold and hence
\begin{align*}
\chi(\alpha,\epsilon;\bar{\mu})
&=\inf_{\zeta>0,\,\eta,t\in\mathbb{R}} D(\zeta,\eta,t),\\
D(\zeta,\eta,t)
&\triangleq t+\eta\bar{\mu}+\zeta\epsilon
+\zeta\ln\mathbb{E}_G\left[\exp\left(\frac{h_t(X)-\eta X}{\zeta}\right)\right].
\end{align*}
The objective is jointly convex, as a supremum of functions affine in $(\zeta,\eta)$ and convex in $t$. Gaussian exponential integrability makes it smooth for $\zeta>0$. Its derivatives, evaluated under $f_{\zeta,\eta,t}$, are
\begin{align*}
\frac{\partial D}{\partial\eta}=\bar{\mu}-\mathbb{E}_f[X],\qquad
\frac{\partial D}{\partial\zeta}=\epsilon-\mathrm{KL}(f\|\varphi_{\mathrm{N}}),\qquad
\frac{\partial D}{\partial t}=1-\frac{\mathbb{P}_f(X>t)}{1-\alpha}.
\end{align*}
At a minimizer with $\zeta>0$, these equations give the mean anchor, binding KL budget, and tail mass. Setting $x_\alpha=t$, $\lambda_1=-\eta/\zeta$, and $\lambda_2=\lambda_1+1/((1-\alpha)\zeta)$ recovers \eqref{eq:N_match}--\eqref{eq:N_kl}; continuity at $t$ follows from the positive-part exponential. Conversely, an admissible calibration triple gives a dual minimizer with $\zeta=((1-\alpha)(\lambda_2-\lambda_1))^{-1}$, $\eta=-\zeta\lambda_1$, and $t=x_\alpha$: the entropy inequality and the ES threshold representation are both equalities at its density. Thus the primal certificate and the numerical dual give the same density and value. The normalizing constant $Z$ is evaluated through \eqref{eq:J0}; numerical output must still satisfy the three calibration equations to the desired tolerance.

It remains to show that \eqref{eq:N_match}--\eqref{eq:N_kl} has exactly one solution with $\lambda_1<\lambda_2$. Write $g(t)\triangleq t+v_t(\bar{\mu},\epsilon)$, so that $\chi(\alpha,\epsilon;\bar{\mu})=\inf_{t\in\mathbb{R}}g(t)$ by the minimax identity above. The function $g$ is finite and convex on $\mathbb{R}$, hence continuous. It is also coercive: for any $P\in\mathcal{P}$, $h_t\geq0$ gives $g(t)\geq t$, while $(x-t)_+\geq x-t$ gives $g(t)\geq t+(\bar{\mu}-t)/(1-\alpha)=(\bar{\mu}-\alpha t)/(1-\alpha)$. Hence $g$ attains its minimum at some $t^*$, and $g(t^*)=\chi(\alpha,\epsilon;\bar{\mu})$. The fixed-threshold duality at $t^*$ supplies $\eta^*\in\mathbb{R}$ and $\zeta^*>0$ with $D(\zeta^*,\eta^*,t^*)=g(t^*)$. Since the supremum defining $\zeta\ln Z$ dominates its restriction to $\mathcal{P}$, weak duality gives $D(\zeta,\eta,t)\geq g(t)\geq g(t^*)$ for all $\zeta>0$ and $\eta,t\in\mathbb{R}$. Therefore $(\zeta^*,\eta^*,t^*)$ minimizes the differentiable function $D$ over the open set $(0,\infty)\times\mathbb{R}^2$, and its gradient vanishes there. By the correspondence just established, $x_\alpha=t^*$, $\lambda_1=-\eta^*/\zeta^*$, and $\lambda_2=\lambda_1+1/((1-\alpha)\zeta^*)>\lambda_1$ solve \eqref{eq:N_match}--\eqref{eq:N_kl}. For uniqueness, every solution with $\lambda_1<\lambda_2$ yields, by the first part of the proof, the unique optimal density $\tilde{f}^*$. That density determines the triple: $x_\alpha$ is its $\alpha$-quantile, which is unique because $\tilde{f}^*>0$, and $\lambda_1$ and $\lambda_2$ are the slopes of $x\mapsto\ln\big(\tilde{f}^*(x)/\varphi_{\mathrm{N}}(x)\big)$ on $(-\infty,x_\alpha)$ and $(x_\alpha,\infty)$.
\Halmos

For the limits stated in Section~\ref{PRRM-case-KL-Gaussian}, fix $\alpha\in(0,1)$ and set $\bar{\mu}=0$. If $\epsilon_n\downarrow0$ and $P_n$ is an optimizer at radius $\epsilon_n$, the uniform second-moment bound above makes $(P_n)$ relatively compact in $W_1$. Lower semicontinuity of KL forces every limit to equal $G=\mathsf{N}(0,1)$, so $P_n\to G$ in $W_1$. The quantile representation gives $|\mathrm{ES}_\alpha(P_n)-\mathrm{ES}_\alpha(G)|\leq W_1(P_n,G)/(1-\alpha)$, proving $\chi(\alpha,\epsilon)\to\mathrm{ES}_\alpha(G)$ as $\epsilon\downarrow0$. For the other limit, $P_v=\mathsf{N}(0,v^2)$, $v\geq1$, preserves the mean and has $\mathrm{KL}(P_v\|G)=(v^2-1-\ln v^2)/2$ and $\mathrm{ES}_\alpha(P_v)=v\,\mathrm{ES}_\alpha(G)$. Since $\mathrm{ES}_\alpha(G)>0$ and the feasible sets increase with $\epsilon$, these laws imply $\chi(\alpha,\epsilon)\to\infty$ as $\epsilon\to\infty$.

\section{Proofs for the Portfolio Application}\label{appendix:port_proofs}

\subsection{Proof of Lemma~\ref{lem:KL_proj}}

($\subseteq$) Let $P\in\mathcal{P}_\epsilon$ and let $F$, with density $f$, be the law of $X_w=w^\top Y$ under $P$; the density exists because $P\ll\widehat{P}$ implies that the projected law is absolutely continuous with respect to $\mathsf{N}(\mu_w,\sigma_w^2)$. The mean condition is immediate: $\int x\,f(x)\,\mathrm{d}x=\mathbb{E}_P[w^\top Y]=w^\top\hat{\mu}=\mu_w$. The entropy chain rule \eqref{eq:KL_chain} gives $\mathrm{KL}(f\|g_{\mu_w,\sigma_w})\leq\mathrm{KL}(P\|\widehat{P})\leq\epsilon$; see \citet{cover2006elements,dupuis1997weak}. Finally, the Donsker--Varadhan inequality bounds the second moment of $f$ uniformly over the KL ball (as in Appendix~\ref{proof_theo_KL_normal}), so $F\in\widetilde{\mathcal{F}}_0$ with $p=2$, and $F\in\widetilde{\mathcal{F}}_{1\mathrm{N}}(\mu_w,\sigma_w;\mu_w,\epsilon)$.

($\supseteq$) Let $F\in\widetilde{\mathcal{F}}_{1\mathrm{N}}(\mu_w,\sigma_w;\mu_w,\epsilon)$ with density $f$, set $\varrho(t)\triangleq f(t)/g_{\mu_w,\sigma_w}(t)$, and define $P$ by the density
\begin{align}
\frac{\mathrm{d}P}{\mathrm{d}\widehat{P}}(y)\triangleq \varrho\big(w^\top y\big).
\label{eq:reweighting}
\end{align}
Then: (i) $P$ is a probability measure, since $\mathbb{E}_{\widehat{P}}[\varrho(w^\top Y)]=\int \varrho(t)\,g_{\mu_w,\sigma_w}(t)\,\mathrm{d}t=1$; (ii) the law of $w^\top Y$ under $P$ has density $\varrho\,g_{\mu_w,\sigma_w}=f$; (iii) the divergence is preserved,
\begin{align*}
\mathrm{KL}(P\|\widehat{P})=\mathbb{E}_P\big[\ln \varrho(w^\top Y)\big]
=\int_{\mathbb{R}} f(t)\ln \varrho(t)\,\mathrm{d}t=\mathrm{KL}\big(f\,\|\,g_{\mu_w,\sigma_w}\big)\leq\epsilon .
\end{align*}
It remains to verify the mean anchor. Under the Gaussian nominal, the conditional mean is affine in the conditioning value:
\begin{align}
\mathbb{E}_{\widehat{P}}\big[Y\mid w^\top Y=t\big]
=\hat{\mu}+\frac{\widehat{\Sigma} w}{\sigma_w^{2}}\,(t-\mu_w).
\label{eq:cond_linear}
\end{align}
The conditional Gaussian second moment grows at most quadratically in $t$, so the finite second moment of $f$ implies $\mathbb{E}_P[\|Y\|_2^2]<\infty$. Since $\mathrm{d}P/\mathrm{d}\widehat{P}$ is a function of $w^\top y$ alone, the tower property then gives
\begin{align*}
\mathbb{E}_P[Y]
&=\mathbb{E}_{\widehat{P}}\big[\mathbb{E}_{\widehat{P}}[Y\mid w^\top Y]\,
\varrho(w^\top Y)\big]\\
&=\int_{\mathbb{R}}\Big(\hat{\mu}+\frac{\widehat{\Sigma} w}{\sigma_w^2}
(t-\mu_w)\Big)f(t)\,\mathrm{d}t\\
&=\hat{\mu}+\frac{\widehat{\Sigma} w}{\sigma_w^2}
\Big({\textstyle\int}t f(t)\,\mathrm{d}t-\mu_w\Big)=\hat{\mu},
\end{align*}
using the mean anchor of $f$. Hence $P\in\mathcal{P}_\epsilon$ and its projection is $F$.
\Halmos

\subsection{Proof of Proposition~\ref{prop:port_convex}}

For fixed $P$ and $\varepsilon\in[0,1]$, $X_{\varepsilon w+(1-\varepsilon)w'}=\varepsilon X_w+(1-\varepsilon)X_{w'}$ pointwise, so subadditivity and positive homogeneity of any coherent risk measure $R$ give $R(X_{\varepsilon w+(1-\varepsilon)w'};P)\leq\varepsilon\,R(X_w;P)+(1-\varepsilon)\,R(X_{w'};P)$; a pointwise supremum of convex functions is convex, and $V$ is such a supremum over the $w$-independent set $\mathcal{P}_\epsilon$. The argument uses only that $\mathcal{P}_\epsilon$ does not depend on $w$, so it holds for an arbitrary nominal $\widehat{P}$. \Halmos

\subsection{Proof of Theorem~\ref{thm:KL_port}}

The reduction \eqref{eq:port_socp} restates \eqref{eq:KL_port_prob} through \eqref{eq:V_socp}, which combines \eqref{eq:law_inv_reduction}, Lemma~\ref{lem:KL_proj}, and Lemma~\ref{lem:standardize} with $\bar{\mu}=0$. The coefficient $\chi(\alpha,\epsilon)$ is finite and positive. Introducing a scalar $v$ with $\|\widehat{\Sigma}^{1/2}w\|_2\leq v$ converts the objective to the linear function $w^\top\hat{\mu}+\chi(\alpha,\epsilon)v$; together with the assumed conic representation of $\mathcal{W}$, this is an SOCP. Continuity of the objective and compactness of $\mathcal{W}$ give attainment. Statements (i)--(ii) use the calibration triple of Theorem~\ref{theo-KL-normal} with $\bar{\mu}=0$, which exists and is unique because $\epsilon>0$.

(i) By Theorem~\ref{theo-KL-normal} and Lemma~\ref{lem:standardize}, the projected worst case $F^*_w$ is unique and its KL constraint is active, $\mathrm{KL}(f^*_w\|g_{\mu_w,\sigma_w})=\epsilon$. For a fixed $w\in\mathcal{W}$, Lemma~\ref{lem:KL_proj} supplies a joint law attaining $V(w)$. Let $P\in\mathcal{P}_\epsilon$ be any such maximizer; its projection is necessarily $F^*_w$. The chain rule \eqref{eq:KL_chain} then reads 
\begin{align*}
\epsilon\geq\mathrm{KL}(P\|\widehat{P})=\epsilon+\int\mathrm{KL}(P(\cdot\mid t)\|\widehat{P}(\cdot\mid t))\,f^*_w(t)\,\mathrm{d}t,
\end{align*}
 forcing $P(\cdot\mid t)=\widehat{P}(\cdot\mid t)$ for $F_w^*$-almost every $t$. Hence $P$ is exactly the reweighting \eqref{eq:reweighting} with $\varrho=f^*_w/g_{\mu_w,\sigma_w}$, and its conditional laws agree with the nominal ones.

(ii) The projection and standardization arguments give \eqref{eq:V_socp} for every $w\neq0$. The first expression therefore differentiates it directly ($\sigma_w>0$ for every $w\neq0$ since $\widehat{\Sigma}\succ0$); the second substitutes $\chi(\alpha,\epsilon)=(V(w)-\mu_w)/\sigma_w$. For the Danskin form, note that for the unique maximizer $P^*_w$ the function $w'\mapsto\mathrm{ES}_\alpha(w'^\top Y;P^*_w)$ is differentiable at $w$ with gradient $(1-\alpha)^{-1}\mathbb{E}_{P^*_w}[Y\mathbf{1}_{\{w^\top Y\geq x_\alpha(w)\}}]$, because $w^\top Y$ has a continuous positive density under $P^*_w$; by (i), the conditional mean \eqref{eq:cond_linear} is preserved under $P^*_w$, so
\begin{align*}
(1-\alpha)^{-1}\,\mathbb{E}_{P^*_w}\big[Y\,\mathbf{1}_{\{w^\top Y\geq x_\alpha(w)\}}\big]
&=\hat{\mu}+\frac{\widehat{\Sigma} w}{\sigma_w^2}\big(\mathbb{E}_{F^*_w}[X_w\mid X_w\geq x_\alpha(w)]-\mu_w\big)\\
&=\hat{\mu}+\frac{\widehat{\Sigma} w}{\sigma_w^2}\big(V(w)-\mu_w\big),
\end{align*}
using $\mathbb{E}_{F^*_w}[X_w\mid X_w\geq x_\alpha(w)]=\mathrm{ES}_\alpha(F^*_w)=V(w)$, where $X_w=w^\top Y$. This matches the direct gradient, confirming \eqref{eq:danskin_grad}.
\Halmos

\section{Projection Algorithms}\label{appendix:algorithms}
We give implementation and convergence details for the quantile algorithm in Section~\ref{subsec:Q_projection} and its density counterpart for bounded-support models in Section~\ref{sec:cases-F}. Both use a proximal method of multipliers on smooth finite-dimensional discretizations. All grid quadrature weights are positive. Projections and gradients use the corresponding weighted inner product, so the quantile projection is weighted isotonic regression and the density projection satisfies the weighted normalization constraint. Numerical gradients must be derivatives of the chosen discrete objective; discretizing a continuum derivative independently can give a different scheme.

Write a discretized problem as $\min_{z\in C}\mathcal{T}(z)$ subject to $\mathcal{M}(z)=0$ and $\mathcal{N}(z)\leq0$, where $C$ is the quantile cone or density simplex, $\mathcal{M}$ is affine, and $\mathcal{T},\mathcal{N}_i$ are finite continuously differentiable convex functions with locally Lipschitz gradients. In density space it is enough that these functions have such extensions to a neighborhood of $C$. Assume the discrete problem has a KKT point. These assumptions specify the algorithmic scope and do not follow merely from convexity of the continuum ambiguity set.

Fix $\tau,\omega>0$. Both algorithms minimize the proximal augmented objective $\mathcal{J}_k$ defined in \eqref{eq:Q_augmented_objective}, with the substitutions in Appendix~\ref{subsec:F_projection} for densities. Equality penalties are bounded below by $-(\gamma_i^k)^2/(2\tau)$ and inequality penalties by $-(\lambda_i^k)^2/(2\tau)$. In quantile space the remaining objective is linear; in density space it is bounded on the simplex. Thus the proximal term makes the inner problem coercive and strongly convex, with a unique minimizer. The effective equality multiplier in its gradient is $\gamma_i^k+\tau\mathcal{M}_i(z)$, and the effective inequality multiplier is $(\lambda_i^k+\tau\mathcal{N}_i(z))_+$, with $(\eta,\zeta)$ replacing $(\gamma,\lambda)$ for densities.

Use projected gradient with backtracking on the smooth inner objective. Start each inner solve with a positive trial step size. At inner iteration $\ell$, given the current iterate $z$ and trial step size $\delta_\ell$, form $v=\operatorname{Proj}_C(z-\delta_\ell\nabla\mathcal{J}_k(z))$ and halve $\delta_\ell$ until
\begin{align}
\mathcal{J}_k(v)\leq\mathcal{J}_k(z)+\langle\nabla\mathcal{J}_k(z),v-z\rangle+\frac{\|v-z\|^2}{2\delta_\ell}.
\end{align}
Use the accepted $\delta_\ell$ as the initial trial value for $\delta_{\ell+1}$, reducing it further if needed. All norms and inner products use the grid weights. Local Lipschitz continuity on the bounded inner level set ensures acceptance for sufficiently small steps. Strong convexity then gives convergence of the inner projected-gradient sequence. To specify its stopping condition, define at an accepted step
\begin{align}
e=\nabla\mathcal{J}_k(v)-\nabla\mathcal{J}_k(z)+(z-v)/\delta_\ell.
\label{eq:inner_residual_v37}
\end{align}
The projection inequality gives $e\in\nabla\mathcal{J}_k(v)+N_C(v)$. Accept $z^{k+1}=v$ when $\|e\|\leq\varepsilon_k$, where $\varepsilon_k\geq0$ and $\sum_k\varepsilon_k<\infty$. For example, use a positive tolerance proportional to $(k+1)^{-2}$. The residual tends to zero along the inner iterations, so every positive tolerance is reached in finitely many steps. The case $\varepsilon_k=0$ denotes an exact inner solve, not a claim of finite termination of projected gradient. Stop the outer loop only after checking primal feasibility, complementary slackness, and the original Lagrangian projected-gradient residual; a small objective change alone is insufficient.

\subsection{Quantile-Space Projection}
The residuals $\mathcal{M}_i,\mathcal{N}_i$ and the augmented objective are defined in Section~\ref{subsec:Q_projection}. On ordered grid nodes $0<u_1<\cdots<u_n<1$ with quadrature weights $\Delta_j>0$, the discrete projection of an input vector $h=(h_1,\ldots,h_n)$ is
\begin{align*}
\operatorname{Proj}_{\mathcal{Q}_0}(h)=\arg\min_{Q_1\leq\cdots\leq Q_n}\frac{1}{2}\sum_{j=1}^n\Delta_j\mathbin{\cdot}(Q_j-h_j)^2.
\end{align*}
This weighted isotonic regression has a unique solution and preserves monotonicity at every inner iteration. The grid weights also define the inner product used for the discrete gradient.
\begin{algorithm}[H]
\caption{Proximal Method of Multipliers in Quantile Space}
\label{alg:dual_ascent_Q}
\footnotesize
\begin{algorithmic}[1]
\Require A fixed smooth convex discretization with a KKT point; $Q^0\in\mathcal{Q}_0$, $\gamma^0$, $\lambda^0\geq0$, $\tau,\omega>0$, and summable inner tolerances $\varepsilon_k$.
\For{$k=0,1,\ldots$ until the KKT residuals meet the requested tolerances}
  \State Form $\mathcal{J}_k^Q$ from \eqref{eq:Q_augmented_objective}.
  \State Starting at $Q^k$, apply \eqref{eq:Q_projected_step} with weighted isotonic regression and backtracking until \eqref{eq:inner_residual_v37} has norm at most $\varepsilon_k$; set the accepted point to $Q^{k+1}$.
  \State $\gamma_i^{k+1}=\gamma_i^k+\tau\mathcal{M}_i(Q^{k+1})$, $i\in\mathcal{I}_1$.
  \State $\lambda_i^{k+1}=(\lambda_i^k+\tau\mathcal{N}_i(Q^{k+1}))_+$, $i\in\mathcal{I}_2$.
\EndFor
\State \Return $Q^k,\gamma^k,\lambda^k$ and the KKT residuals.
\end{algorithmic}
\end{algorithm}

\subsection{Distribution-Space Projection}\label{subsec:F_projection}
For smooth bounded-support instances of the distribution problem \eqref{prob:F_WCSRM} without an explicit solution, we apply the proximal method of multipliers from Section~\ref{subsec:Q_projection} directly to the density. We work with $L^2$ densities and write
\begin{align*}
\mathsf{D}&\triangleq\left\{f\in L^2[a,b]:f\geq0\text{ a.e.},\ \int_a^bf(x)\,\mathrm{d}x=1\right\},
&F_f(x)&\triangleq\int_a^xf(y)\,\mathrm{d}y,\\
\mathcal{M}_j(f)&\triangleq\int_a^bk_j(x)f(x)\,\mathrm{d}x-c_j,
&\mathcal{N}_j(f)&\triangleq\int_a^bK_j(x,f(x))\,\mathrm{d}x-d_j.
\end{align*}
The minimization objective is $\mathcal{T}(f)=\int_a^b\Phi(F_f(x))\,\mathrm{d}x-b$. Define $\mathcal{J}_k^F$ by replacing $Q$, its objective, its residuals, and $(\gamma^k,\lambda^k)$ in \eqref{eq:Q_augmented_objective} by $f$, $\mathcal{T}$, the density residuals, and $(\eta^k,\zeta^k)$. Fubini's theorem gives its $L^2$ gradient:
\begin{align}
\nabla\mathcal{J}_k^F(f)(y)
&=\int_y^b\phi(F_f(x))\,\mathrm{d}x
+\sum_{j\in\mathcal{I}_1}\big(\eta_j^k+\tau\mathcal{M}_j(f)\big)k_j(y)\notag\\
&\quad+\sum_{j\in\mathcal{I}_2}\big(\zeta_j^k+\tau\mathcal{N}_j(f)\big)_+\left.\frac{\partial K_j(y,z)}{\partial z}\right|_{z=f(y)}
+\omega\mathbin{\cdot}\big(f(y)-f^k(y)\big),
\label{eq:F_density_gradient}
\end{align}
Here $z$ is the density argument of $K_j$. The inner update is
\begin{align}
f^{k,\ell+1}=\operatorname{Proj}_{\mathsf{D}}\big(f^{k,\ell}-\delta_\ell\nabla\mathcal{J}_k^F(f^{k,\ell})\big).
\label{eq:F_projected_step}
\end{align}
This projects a gradient step for the constrained inner problem. Projecting an unconstrained solution of the zero-residual equation need not solve that problem.

For any $h\in L^2[a,b]$, the required projection is
\begin{align}
\operatorname{Proj}_{\mathsf{D}}(h)(x)=(h(x)+\varpi)_+,
\qquad \int_a^b(h(x)+\varpi)_+\,\mathrm{d}x=1,
\label{eq:F_density_projection}
\end{align}
with $\varpi$ obtained by scalar bisection. Every inner iterate is therefore a probability density. After the inner solve, set $F^{k+1}=F_{f^{k+1}}$ and update
\begin{align}
\eta_j^{k+1}&=\eta_j^k+\tau\mathcal{M}_j(f^{k+1}),&&j\in\mathcal{I}_1,\notag\\
\zeta_j^{k+1}&=\big(\zeta_j^k+\tau\mathcal{N}_j(f^{k+1})\big)_+,&&j\in\mathcal{I}_2.
\end{align}
At a joint fixed point, the proximal term disappears and the effective multipliers equal $(\eta,\zeta)$. If $d_f\mathcal{L}$ denotes the density gradient of the original minimization Lagrangian, the fixed-point condition is $-d_f\mathcal{L}\in N_{\mathsf{D}}(f)$. Equivalently, $d_f\mathcal{L}$ equals a constant on $\{f>0\}$ and is at least that constant on $\{f=0\}$, a.e. Under the regularity needed for integration by parts, $(d_f\mathcal{L})'=\rho^*$ as in \eqref{optimal-p-simp}, and
\begin{align*}
\int_a^bd_f\mathcal{L}(y)\mathbin{\cdot}(v(y)-f(y))\,\mathrm{d}y
=-\int_a^b\rho^*(x)\mathbin{\cdot}(F_v(x)-F_f(x))\,\mathrm{d}x.
\end{align*}
Thus this is precisely the CDF normal-cone condition, together with feasibility and complementary slackness. It does not require every zero-density interval to be a flat region at CDF level $0$ or $1$.

The ordinary projected-gradient inner solver does not cover the singular entropy derivative at zero density or the kink of the ES spectrum without additional treatment; the KL constructions in Sections~\ref{PRRM-case-KL} and~\ref{PRRM-case-KL-Gaussian} are certified analytically and do not rely on this algorithm.

On a fixed grid, use the discrete negative-SRM objective and its density gradient to form $\mathcal{J}_k^F$. In particular, the adjoint of the discrete cumulative-sum operator supplies the first term in the discrete counterpart of \eqref{eq:F_density_gradient}. Algorithm~\ref{alg:dual_ascent_F} combines these updates with the backtracking and inner stopping rule specified above.
\begin{algorithm}[H]
\caption{Proximal Method of Multipliers in Density Space}
\label{alg:dual_ascent_F}
\footnotesize
\begin{algorithmic}[1]
\Require A fixed smooth convex discretization with a KKT point; $f^0\in\mathsf{D}$, $\eta^0$, $\zeta^0\geq0$, $\tau,\omega>0$, and summable inner tolerances $\varepsilon_k$.
\For{$k=0,1,\ldots$ until the KKT residuals meet the requested tolerances}
  \State Form $\mathcal{J}_k^F$ from the density objective and residuals defined above.
  \State Starting at $f^k$, apply \eqref{eq:F_projected_step} with weighted shift-and-truncate projection \eqref{eq:F_density_projection} and backtracking until \eqref{eq:inner_residual_v37} has norm at most $\varepsilon_k$; set the accepted point to $f^{k+1}$.
  \State Set $F^{k+1}=F_{f^{k+1}}$ using the same discrete integration operator as in the objective.
  \State $\eta_j^{k+1}=\eta_j^k+\tau\mathcal{M}_j(f^{k+1})$, $j\in\mathcal{I}_1$.
  \State $\zeta_j^{k+1}=(\zeta_j^k+\tau\mathcal{N}_j(f^{k+1}))_+$, $j\in\mathcal{I}_2$.
\EndFor
\State \Return $F^k,f^k,\eta^k,\zeta^k$ and the KKT residuals.
\end{algorithmic}
\end{algorithm}

\subsection{Convergence on a Fixed Grid}\label{app:fixed-grid-convergence}

We prove the fixed-grid convergence claim of Section~\ref{subsec:Q_projection} for Algorithm~\ref{alg:dual_ascent_Q} and its density counterpart, Algorithm~\ref{alg:dual_ascent_F}, under the assumptions stated at the start of this appendix. Let $\mathbf{z}=(z,\gamma,\lambda)$, and define the KKT operator, with adjoints taken in the chosen weighted inner products,
\begin{align*}
\mathcal{A}(\mathbf{z})=\begin{pmatrix}
\nabla\mathcal{T}(z)+\mathcal{M}'{}^\top\gamma+\sum_i\lambda_i\nabla\mathcal{N}_i(z)+N_C(z)\\
-\mathcal{M}(z)\\
-\mathcal{N}(z)+N_{\mathbb{R}_+^n}(\lambda)
\end{pmatrix}.
\end{align*}
Convexity and nonnegative inequality multipliers make this operator monotone. For an inequality term, the two-point monotonicity expression is a sum of two nonnegative convexity gaps weighted by the two multipliers; the affine equality terms cancel. Its graph is closed under the stated smoothness assumptions. The inner stopping rule and outer updates imply
\begin{align*}
(e_k,0,0)-D\mathbin{\cdot}(\mathbf{z}^{k+1}-\mathbf{z}^k)\in\mathcal{A}(\mathbf{z}^{k+1}),
\qquad D=\operatorname{diag}(\omega I,\tau^{-1}I,\tau^{-1}I),
\quad\|e_k\|\leq\varepsilon_k.
\end{align*}
For any KKT point $\mathbf{z}^*$, monotonicity therefore yields
\begin{align*}
\|\mathbf{z}^{k+1}-\mathbf{z}^*\|_D^2
\leq\|\mathbf{z}^k-\mathbf{z}^*\|_D^2-\|\mathbf{z}^{k+1}-\mathbf{z}^k\|_D^2
+2\langle e_k,z^{k+1}-z^*\rangle.
\end{align*}
Cauchy--Schwarz first bounds the distance increase by $2\varepsilon_k/\sqrt{\omega}$; summability makes the iterates bounded. Summing the displayed inequality then gives $\mathbf{z}^{k+1}-\mathbf{z}^k\to0$. Since $e_k\to0$, every cluster point solves the KKT system by closedness of the graph. The distance to each KKT point has a limit by the same summable-error inequality, so the entire finite-dimensional sequence converges to a KKT point. This proves convergence on the chosen grid, not consistency as the grid is refined. In density space, entropy terms at zero and kinks of $\Phi$, as with an ES step spectrum, fall outside the stated smoothness assumptions and require a different inner solver or a separately justified smoothing scheme. In quantile space, the spectral objective is linear in $Q$, so jumps in $\phi$ do not create this difficulty; the constraint functionals must still satisfy the stated smoothness assumptions.

\subsection{Numerical Checks}

The following tests check the implementations of Algorithms~\ref{alg:dual_ascent_Q} and~\ref{alg:dual_ascent_F} on smooth discretizations of the quantile and distribution problems. The quantile test permits comparison with an explicit discrete optimizer; the density test checks feasibility and stationarity. For both tests, take $\phi(u)=2u$, $\tau=1$, $\omega=0.5$, zero initial multipliers, and inner tolerances $2\times10^{-3}(k+1)^{-1.6}$. The quantile test has mean zero and variance at most one, starts at $Q^0=0$, and uses 80 equal-weight midpoint nodes. Its discrete optimizer is the centered, variance-normalized spectrum; the computed quantile differs from it by at most $1.82\times10^{-7}$ at a node. The density test uses $[a,b]=[0,1]$, mean $1/2$, $\int_0^1f^2\leq2$, and $f^0=1$. With cell width $h=1/n$, it evaluates $F_i=h\mathbin{\cdot}(\sum_{j<i}f_j+f_i/2)$ and $\mathcal{T}_h=h\sum_iF_i^2-1$, and differentiates this same discrete objective. All runs stop when feasibility, complementary slackness, and the unit-step projected-gradient residual are below $3\times10^{-7}$.
\begin{table}[H]
\centering\small
\caption{Checks of the smooth-grid procedures from zero initial multipliers.}
\label{tab:projection-check}
\begin{tabular}{lrrrr}
\toprule
Model & Grid size & Outer iterations & Risk value & Stationarity residual\\
\midrule
Quantile & 80 & 60 & 0.57730534 & $2.94\times10^{-7}$\\
Density & 80 & 261 & 0.71870715 & $2.99\times10^{-7}$\\
Density & 160 & 261 & 0.71870693 & $2.98\times10^{-7}$\\
\bottomrule
\end{tabular}
\end{table}
For comparison, the feasible continuous density $1+\cos(2\pi x)$ has risk $2/3+3/(8\pi^2)\approx0.70466$, whereas the uniform density has risk $2/3$. At the uniform density, the unaugmented objective gradient is $1-y^2$, which is not constant; the corrected density update therefore does not leave this nonoptimal density fixed. The two grid values in Table~\ref{tab:projection-check} are numerical checks, not a grid-refinement theorem.

\section{Proofs for the Central- and Geometric-Moment Models}\label{appendix:analytical_examples}

This appendix verifies the two moment-model characterizations stated in Sections~\ref{subsec:central_moment} and~\ref{subsec:geometric_moment}. The central-moment proof uses pointwise stationarity in quantile space; the geometric-moment proof verifies calibration, support inclusion, and the CDF boundary signs.

\subsection{Proof of Theorem~\ref{theo-1,k}}\label{proof_theo_1_k}

We verify the assumptions and derive the closed-form solution for the ambiguity set $\mathcal{Q}_{1\kappa}$ in \eqref{eq:Q_1kappa}. Assume that $m_\kappa>0$. In this proof, the underlying quantile space is $L^\kappa[0,1]$. The constant quantile $\bar{Q}(u)\equiv\mu$ is feasible: it is nondecreasing, has mean $\mu$, and has zero central moment.
Since $m_\kappa>0$, the same function also satisfies the central moment inequality strictly, and hence gives the Slater-type condition for the inequality constraint.
The equality qualification also holds: the constant quantiles $Q_\delta(u)\equiv\mu+\delta$ satisfy the moment inequality strictly whenever $|\delta|^\kappa<m_\kappa$, so their attainable means contain an open neighborhood of $\mu$.

The weak compactness follows from the central moment bound. For every feasible $Q\in\mathcal{Q}_{1\kappa}$, $|Q(u)|^\kappa\leq2^{\kappa-1}\big(|Q(u)-\mu|^\kappa+|\mu|^\kappa\big)$,
and therefore $\|Q\|_\kappa^\kappa\leq 2^{\kappa-1}(m_\kappa+|\mu|^\kappa)$.

Thus $\mathcal{Q}_{1\kappa}$ is bounded in $L^\kappa[0,1]$. Since $1<\kappa<\infty$, the space $L^\kappa[0,1]$ is reflexive. The mean constraint is closed and affine, the central moment sublevel set is closed and convex, and $\mathcal{Q}_0$ is a closed convex cone in $L^\kappa[0,1]$. Hence all these sets are weakly closed, and $\mathcal{Q}_{1\kappa}$ is weakly compact.

The differentiability condition also holds: for $H_1(u,y)=|y-\mu|^\kappa$, we have $\partial H_1(u,y)/\partial y=\kappa |y-\mu|^{\kappa-1}\mathrm{sgn}(y-\mu)$.
Let $q=\kappa/(\kappa-1)$. For any feasible $Q$, the $q$th power of the derivative's $L^q$ norm is $\kappa^q\int_0^1|Q(u)-\mu|^\kappa\,\mathrm{d}u\leq\kappa^q m_\kappa<\infty$. Thus $u\mapsto \left.(\partial H_1(u,y)/\partial y)\right|_{y=Q(u)}$ belongs to $L^q[0,1]$. Moreover, $|H_1(u,y)|\leq C(1+|y|^\kappa)$ and $|\partial H_1(u,y)/\partial y|\leq C(1+|y|^{\kappa-1})$ for a constant $C$ depending only on $\mu$ and $\kappa$. These bounds verify the growth and first-variation requirements of Assumption~\ref{assump:Q_dual}(ii). Thus both Assumptions~\ref{assump:Q_regular} and~\ref{assump:Q_dual} hold for $\mathcal{Q}_{1\kappa}$.

We now construct a candidate using Corollary~\ref{corollary-normal-QRRM}. With mean multiplier $\gamma_1\in\mathbb{R}$ and central-moment multiplier $\lambda_1\geq0$, pointwise stationarity requires 
\begin{align*}
\phi(u)-\gamma_1-\kappa\lambda_1|Q(u)-\mu|^{\kappa-1}\mathrm{sgn}\big(Q(u)-\mu\big)=0
\end{align*}
 a.e. Since $\phi$ is not a.e.\ constant, $\lambda_1=0$ is impossible. Solving for $Q$ gives the quantile formula in the theorem, with $(\gamma_1,\lambda_1)$ in place of the calibrated multipliers.

The mean constraint is then precisely the centering equation stated in the theorem.
The left-hand side is continuous in $\gamma_1$ (by dominated convergence, using $\phi\in L^{\kappa/(\kappa-1)}[0,1]$) and strictly decreasing, with limits $+\infty$ and $-\infty$ as $\gamma_1\to-\infty$ and $\gamma_1\to+\infty$, respectively. Hence the equation admits a unique solution, denoted by $\gamma_1^*$.

Since $\lambda_1>0$, complementary slackness requires $\int_0^1|Q(u)-\mu|^\kappa\,\mathrm{d}u=m_\kappa$. Substituting the quantile formula gives 
\begin{align*}
(\kappa\lambda_1)^{-\kappa/(\kappa-1)}\int_0^1|\phi(u)-\gamma_1^*|^{\kappa/(\kappa-1)}\,\mathrm{d}u=m_\kappa,
\end{align*}
 which determines $\lambda_1^*>0$ and hence $Q^*$ as stated in Theorem~\ref{theo-1,k}.

It remains to verify admissibility and compute the value. Since $\phi$ is nondecreasing and the map $t\mapsto |t|^{1/(\kappa-1)}\mathrm{sgn}(t)$ is increasing, the resulting $Q^*$ is nondecreasing a.e. Hence $Q^*\in\mathcal{Q}_0$. Together with the mean constraint and the active central moment constraint verified above, this shows that $Q^*$ is feasible for $\mathcal{Q}_{1\kappa}$. Corollary~\ref{corollary-normal-QRRM} then implies that $Q^*$ is optimal.

Finally, using the defining equation of $\gamma_1^*$, we have
\begin{align*}
\int_0^1\phi(u)|\phi(u)-\gamma_1^*|^{\frac{1}{\kappa-1}}\mathrm{sgn}\big(\phi(u)-\gamma_1^*\big)\,\mathrm{d}u
=\int_0^1|\phi(u)-\gamma_1^*|^{\frac{\kappa}{\kappa-1}}\,\mathrm{d}u.
\end{align*}
Together with $\int_0^1\phi(u)\,\mathrm{d}u=1$ and the expression for $\lambda_1^*$, this gives the stated worst-case value.
To prove uniqueness, note that the constructed certificate has $\lambda_1^*>0$. Strong duality therefore forces every optimizer to exhaust the moment budget. If two distinct optimizers existed, their average would remain feasible and optimal, whereas strict convexity of the $L^\kappa$ moment functional would give the average a strictly smaller central moment. This contradicts binding of the moment budget at every optimizer. This proves Theorem~\ref{theo-1,k}.
\Halmos

\subsection{Proofs for the Geometric-Moment Model}\label{appendix:geometric_moment}

We prove the feasibility statement in Lemma~\ref{lem:nonempty_geometric} and the conditional optimality result in Theorem~\ref{theo-log}, for the model in Section~\ref{subsec:geometric_moment}.

\textbf{\emph{Proof of Lemma~\ref{lem:nonempty_geometric}.}} For any probability density with mean $\mu$, Jensen's inequality gives $\int_a^b\ln(w_0x)f(x)\,\mathrm{d}x\leq\ln(w_0\mu)$.
Equality requires a point mass at $\mu$, so the inequality is strict for every absolutely continuous law; consequently, $c<\ln(w_0\mu)$ is necessary. Conversely, take the uniform density on $[\mu-\delta,\mu+\delta]\subset(a,b)$. It has mean $\mu$, belongs to $L^r[a,b]$ for every finite $r$, and its logarithmic moment tends to $\ln(w_0\mu)$ as $\delta\downarrow0$. For $c<\ln(w_0\mu)$, choosing $\delta$ small enough therefore gives a feasible density with strict slack. Small translations of this interval preserve the strict inequality and the support inclusion while varying the mean on both sides of $\mu$, which verifies the equality part of the relative Slater condition. As $\delta\downarrow0$, these densities approach the Jensen bound, although no density attains it.
\Halmos

\textbf{\emph{Proof of Theorem~\ref{theo-log}.}} Writing the geometric-moment lower bound as an integral inequality gives the kernel $\ell(x)=-\ln(w_0x)$, with $\ell'(x)=-1/x$ on $[a,b]\subset(0,\infty)$. The residual is therefore $\rho(x)=-\phi(F(x))+\eta_1-\zeta_1/x$. For $\eta_1>\phi_1$ and $\zeta_1>0$, the stationary candidate in \eqref{eq:geometric_stationary_shape} has the increasing quantile $Q(u)=\zeta_1/(\eta_1-\phi(u))$, $0\leq u\leq1$, whose endpoint values give the stated support.

The identity 
\begin{align*}
\int_0^1[\eta_1-\phi(u)]^{-1}\,\mathrm{d}u=\mathsf{m}(\eta_1)/(\nu\eta_1)
\end{align*}
 reduces the mean constraint to $\zeta_1=\nu\mu\eta_1/\mathsf{m}(\eta_1)$. The binding geometric moment requires 
\begin{align*}
c=\ln(w_0\zeta_1)-\int_0^1\ln(\eta_1-\phi(u))\,\mathrm{d}u.
\end{align*}
 Since $\phi(u)=\phi_0e^{\nu u}$ and $0<\phi_0/\eta_1<\phi_1/\eta_1<1$, the substitution $z=\phi(u)/\eta_1$ gives
\begin{align*}
\int_0^1\ln(\eta_1-\phi(u))\,\mathrm{d}u
=\ln\eta_1-\frac{1}{\nu}\left(\mathrm{Li}_2\left(\frac{\phi_1}{\eta_1}\right)-\mathrm{Li}_2\left(\frac{\phi_0}{\eta_1}\right)\right).
\end{align*}
Substitution of the mean relation yields \eqref{eq:geo_eta_append}. Its left-hand side is continuous on $(\phi_1,\infty)$, tends to $-\infty$ as $\eta_1\downarrow\phi_1$, and tends to $\ln(w_0\mu)$ as $\eta_1\to\infty$. It therefore attains every $c<\ln(w_0\mu)$.

Choose such a root $\eta_1^*$ and define $\zeta_1^*$, the endpoints, and $F^*$ as in the theorem. This construction gives a continuous nondecreasing CDF with the required mean and binding geometric moment. On its support, $f^*(x)=\zeta_1^*/[\nu x^2\mathbin{\cdot}(\eta_1^*-\zeta_1^*/x)]$ is bounded because $x\geq x_L^*>0$ and $\eta_1^*-\zeta_1^*/x\geq\phi_0>0$. Extended by zero, the density belongs to every finite $L^r[a,b]$. Thus, when the support is contained in $[a,b]$, $F^*$ is feasible in the stated Sobolev class.

The map $x\mapsto\eta_1^*-\zeta_1^*/x$ is increasing and equals $\phi(0)$ at $x_L^*$ and $\phi(1)$ at $x_U^*$. Consequently, $\rho(x)\leq0$ on the left flat region and $\rho(x)\geq0$ on the right. Together with interior stationarity and complementary slackness, these signs give optimality by Corollary~\ref{corollary-normal-PRRM}. Strict increase of the spectrum gives uniqueness, so every calibration root satisfying the support condition produces the same optimal CDF.

Finally, integrating $\phi(u)Q^*(u)$ with $z=\phi(u)$ and $\mathrm{d}z=\nu z\,\mathrm{d}u$ gives
\begin{align*}
\mathcal{R}_{\phi}(\mathcal{F}_{1\ln})
=\frac{\zeta_1^*}{\nu}\int_{\phi_0}^{\phi_1}\frac{1}{\eta_1^*-z}\,\mathrm{d}z
=\frac{\zeta_1^*}{\nu}\ln\frac{x_U^*}{x_L^*}.
\end{align*}
\Halmos

\clearpage
\bibliographystyle{abbrvnat}
\bibliography{wc-srm}
\end{document}